\documentclass[a4paper,10pt]{article}
\usepackage[utf8]{inputenc}
\usepackage{jheppubMod}
\usepackage{amsmath,amsfonts,amssymb,graphicx,MnSymbol,subfigure}
\usepackage{hyperref} 
\usepackage{tikz}
\usetikzlibrary{shapes.geometric, arrows}
\usepackage[toc,page]{appendix}
\usepackage{verbatim}

\title{Hidden valley scenario sensitivity in the CMS muon endcap detector}

\abstract{We study the sensitivity of the CMS search to displaced showers arising from the decays of long-lived particles in the muon system, within the framework of Hidden Valley scenarios. To establish our simulation setup, we employ a parameterization of Hidden Valley theory space and adopt a hybrid strategy where the lifetime is treated as a free parameter to provide model-independent Hidden Valley quark production cross-section upper limits. Our results indicate that the CMS search is broadly sensitive to variations in Hidden Valley parameters such as the overall scale and relevant mass ratios, while showing comparatively weak dependence on the number of Hidden Valley colors or flavors. The exact quantitative results we derive depend on the underlying hadronization model employed and thus, we urge caution in interpreting the results. Within these limitations, we also establish model-dependent limits on the maximum strength of the mediator coupling to Hidden Valley quarks. Our approach illustrates how one can move beyond simplified model strategies by employing a first-principles-inspired theory setup, while highlighting the theoretical uncertainties inherent in modeling Hidden Valley phenomenology.}
\author[a]{Wei Liu,}
\author[b]{Joshua Lockyer,}
\author[b]{Suchita Kulkarni}
\affiliation[a]{Department of Applied Physics and MIIT Key Laboratory of Semiconductor Microstructure and Quantum Sensing, Nanjing University of Science and Technology, Nanjing 210094, China}
\affiliation[b]{Institute of Physics, NAWI Graz, University of Graz, Universit\"atsplatz 5, A-8010 Graz, Austria}
\emailAdd{wei.liu@njust.edu.cn}
\emailAdd{joshua.lockyer@uni-graz.at}
\emailAdd{suchita.kulkarni@uni-graz.at}

\date{March 2024}

\newcommand{\newc}{\newcommand}

\newcommand{\nc}{{N_C}}                
\newcommand{\nf}{{N_F}}                
\newcommand{\mpl}{m_{\pi}/\Lambda}               
\newcommand{\fc}{{\nf/\nc}}                
\newcommand{\pid}{{\pi}}                
\newcommand{\pidon}{{\pi^{0}}}                
\newcommand{\pidoff}{{\pi^{\pm}}}                
\newcommand{\rhod}{{\rho}}                
\newcommand{\rhodon}{{\rho^{0}}}                
\newcommand{\rhodoff}{{\rho^{\pm}}}                
\newcommand{\mzp}{m_{Z_D}}
\newcommand{\ld}{\Lambda}

\newc{\UV}{{\mathchoice{}{}{\scriptscriptstyle}{}UV}}
\newc{\IR}{{\mathchoice{}{}{\scriptscriptstyle}{} IR}}

\newc{\ntinit}{N_{\rm{tot}}^{0}}
\newc{\ntfin}{N_{\rm{tot}}}

\newc{\npinit}{N_{\pid}^{0}}
\newc{\nponinit}{N_{\pidon}^{0}}
\newc{\npoffinit}{N_{\pidoff}^{0}}
\newc{\nrinit}{N_{\rhod}^{0}}
\newc{\nroninit}{N_{\rhodon}^{0}}
\newc{\nroffinit}{N_{\rhodoff}^{0}}

\newc{\npfin}{N_{\pid}}
\newc{\nponfin}{N_{\pidon}}
\newc{\npofffin}{N_{\pidoff}}
\newc{\nrfin}{N_{\rhod}}
\newc{\nronfin}{N_{\rhodon}}
\newc{\nrofffin}{N_{\rhodoff}}

\newc{\sk}[1]{\textcolor{blue}{#1}}
\newc{\skfw}[1]{\textcolor{green}{(For Wei: #1)}}
\newc{\skfj}[1]{\textcolor{teal}{(For Joshua: #1)}}

\newc{\wl}[1]{\textcolor{magenta}{(Wei: #1)}}
\newc{\jl}[1]{\textcolor{blue}{(Joshua: #1)}}

\begin{document}
\maketitle

\section{Introduction}
Beyond the Standard Model (SM) scenarios featuring a new non-Abelian gauge group in addition to the SM contents are theoretically, phenomenologically and experimentally interesting possibilities. These extensions are weakly coupled to the SM and, in the chirally-broken regime, present bound states resulting from confinement within the non-Abelian gauge group. Their origin lies in the Hidden Valley~(HV) setup~\cite{Strassler:2006im, Strassler:2006qa}, but they also occur in the context of Twin Higgs and its variations~\cite{Chacko:2005pe,Craig:2015pha, Craig:2014aea, Craig:2016kue, Craig:2014roa}, and have been shown to contain interesting dark matter phenomenology, see e.g.~\cite{Butterworth:2021jto,Pomper:2024otb,Bernreuther:2023kcg, Zierler:2022uez, Berlin:2018tvf, Choi:2018iit,Hochberg:2015vrg,Katz:2020ywn,Hur:2007uz,Hur:2011sv,Frandsen:2011kt,Buckley:2012ky,Hambye:2013dgv,Bai:2013iqa,Bai:2013xga,Cline:2013zca,Boddy:2014yra,Antipin:2015xia,GarciaGarcia:2015fol,Dienes:2016vei,Lonsdale:2017mzg,Davoudiasl:2017zws,Choi:2018iit,Khlopov:2010pq}. These HV scenarios are often also known as ``interacting'' dark sector (DS) scenarios and will thus be referred to as HV/DS scenarios throughout this work. 

A particularly fascinating HV/DS experimental signature arises if the energy of the probe is high compared to the scale of chiral symmetry breaking. This can be realized at high energy experiments such as the LHC, where the production of HV/DS quarks via the SM -- HV/DS portal is followed by a rapid HV/DS parton shower and hadronization process which leads to the formation of bound states. Decays of some of these bound states via the portal produces visible final states similar to the SM QCD jets~\cite{Strassler:2006im, Strassler:2006qa}. Such HV/DS jets can however be characteristically very different than the traditional SM QCD jets and are collectively known as anomalous jets or darkshowers. The corresponding experimental signatures include semi-visible jets~\cite{Cohen:2015toa,Cohen:2017pzm,Beauchesne:2017yhh}, lepton-jets~\cite{ArkaniHamed:2008qp,Baumgart:2009tn,Chan:2011aa,Buschmann:2015awa}, emerging jets~\cite{Schwaller:2015gea,Renner:2018fhh} and also contain some extreme signatures such as soft-unclustered energy patters~\cite{Strassler:2008bv, Hofman:2008ar,Hatta:2008tx,Knapen:2016hky,Harnik:2008ax}. Results from the first experimental searches for semi-visible jets~\cite{CMS:2021dzg,ATLAS:2023swa,ATLAS:2023kao,ATLAS:2025kuz}, emerging jets~\cite{CMS:2018bvr, CMS:2024rea,CMS:2024gxp,ATLAS:2025bsz}, displaced dimuons~\cite{CMS-PAS-EXO-24-008} and soft-unclustered energy patterns~\cite{CMS:2024nca} are also available. For a review on strongly-coupled theories see e.g.~\cite{Albouy:2022cin,Kribs:2016cew,Cacciapaglia:2020kgq}. 

A characteristic feature of HV/DS scenarios is their suppressed coupling with the SM particles. Due to this suppressed coupling, one or more HV/DS mesons can be long-lived, featuring displaced vertex like signatures due to their decays. Generically, such long-lived mesons can arise in a variety of models~\cite{Strassler:2006ri, Han:2007ae, Schwaller:2015gea, Renner:2018fhh, Bernreuther:2022jlj, Carmona:2024tkg, Cheng:2024hvq, Cheng:2021kjg, Born:2023vll} and the corresponding LHC searches are being carried out in the form of emerging jets signatures~\cite{CMS:2018bvr, CMS:2024rea,CMS:2024gxp}. These searches are most appropriate for small meson lifetimes where meson decays are within the tracker ($c\tau^{\rm lab}_{\rm LLP} \sim 10\, \textrm{--}\, 100\,\rm{mm}$).

The theoretical setup often used to devise and search for these anomalous signatures is a ``simplified-model'' parametrization of the HV/DS sector,  which is also adopted in phenomenological studies. Given the multidimensional HV/DS parameter space such an approach may be justified. In this simplified-model parametrization, the HV/DS parameters such as the masses of bound states are often fixed. The results are then presented, e.g., as a function of the mediator mass and the ``invisible meson fraction'' ($r_{\rm inv}$) for prompt final states, or as a function of the HV/DS meson lifetime and the mediator mass for emerging jet search results. Some examples of these strategies can be found in \cite{Cohen:2015toa,CMS:2021dzg,ATLAS:2023swa,ATLAS:2023kao,Bernreuther:2020vhm,Bernreuther:2019pfb,Lu:2023gjk} for prompt decays and in ~\cite{CMS:2018bvr, CMS:2024rea,Schwaller:2015gea} for displaced decays. 

Fixing the HV/DS parameters often independently of each other however neglects their complex interplay, which may affect the search sensitivity or optimization strategies. Analyzing such an interplay also has the possibility to further our understanding of viable HV/DS characteristics, which will help model building efforts. Finally, it may also shed light on the non-trivial question of setting hadronization parameters or at the very least the effect the lack of knowledge of them has on HV/DS searches. 

With this strategy in mind, in this work, we go beyond the existing phenomenological studies and experimental search (re-)interpretations in a few different ways. First, we consider scenarios where the HV/DS mesons are extremely long-lived. In such situations, reconstruction of a displaced vertex, specifically for a hadronically decaying long-lived meson, is challenging. We therefore reinterpret the CMS displaced shower search~\cite{CMS:2021juv} for decays of long-lived particles (LLP) in the muon system. These decays create a cluster of visible hits in the cathode strip chambers (CSC) and are appropriately called CSC clusters, which are considered a smoking gun signature of the model of interest. Typically, such a search is sensitive to meter scale LLP lifetimes in the lab-frame ($c\tau^{\rm lab}_{\rm LLP} = \mathcal{O} (\textrm{few meters})$). In a variety of studies~\cite{Cottin:2022nwp, Liu:2024fey, Baumgart:2009tn, Fitzpatrick:2023xks, Haisch:2023rqs, Bhattacherjee:2025pxg}, it has been shown that this search leads to powerful limits for a range of new physics scenarios, making it an attractive candidate for our HV/DS studies.

Second, rather than constraining the properties of the mediator sector, we aim to learn possible characteristics of the HV/DS sector via our reinterpretation. We demonstrate for the first time a phenomenological sensitivity of the displaced‐shower search to HV/DS parameters, with the caveat that hadronization modeling uncertainties limit direct translation into underlying theory parameters. While these sensitivity studies can help us map out viable regions of Hidden Valley parameter space, they do not guarantee that fundamental theory parameters can be uniquely extracted, given the freedom to retune hadronization to mimic many different dark sector behaviors.

Third, although the CMS muon system search has been reinterpreted in the context of some HV/DS scenarios~\cite{CMS:2024bvl, Mitridate:2023tbj}, within our work, we use theoretically guided wisdom to set the HV/DS parameters\footnote{In addition, ref.\cite{Mitridate:2023tbj} uses $\nf = 1$ scenario -- a choice we specifically do not consider as it does not lead to chiral symmetry breaking and hence is outside the scope of {\tt PYTHIA} simulations.}. Such theoretically guided investigations, coupled together with our aim of constraining the HV/DS parameters paves the way for a systematic experimental exploration of these theories. 

Finally, we adopt a model agnostic approach in which we do not consider exclusions derived by a specific choice of HV/DS flavor symmetry breaking pattern or coupling values but rather provide cross-section upper-limits which are model independent and can be used for constraining a class of different models. We comment on possible theory scenarios where our limits are applicable and justify the use our parameterizations. Finally, we provide model dependent results, and detail their limitations. Our approach and our results make important observations about the search sensitivity to long-lived meson characteristics and therefore underlying theory space. We also closely observe the effect of having multiple long-lived mesons on the search sensitivity. 

The plan of the paper is the following, after introducing the theoretical models and discussing some LLP characteristics in section~\ref{sec:models}, in section~\ref{sec:search_details} we discuss the details of the CMS search used in this work. In section~\ref{sec:eff_dependence}, we detail the variation of the efficiency as a function of the HV/DS theory parameters and in section~\ref{sec:sensitivity} we demonstrate the search sensitivity, before concluding in section~\ref{sec:conclusion}. We provide a number of helpful appendices detailing more technical discussions concerning HV/DS theory space and simulation results.
\section{Dark Shower Models}
\label{sec:models}
We will consider models where HV/DS pions are unstable and hence are the LLPs of interest to us. There are a number of UV complete HV/DS models, which feature unstable pions~\cite{Cheng:2021kjg,Cheng:2024aco,Cheng:2024hvq}. Our aim here is to sketch a simple HV/DS model which can justify our choice of free parameters. We do not intend to work out entire model phenomenology, but rather provide a template model. 

The broad class of HV/DS models we consider exhibit a $SU(\nc)\times U(1)_D$ extension of the SM. These models contain Dirac fermions in the fundamental representation and feature a $SU(\nf)_L\times SU(\nf)_R$ global flavor symmetry in the chirally-unbroken phase. In the chirally-broken phase the leftover symmetry is the diagonal subgroup of the global flavor group denoted by $SU(\nf)_V$. The HV/DS pions and rho mesons are in the adjoint representation of this $SU(\nf)_V$ group and hence are arranged in $N^2_F - 1$ multiplets.

The $U(1)_D$ group plays a two-fold role here. First, the Abelian Higgs mechanism associated with the breaking of $U(1)_D$ symmetry gives mass to the $U(1)_D$ gauge boson $Z_D$ as well as the chiral HV/DS quarks. Second, the $Z_D$ serves as a mediator between the HV/DS quarks and the SM matter content, giving rise to $pp \to Z_D \to q_D \bar{q}_D$ processes which of of interest to us. Furthermore, we assume a leptophobic $Z_D$, in line with the HV/DS literature considered elsewhere, see e.g.~\cite{Albouy:2022cin} where an overview of such models is presented.

For these class of models, the relevant terms in the Lagrangian are
\begin{eqnarray}
    \mathcal{L} &\supset& \mathcal{Q}_V^{\rm SM} \kappa_D   Z^\mu_D \bar{q} \gamma_\mu q + \mathcal{Q}_A^{\rm SM} \kappa_D   Z^\mu_D \bar{q} \gamma_\mu\gamma_5 q + \mathcal{Q}_V^D  \kappa_{\rm D}  Z^\mu_D \bar{q}_D \gamma_\mu q_D +\mathcal{Q}_A^D  \kappa_{\rm D}  Z^\mu_D \bar{q}_D \gamma_\mu\gamma_5 q_D \nonumber\\ 
    &+& Y_{ij} \phi_D \bar{q}_D q_D + (D_\mu \phi)^\dagger (D^\mu \phi),
\label{eq:lagrangian}
\end{eqnarray}

where $\mathcal{Q}_{V,A}^D$ and $\mathcal{Q}_{V,A}^{\rm SM}$ are $\nf\times \nf$ and $6\times 6$ the dark and SM $U(1)_D$ vector and axial-vector charge matrices respectively, $Y_{ij}$ is the dark quark Yukawa matrix associated with dark Higgs ($\phi$). Both the charge and Yukawa matrices are assumed to be real and diagonal for simplicity. $D_\mu$ is the appropriate $U(1)_D$ covariant derivative.  $\kappa_D$ is the $U(1)_D$ gauge coupling. 

In the chirally--broken phase, the dark sector contains $N_F^2$ pions, of which $\nf$ are diagonal ($\pi^0_{D,i}$) and $\nf(\nf-1)$ are off--diagonal ($\pi^\pm_{D,ij}$). In the absence of explicit breaking terms, all of these states would be stable pseudo--Nambu--Goldstone bosons. In practice, however, several ingredients of the Lagrangian in Eq.~\eqref{eq:lagrangian} break the global $SU(\nf)_L \times SU(\nf)_R$ flavour symmetry and govern which pions are unstable:

\begin{itemize}
    \item \emph{Yukawa couplings} If the Yukawa matrix $Y_{ij}$ is not proportional to the identity, dark quark masses are non--degenerate. This lifts the degeneracy of the diagonal pions and induces their mixing with the longitudinal component of the $U(1)_D$ gauge boson ($Z_D$). As a result, diagonal pions can decay into SM fermion pairs. 

    \item \emph{Charge assignments} Non--universal entries in $\mathcal{Q}^{D}_{V, A}$ lead directly to symmetry breaking. If only non--universal vector charges are considered, the pions can decay via $\pi_D \to Z^\prime Z^\prime$ couplings. The resulting lifetime will likely be very long given the heavy  $Z^\prime$. Non--universal entries in $\mathcal{Q}^{D}_A$ will lead to $\pi_D$ mixing with the longitudinal component of the $Z^\prime$, and decay via two body final state, which leads to much smaller lifetime. For $t^{\rm th}$ diagonal pion, this decay is proportional to ${\rm Tr}[(\mathcal{Q}^D_A)\cdot T_i]$. Thus if both $\mathcal{Q}^{D}_{V, A}$ matrices were to be diagonal, the pions would remain stable. 

    \item \emph{Kinetic mixing} While kinetic mixing $\kappa_D$ does not lead to any flavour symmetry breaking, it connects the dark quarks to SM fermions through $Z_D$. In combination with flavour breaking, this provides the physical channel by which unstable pions decay to SM final states. 
\end{itemize}

The flavour symmetry breaking patterns can be understood as follows. For an $SU(\nc)$ gauge theory with $\nf$ light flavors, the global chiral symmetry in the massless limit is $SU(\nf)_L \times SU(\nf)_R \to SU(\nf)_V$, leading to the usual $\nf^2 - 1$ pions.  If the dark quark masses are not degenerate, the explicit mass terms break $SU(\nf)_V$ further. As a rule of thumb, for every flavor whose mass is distinct from the others, the unbroken subgroup reduces by one rank: $SU(\nf)_V \to SU(\nf-1)_V \times U(1)$, for a single heavy or non-degenerate flavor. More generally, if the $\nf$ flavors are split into groups of equal mass, the residual flavor symmetry is $SU(N_{f_1}) \times SU(N_{f_2}) \times \cdots \times U(1)^{k-1}$, where $N_{f_1} + N_{f_2} + \cdots = \nf$, and $k$ is the number of distinct mass values. The pattern of pion multiplets then reorganizes into irreps of this subgroup, with only the exact multiplets remaining mass-degenerate. The off-diagonal pions are stable under the residual $U(1)$ symmetry. 

A similar logic applies when the $U(1)_D$ charges of the dark quarks are not universal. In the charge-degenerate case, the pions are exact eigenstates of the full $SU(\nf)_V$. Non-universal charge assignments explicitly break $SU(\nf)_V$ down to the subgroup that commutes with the charge matrix. For example, if one flavor carries a different $U(1)_D$ charge, the residual symmetry is again $SU(\nf)_V \to SU(\nf-1)_V \times U(1)$, while more general charge splittings decompose $SU(\nf)_V$ into products of $SU(N_{f_i})$ factors with additional $U(1)$'s, exactly paralleling the case of non-degenerate masses. The stability of the diagonal pions is therefore directly tied to whether the corresponding flavor subgroup survives the breaking. 

The resulting lifetime of a diagonal dark pion can be estimated by analogy with the SM charged pion:
\begin{equation}
    \Gamma_{\pi^0_{D,i}} \;\approx\; 
    \left(\frac{\kappa_D}{m_{Z_D}}\right)^4
    \frac{\mathrm{Tr}\!\left[(\mathcal{Q}^D_A)\cdot T_i\right]^2 (\mathcal{Q}^{SM}_A)^2}{2\pi (4\pi)^3}
    \, N_c \, \tilde{\Lambda}^2 \, m_{\pi_D} \, m_q^2
    \sqrt{1-\frac{4m_q^2}{m_{\pi_D}^2}},
    \label{eq:pion_lifetime}
\end{equation}
where $f_{\pi_D} \sim \sqrt{N_c}\,\tilde{\Lambda}/(4\pi)$ has been used, $\tilde{\Lambda}$ denotes the chiral symmetry breaking scale (of order $\Lambda$), and $m_q$ is the mass of the SM quark in the final state. A nonzero trace factor ensures that the corresponding pion is unstable; its magnitude is controlled by the differences in axial charges.  Eq.~\eqref{eq:pion_lifetime} makes the dependence on symmetry--breaking structures explicit. 

Finally, note the distinction between production and decay: the inclusive dark quark production cross section depends on the sum of squared charges,
\begin{equation}
    \sigma(pp \to Z_D \to q_D \bar{q}_D) \;\propto\;
    \sum_{i=1}^{\nf} \Big[ (\mathcal{Q}^D_{V,i})^2 + (\mathcal{Q}^D_{A,i})^2 \Big] \, \kappa_D^2,
\end{equation}
whereas the pion lifetime depends on differences of axial charges. This separation implies that in phenomenological studies one may effectively treat the pion lifetime as a free parameter, while recognizing that it originates from explicit flavour symmetry breaking in the Lagrangian. To simplify our notation, we will denote $(g^{tot}_{D})^2 = \sum_{\nf}\left((\mathcal{Q}_{V,i}^D)^2+(\mathcal{Q}_{A,i}^D)^2)\right)\kappa^2_D$. This combination enters the HV/DS quark production cross section. We note here that the sum over different flavors can only be simplified if all HV/DS quarks have same $U(1)_D$ charges. As our model fundamentally requires different axial vector couplings to trigger the HV/DS flavor symmetry breaking, we can not simplify the sum. It is therefore clear that we can only constrain $g^{tot}_{D}$~\footnote{Note that the expression for $g_{D}$ contains explicit sum over HV/DS quark charges. These need to be different to trigger flavor symmetry breaking and hence the sum can not be simplified. This justifies our rather cumbersome notation in favor of clarity.} unless specific charges are chosen. We use a similar expression for $g_{SM}$, except we simplify the sum over flavors by assuming that all SM quarks have same $U(1)_D$ charges. We remind the reader that these choices do not affect our model independent limits but only the model dependent exclusions.

A running coupling akin to SM QCD is an important consequence of such models. The running coupling computed at a fixed order in perturbation theory, usually at one, two-loop or three-loops in event generators, can become arbitrarily large at low energies signifying a breakdown of perturbation theory. We define a scale $\ld$, which signifies this breakdown. In particular, we set $\ld$ to be the energy scale at which the perturbative one-loop coupling runs to infinity.

In order for this setup to feature a chirally-broken phase, we restrict $\fc \lesssim 2.6$, beyond which the theories enter conformal window\footnote{The lower boundary of conformal window defined here to be $\fc \lesssim 2.6$ is not exactly known. We set it to $\fc$ where perturbative two-loop running coupling develops a fixed-point.}. In the chirally-broken phase, these theories contain a tower of spectrum. Neglecting any states heavier than the rho mesons, these theories consist of $N^2_F-1$ HV/DS pions and rho mesons, along with one spin-0 and one spin-1 singlet. Our assumption of neglecting heavier states is particularly well suited in the parameter space considered in this work. Keeping the spin-0 singlet ($\eta^\prime_D$) meson mass degenerate with the HV/DS pions, ($\pid$), we effectively have $N^2_F$ HV/DS pions\footnote{The spin-0 singlet is close to HV/DS pion mass in the large-$\nc$ limit where the contribution of axial anomaly to the $\eta^\prime$ mass is negligible.}. In total $\nf$ of these HV/DS pions are diagonal ($\pidon$) and $\nf(\nf - 1)$ are off-diagonal ($\pidoff$)\footnote{The $0, \pm$ denote $U(1)_D$ charges in analogy with the SM convention.}. Similarly for the spin-1 mesons, by ignoring the $\omega_D$ -- $\rhod$ mass difference, we have $N^2_F$ HV/DS rho mesons, ($\rhod$), with $\nf$ diagonal ($\rhodon$) and $\nf(\nf - 1)$ off-diagonal rho $(\rhodoff$) mesons. 

We assume that the HV/DS rho and pion masses are independent of $\nc, \nf$. They are set according to the Snowmass fits~\cite{Albouy:2022cin} derived using results from~\cite{Fischer:2006ub} for $m_\pid/\ld \lesssim 2$
\begin{eqnarray}
\frac{m_{\pid}}{\ld}=5.5 \sqrt{\frac{m_{q_D}}{\ld}}, \qquad\qquad\qquad
\frac{m_{\rhod}}{\ld}=\sqrt{5.76+1.5\frac{m_{\pid}^2}{\ld^2}},
\label{eqn:meson_masses}
\end{eqnarray}
where $\Lambda$ is the confinement scale, here conveniently equated with the scale at which one loop running coupling diverges. We note here that these fits can further be improved e.g., using results given in~\cite{DeGrand:2019vbx}. We make no such improvements as our aim is to understand the implications of such lattice fits and in particular use the methodology derived in the Snowmass report.

The assumed $\nc, \nf$ independence of pion and rho meson masses is a good approximation in a large region of HV/DS parameter space~\cite{Castagnini:2015ejr}, however for $\fc \sim 2.6$, lattice investigations suggest a non-QCD like spectrum~\cite{LatticeStrongDynamics:2018hun, LSD:2023uzj,LatKMI:2013bhp}. We ignore this complication here. 

Among the dark meson decays, depending on the $\rhod$ -- $\pid$ mass hierarchy, HV/DS rho mesons may decay to HV/DS pions or the SM final state. For simplicity we only consider $\mpl < 1.5$ where $m_\rhod > 2 m_\pid$ and hence $\rhod \to \pid \pid$ are allowed. 

\begin{figure}[h!]
\centering
\includegraphics[width=0.39\textwidth]{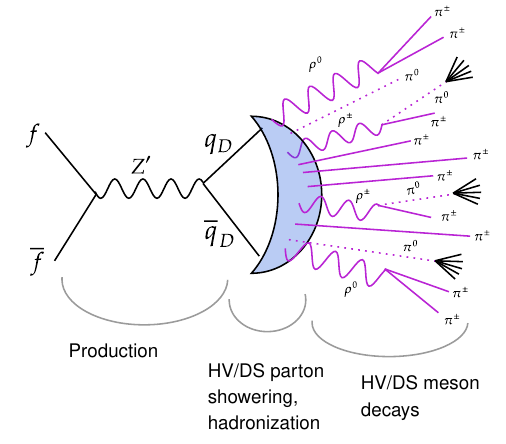}
\caption{A sketch of the production and decay mechanisms for HV/DS quarks and subsequently HV/DS mesons considered in this work.}
\label{fig:feynmann_sketch}
\end{figure}

The resulting process is sketched in fig.~\ref{fig:feynmann_sketch} for a first conceptual understanding. As can be imagined, some of the HV/DS quarks will decay within the calorimeter, while others may decay in the muon system. This is an important point whose consequences will be discussed in the later sections.

\begin{table}[h!]
\centering
\begin{tabular}{ |c|c|c|c|c|c|c| } 
\hline
Benchmark name &$\nc$ & $\fc$ & $\mpl$ & $\ld\,\rm{GeV}$ & Stable mesons &  Meson decay modes \\
\hline
one-$\pidon$ decay & 5 & 0.4 - 2.5 & 0.2 - 1.5 & 2 - 60 & All $\pidoff$ & $\rhodon \to \pidoff \pidoff$\\
& &  &  &  & $\nf -1$ $\pidon$ &  $\rhodoff \to \pidoff \pidon$\\
& &  &  &  &  &  $\pidon \to q \bar{q}$\\
\hline
all-$\pidon$ decay  & 5 & 0.4 - 2.5 & 0.2 - 1.5 & 2 - 60 & All $\pidoff$ &$\rhodon \to \pidoff \pidoff$\\ 
& &  &  &  & 0 $\pidon$ & $\rhodoff \to \pidoff \pidon$\\
& &  &  &  &  &  $\pidon \to q \bar{q}$\\
\hline
\end{tabular}
\caption{Table of the parameters for the two benchmark models considered in this work. Although we depict $\pid$ decays as $\pidon \to q \bar{q}$, we model the $\pidon$ decays such that it always decays to the heaviest available $q\bar{q}$ pair. The parameters of mediator sector are not mentioned. For each of the benchmarks, we consider two mediator masses $\mzp = 2, 3.5\,\rm{TeV}$.}
\label{tab:benchmarks}
\end{table}
The above class of models offer a large parameter space for study. In order to simplify our approach, in this work, we restrict ourselves to scenarios when $\rhod \to \pid \pid$ is allowed. Further, we assume that via appropriate $U(1)_D$ charge assignments any number of diagonal pions ($\pidon$) can be made unstable, however the off-diagonal pions ($\pidoff$) remain stable due to HV/DS flavor symmetry. For concrete simulations, we consider either one diagonal HV/DS pion decay or all diagonal HV/DS pion decays to illustrate a broad class of HV/DS models. The model details and decay modes are outlined in table~\ref{tab:benchmarks}. It is important to note that the $\ld$ ranges are chosen such that the process always leads to a multi-HV hadron final state i.e. a $2\to n$ HV final state is simulated rather than a $2\to 2$. 

Finally, we briefly comment on the effect of $\nc, \fc, \ld$ and $\mpl$ on HV/DS parton shower and hadronization processes. Among the free parameters of the theory, the $\fc, \ld$ control the shower and shower cutoff, $\mzp/\ld$ controls meson multiplicity via the length of the shower while $\fc$ controls the meson multiplicity via the running coupling. $\mpl, \ld$ influence the HV/DS meson spectra and control the $\rhod \to \pid \pid$ threshold, in addition, as we will discuss below, $\mpl$ also controls the HV/DS rho to pion production rate through hadronization parameter known as {\tt probVector}. It is therefore, non-trivial to decouple several interdependent effects in the simulations and proper theory relations must be taken into account in order to have a somewhat consistent HV/DS simulation. Clearly, the results may also be influenced by additional hadronization parameters and we comment on this briefly.

\subsection{Characteristic LLP multiplicity}
\label{sec:LLP_mult}

To begin with we discuss some generator level results in order to understand theory characteristics. This is motivated by two observations. First, we are interested in simulating theories up to $\fc \sim 2.6$, which contains regions with $\nf > 8$, going beyond the current {\tt PYTHIA8} Hidden Valley defaults. Second, we systematically account for $\rhod$ production, whose decays have non-trivial effects for the $\pidon$ multiplicity as we show below. We therefore demonstrate the validity of our simulation setup and effects of modeling hadronization parameter {\tt probVector} on the relative fraction of HV/DS rho to pion production and finally, we discuss the resulting $\pidon$ multiplicities. Within this subsection, we discuss the salient results of our analysis, while the associated numerical results are presented in Appendix~\ref{app:meson_multiplicity}.

We concentrate on multiplicities resulting from two HV/DS meson stages, first, number of mesons produced during hadronization i.e. just after parton showering denoted by $\left<N^{0}_{\pid, \rhod}\right>$, and second the number of pions ($\left<\npfin\right>$) once the $\rhod$ have decayed. The HV/DS meson multiplicities produced during HV/DS hadronization are proportional to the parton multiplicity, i.e. the amount of HV/DS quarks/gluons emitted during parton showering. On average, the parton multiplicity increases with $\fc$ and so do the total HV/DS meson multiplicities $\left<N_{\rhod}\right>$ and $\left<N_{\pid}\right>$. 

As discussed in Appendix~\ref{app:meson_multiplicity}, for a fixed $\nc, \mpl, \ld$, the total meson multiplicity is directly proportional to $\fc$ while the fraction of diagonal and off-diagonal pions varies as $1/\nf$ and $1-1/\nf$ demonstrating that our simulations are valid for $\nf > 8$. 

Analyzing the HV/DS meson multiplicities as a function of $\mpl$ is non-trivial. For this we must note that the fraction of initial mesons that are labeled spin-0 $\pid$ or spin-1 $\rhod$ by {\tt PYTHIA} is determined by a \texttt{PYTHIA} internal quantity called \texttt{HiddenValley:probVector} defined as $\texttt{probVector} = \left<N_{\rhod}^{0}/N_{\rm tot}^{0}\right>$. Spin-counting of 3 spin-1 and 1 spin-0 state would dictate that $\texttt{probVector}=0.75$, however since $\rhod$ and $\pid$ are non-degenerate, changing their relative masses should change their relative abundances during HV/DS hadronization. To account for this, we must introduce a mass-dependent Boltzmann factor that is a function of $m_{\rhod}/m_{\pid}$ and fit accordingly~\cite{Batini:2024zst,Becattini:1996gy}. The ratio $m_{\rhod}/m_{\pid}$ is inversely proportional to $\mpl$ as seen in eqn.~\eqref{eqn:meson_masses} thus implying that $\texttt{probVector}$ is a function of $\mpl$. Throughout the text we use the exponential fit given by 
\begin{equation}
\texttt{probVector} = \frac{3\exp(-\omega(m_{\rhod}/m_{\pid} - 1))}{1+3\exp(-\omega(m_{\rhod}/m_{\pid} - 1))};
\label{eq:exponential_fit}
\end{equation}
the details of which are described in Appendix~\ref{app:probvec_modelling} alongside a few alternatives. We do not recommend to use these fits for $\mpl \lesssim 0.2$ due to uncertainties in the applicability of Lund hadronization model as explained in the Appendix~\ref{app:probvec_modelling}. We also note here that a fit as considered here will ensure a proper phase space suppression of rho meson production, however the form of the fit is not unique. Among the choices we considered, we find the exponential fit to be the best possibility. We encourage consideration of other fit functions which may match data more suitably. As we show, the exact form of the fit we consider is largely irrelevant to the qualitative behavior of the results, the quantitative results may differ under a different fit choice. Given these fits, the ratio of initial rho to pion multiplicity ($\left<\nrinit/\npinit\right>$) is $0.90$ for $\mpl=0.6$ and $1.98$ for $\mpl=1.4$ for all considered $\fc$.

We are now in a position to determine the composition of meson species i.e. the initial and final (off-) diagonal $\pid, \rhod$, keeping in mind $\texttt{probVector} = \left<N_{\rhod}^{0}\rangle/\langle N_{\rm tot}^{0}\right>$. Solving for all of the initial meson species in terms of the $\texttt{probVector}, \nf$ and $\left<\ntinit\right>$ gives,
\begin{eqnarray}
\label{eq:initial_pimeson}
\langle \nponinit\rangle = \frac{1-\texttt{probVector}}{\nf}\left<\ntinit\right>, && \qquad \langle \npoffinit\rangle=(1-\texttt{probVector})\left(1-\frac{1}{\nf}\right)\left<\ntinit\right>,\\
\langle \nroninit\rangle = \frac{\texttt{probVector}}{\nf}\left<\ntinit\right>, && \qquad \langle \nroffinit\rangle=\texttt{probVector}\left(1-\frac{1}{\nf}\right)\left<\ntinit\right>.
\label{eq:initial_rhomeson}
\end{eqnarray}

While the above discussion explains initial HV/DS meson multiplicities in $\mpl$ -- $\fc$ space, we are interested in the number LLPs which in our case are either a fraction or all diagonal pions. Given the details of the decay modes in table~\ref{tab:benchmarks}, we analyze the final (off-) diagonal HV/DS pions multiplicity once the rho mesons decay to pions. One can use eqns.~\eqref{eq:initial_pimeson} and~\eqref{eq:initial_rhomeson} to write the final-state diagonal pions as $\left<N_{\pidon}\right>=\left<N^0_{\pidon}\right>+\left<N^{0}_{\rhodoff}\right>$, giving,
\begin{equation}
\langle \nponfin\rangle=\left(\frac{1}{\nf}+\texttt{probVector}\left(1-\frac{2}{\nf}\right)\right)\left<\ntinit\right>,
\label{eq:final_onpi}
\end{equation}
\begin{equation}
\langle \npofffin\rangle=\left(\left(1-\frac{1}{\nf}\right)+\frac{2\,\texttt{probVector}}{\nf}\right)\left<\ntinit\right>.
\label{eq:final_offpi}
\end{equation}

Additionally, this yields $\left<\ntfin\right> = (1+\texttt{probVector})\langle \ntinit\rangle$. In appendix~\ref{app:meson_multiplicity}, we show additional plots clearly illustrating the effect of these fits. 

There are two important conclusions of our studies, first due to the presence of the rho mesons, the average diagonal pion multiplicity increases with $\fc$, second majority of $\pid$ after the decay of the HV/DS rho mesons are off-diagonal. This demonstrates the effect of correctly including rho mesons in HV simulations. Second, the dependence of the final pion multiplicities on $\fc$ as well as $\mpl$ for a fixed $\ld$. We stress here that although the observed numerical values of initial and final meson multiplicity for a given $\fc$ and $\mpl$ is a consequence of our {\tt probVector} fit given in eqn.~\eqref{eq:exponential_fit}, the exact form of {\tt probVector} does not matter and qualitatively similar behavior will be observed as long as {\tt probVector} is a proportional to $\mpl$. 

\begin{figure}[h!]
\centering
\includegraphics[width=0.45\textwidth]{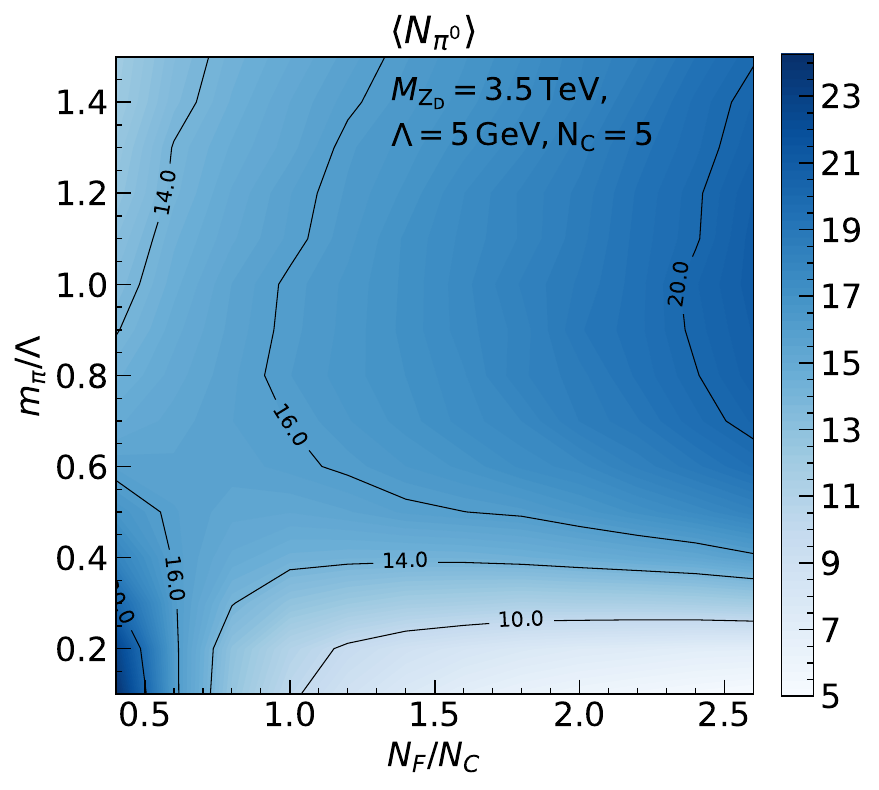}
\includegraphics[width=0.45\textwidth]{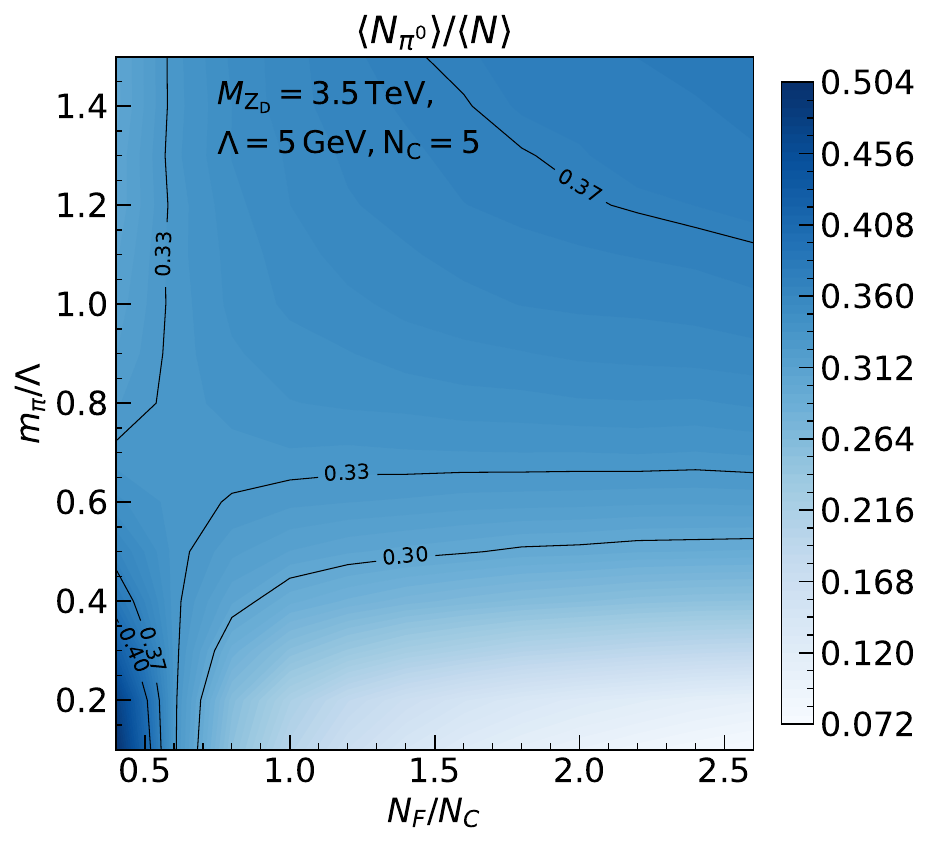}
\caption{(left panel) Final state pion multiplicity as a function of $\mpl$ and $\fc$ for $\nc = 5, \ld = 5\,\rm{GeV}$, (right panel) Ratio of the final state pion multiplicity to total final state meson multiplicity as a function of $\mpl$ and $\fc$ for $\nc = 5, \ld = 5\,\rm{GeV}$.}
\label{fig:mpl_nf_probvec}
\end{figure}
Finally, in fig.~\ref{fig:mpl_nf_probvec} we show the contours of the final state diagonal pion multiplicity (left panel) and ratio of diagonal pion to total final state meson multiplicity (right panel) as a function of $\fc$  and $\mpl$. These results clearly demonstrate an important interplay of {\tt probVector} and $\fc$. We see that the final state diagonal pion multiplicity is a function of both $\fc$ and $\mpl$ and is not simply a phase space effect which is driven by $\mpl$. For example, we see that the final state pion multiplicity decreases with $\mpl$ for $\fc = 0.4$, while it initially increases before flattening out for larger $\fc$, e.g., $\fc = 2.6$. This can be understood by recalling that eqn.~\eqref{eq:final_onpi} contains two competing factors; $\left<\ntinit\right>$ which decreases with $\mpl$ and a pre-factor that increases with $\mpl$ for fixed $\fc>2$. At $\fc=2$, this pre-factor is constant with $\mpl$ since all $\texttt{probVector}$ dependence cancels out. Hence, the net effect at $\fc = 0.4$ is a decrease in the multiplicity. For $\fc = 2.6$, the pre-factor increases with $\mpl$ and so for small $\mpl$ dominates over the rate of decrease of $\left<\ntinit\right>$ with $\mpl$. For larger $\mpl$, $\left<\ntinit\right>$ begins to decrease at a faster rate than the pre-factor leading to a slight decrease at large values of $\mpl$.

Similarly, the ratio of diagonal final state pions to total final state mesons show interesting trends. We observe that for small $\mpl$ and $\fc$, the ratio can reach as high as 50\%, which while this initially drops as the two variables increase, increases once again for large enough $\mpl$ and $\fc$ again reaching as high as 50\%. This will have interesting consequences for the sensitivity of the CMS analysis as the fraction of HV/DS energy deposited in the SM final state will vary according to this ratio. 
\section{CMS Displaced Shower Search}
\label{sec:search_details}
\subsection{Search description}
The CMS analysis we use here is documented in~\cite{CMS:2021juv},\footnote{An update of the search is available in~\cite{CMS:2024bvl} however the associated reinterpretation material does not include a {\tt Delphes} code or module implementation, hence we use the older version of the analysis.}, with the associated reinterpretation material including a {\tt Delphes} implementation provided in~\cite{CMSreinterpretation}. The analysis searches for decays of long-lived particles (LLPs) in the CMS muon system, in particular the muon endcap detector. The analysis is optimized for single or pair production of LLPs. Any LLP decaying in the CMS muon system, specifically in the CMS muon endcap detector, introduces hadronic and electromagnetic showers, giving rise to high hit multiplicity in the localized regions of the detector called the CSC clusters. In particular, a CSC cluster associated with the signal requires $N_{\rm hits} > 130$ and the azimuthal angle between the cluster location and the MET ($\Delta \phi_c$) $ < $ 0.75.  The analysis searches for these CSC clusters by using the muon endcap detector as a sampling calorimeter. The search is sensitive to LLPs decaying to hadrons, taus, electrons, or photons. Decays to muons are not considered as muons rarely produce particle shower which lead to the CSC cluster. 

The search requires the following selection criteria: 
\begin{enumerate}

\item The \emph{missing transverse energy}, defined as the negative vector sum of visible $p_T$ from particles identified in the tracker and calorimeter, is ${\rm MET} > 200~\rm{GeV}$, as trigger requirement. For models under consideration, there are two sources of missing energy. First is the initial-state-radiation and a second more subtle source is related to the presence of multiple LLPs as shown in section~\ref{sec:models}. Given the large LLP multiplicity, it is possible that some LLPs decay within the tracker or the calorimeter, which contributes to the overall $\rm{MET}$. It is expected that this effect will be more important for intermediate lifetimes, with high LLP multiplicities, than for extremely large lifetimes or low LLP multiplicities. It will also depend on the number of LLPs present in the signal. We also consider HL-LHC sensitivity and use an optimistic trigger of ${\rm MET} > 50~\rm{GeV}$ for that purpose~\cite{Cottin:2022nwp}.

\item \emph{No electron~(muon)} with transverse momentum $p_T > 35~(25)$~GeV and pseudorapidity $|\eta| < 2.5~(2.4)$, to remove $W$ and top background.
    
\item \emph{At least one CSC cluster} with $|\Delta\phi_c| < 0.75$ to ensure that it originated from the LLP decay. Here, $\Delta\phi(\mathbf{x}_\text{CSC}, \rm{MET})$ is defined as the azimuthal angle between the missing transverse momentum and the cluster location from the IP. The LLP is then close to the missing energy, which points opposite to the vector sum of the visible $p_T$. 

\item Events with \emph{clusters too close to a jet~(muon) are removed}, for $\Delta R = \sqrt{(\Delta\eta)^2 + (\Delta\phi)^2} < 0.4$. If $\Delta R(\rm{jet/muon, cluster}) < 0.4$, the CSC cluster is matched to so-called punch-through jets/muons with $p_T > 10~(20)$~GeV and it is assumed to be created by LLPs inside the jet, e.g. $K_L$, or muon bremsstrahlung. This requirement may affect our signal as we expect some of the LLPs to decay within the calorimeter thus resembling a SM jet. Such events containing prompt plus displaced collimated LLPs will be vetoed by $\Delta R(\rm{jet/muon, cluster}) < 0.4$.
    
\item The \emph{average time of detector hits} in the CSC cluster, relative to the collision is $-5~\text{ns} < \langle\Delta t_\text{CSC}\rangle < 12.5$~ns, to reject pileup clusters.

\end{enumerate}

Finally, since the CSC clusters are used as calorimeters, the cluster efficiency depends on the amount of electromagnetic and hadronic energy deposited in the muon system. The search therefore provides cluster efficiency as a function of the hadronic and electromagnetic energy deposited in the Muon System.

\subsection{Simulation details}
\label{sec:sim_setup}
Similar to previous section, we use the {\tt Hidden Valley} module of {\tt PYTHIA}~v8.311, to simulate our darkshower signal processes, including the pion decays to the SM quarks. Using {\tt MadGraph} to simulate initial state radiation and matching-merging the resulting events with {\tt PYTHIA} leads to 10\% change in the final results we derive. The parameters used for our simulation are outlined in table \ref{tab:benchmarks}. We note here that with {\tt SeperateFlav} option turned off, {\tt PYTHIA8} has only one PID for all the diagonal pions\footnote{We need to keep {\tt SeperateFlav = off} to be able to simulate $\nf > 8$ theories, as {\tt PYTHIA8} has pion PIDs corresponding to only 8 flavors when {\tt SeperateFlav} is switched on.}, hence to simulate one-$\pidon$ decay we set the decay modes of this diagonal pion PID by appropriate fractions. For the case of one-$\pidon$ decay, we set $1/\nf$ fraction to visible SM final state ($q\bar{q}$), while the remaining fraction is set to PID 53, which is declared to be a stable neutral particle. 

Detector effects are taken into account using {\tt Delphes}~v3.5.1 \cite{deFavereau:2013fsa}. In order for {\tt Delphes} to correctly simulate the HV/DS signal, a number of modifications are necessary. First, we modify the {\tt LLPFilter CSCFilter} module in {\tt Delphes} card to filter out the $\pidon$. In addition, we remove all HV mesons from {\tt SimpleCalorimeter HCal} module. Finally, the {\tt Delphes} implementation as provided by the CMS collaboration computes the total hadronic and electromagnetic energy deposited in the muon detector by summing over the energies of the  LLP decay products. By default, it includes the energy of particle with PID 53, declared as an invisible particle in our simulation. We therefore modify {\tt modules/llpfilter.cc} in order to skip PID 53 while computing the total energy deposited in the muon endcap detector. We cluster jets, which result due to initial-state-radiation or promptly decaying LLPs, using {\tt FastJet}~v3.2.1~\cite{Cacciari:2011ma} with anti-$k_T$ algorithm with a clustering parameter of 0.4.  

The output of the {\tt Delphes} implementation includes number of clusters, $\rm{MET}$ as well as the electromagnetic ($E_{\rm{em}}$) and hadronic ($E_{\rm{had}}$) energy per CSC cluster along with other relevant details. The cluster efficiencies include vetoes against objects that decay in the inner part of the detector and reach the muon endcap detector -- so-called segment and rechit vetoes -- as well as the muon veto, time spread cut, and $N_{\rm{hits}} > 130~(370)$~\cite{Cottin:2022nwp}. Additional requirements including jet veto, time cut, and $\Delta \phi_c(\rm{MET}, \rm{cluster})$ cut efficiencies and cut-based id efficiency is to be computed by the user. We implement the latter requirements to analyze {\tt Delphes} output. 

At this point, we observe an important shortcoming of the existing {\tt Delphes} analysis implementation specifically applicable for our signal. First, given the energy of the LLP, the current {\tt CscClusterEfficiency} module computes the probability of the LLP to create to a CSC cluster with given $N_{\rm hits}$. Accordingly, the efficiency map is also created using the hadronic and leptonic energy deposited per LLP. In our setup since multiple LLPs are collimated, their decays in the endcap detector may create CSC clusters with larger $N_{\rm hits}$, however such effects can not be taken into account. Second, given the presence of multiple LLPs, the total hadronic energy per cluster deposited in the endcap detector may be larger or smaller than the energy ranges in the efficiency map provided by the CMS collaboration. We can not analyze these situations due to the limitation of the information provided. Never-the-less, our results are a first indication of the capability of the analysis to explore HV/DS theories and our work shows a possible strategy for systematic presentation of results. 

\section{Dependence of efficiency on theory parameters}
\label{sec:eff_dependence}

The sensitivity of the analysis to the underlying theory parameters is nominally controlled by the efficiency of the displaced shower search. Variations in hadronization parameters or alternative dark-sector non-perturbative dynamics could change these efficiencies, and hence the derived sensitivities, non-trivially. Within our modeling of hadronization, the efficiency can be expressed as
\begin{eqnarray}
\epsilon_{\rm tot} &\approx& N_{\rm{LLP}}\times\mathcal{P}(4\,\rm{m} < \beta\gamma c\tau_{\rm LLP} < 11 \,\rm{m}) \times \mathcal{P}_{\rm angular}(\phi, \eta)\times \epsilon_{\rm reco}\left(E^{\rm CSC}_{\rm had}\right)
\times \epsilon_{\rm cut}(\rm MET)\\
& \equiv & N_{\rm{LLP}}\times \epsilon_{\rm geo}(\beta\gamma c \tau_{\rm{LLP}})\times \epsilon_{\rm reco}\left(E^{\rm CSC}_{\rm had}\right)\times \epsilon_{\rm cut}(\rm MET),\\
& \equiv & N^{\rm CSC}_{\rm{LLP}}\times  \epsilon_{\rm reco}\left(E^{\rm CSC}_{\rm had}\right)\times \epsilon_{\rm cut}(\rm MET),
\end{eqnarray}
and the corresponding number of signal events is given by $N_{signal} = \sigma_{pp\to q_D\bar{q}_D} \times \epsilon_{\rm total} \times \mathcal{L}$, where $\mathcal{L}$ is the search luminosity. We denote $N_{\rm{LLP}}$ to be the number of decaying $\pidon$ while $N^{\rm CSC}_{\rm{LLP}}$ is the number of $\pidon$ decaying within the muon endcap detector, hence $N^{\rm CSC}_{\rm{LLP}} = N_{\rm{LLP}}\times\epsilon_{\rm geo}$. Finally $c \tau_{\rm{LLP}}$ denotes the lifetime of the unstable $\pidon$. We make a deliberate effort to differentiate between $\pidon$ properties and the properties of LLP as not all our $\pidon$ are unstable for the two benchmark scenarios outlined in table~\ref{tab:benchmarks}.

The search efficiency is therefore determined by four factors, 
\begin{itemize}
    \item Number of unstable HV/DS pions ($N_{\rm LLP}$), determined by the shower characteristics, e.g., HV/DS QCD scale, $\ld$, $\mpl$ and $\fc$. As $N_{\rm LLP}$ depends not only on the theory parameters but also on hadronization parameters, in appendix~\ref{app:hadronization_parameters} we also demonstrate the change in $N_{\rm LLP}$ against variations in several Pythia HV-hadronization parameters.
    \item Geometrical efficiency ($\epsilon_{geo}$) ensuring that the HV/DS pions decay inside the muon endcap detector. It is a function of the pion angular distribution as well as the lab frame decay length $\beta\gamma c\tau_{\rm LLP}$.
    \item Reconstruction efficiency ($ \epsilon_{reco}$), which is mainly a function of deposited hadronic energy per CSC cluster\footnote{Since our LLP decays to quarks, we expect very little to no electromagnetic energy in the events and hence consider the resulting efficiency to be only a function of $E_{\rm had}$.} ($E^{\rm CSC}_{\rm had}$) and the lab frame decay length of HV/DS pion ($\beta\gamma c\tau_{\rm LLP}$).
    \item Selection efficiency ($\epsilon_{cut}$) which is primarily influenced by the required missing energy ($\rm{MET}$) of the event. 
\end{itemize} 

In order to understand the change in the efficiency as a function of theory parameters, we need to analyze components of the efficiency, $N_{\rm LLP}$, $\epsilon_{\rm geo} \equiv \mathcal{P}(4\,\rm{m} < \beta\gamma c\tau_{\rm LLP} < 11 \,\rm{m}) \times \mathcal{P}_{\rm angular}(\phi, \eta)$, $\epsilon_{reco}$ and $\epsilon_{cut}$ one by one. $\epsilon_{\rm geo}$ mainly depends on the lab frame decay length, which is a function of the boost factor $\beta \gamma = p_{\pidon}/m_{\pidon}$ for fixed $c \tau_{\rm LLP}$.  The angular distribution of the HV/DS pions is roughly the same within our theory parameters, hence $\mathcal{P}_{\rm angular}$ is roughly constant. $\epsilon_{reco}$ has a mild dependence on the location within the detector, but dominantly depends on $E^{\rm CSC}_{\rm had}$ -- the hadronic energy deposited per HV/DS pion decay inside the muon endcap. 

For small to moderate lifetimes, the requirement for the HV/DS pions to decay inside the muon endcap generically selects energetic and thus highly boosted pions, such that

\begin{align}
E^{\rm{CSC}}_{\rm had} = \sqrt{m_{\pidon}^2 + p_{\rm LLP}^2} \approx p_{\rm LLP} \approx \beta \gamma \times m_{\pidon} 
\approx \frac{L_{\rm lab}}{c \tau_{\rm LLP}} \times m_{\pidon}
\\
{\rm with}~4~{\rm m} \lesssim L_{\rm lab}= \beta \gamma c \tau_{\rm LLP} \lesssim 11~{\rm m}. 
\label{eq:Ehad_CSC_def}
\end{align}
This fact will be useful later. 

Finally, $\epsilon_{\rm cut}$ is mainly a function of the missing transverse energy, which is roughly constant for $Z_D$ masses we consider here. We therefore show the distribution of $N_{\rm LLP}$, $N_{\rm LLP}^{\rm CSC} = N_{\rm LLP} \times \epsilon_{\rm geo}$, $\beta \gamma$ and $E_{\rm had}^{\rm CSC}$ as a function of theory parameters to demonstrate their effect on final efficiency. Since $\epsilon_{\rm cut}$ is roughly constant for all variations of considered theory parameters, we define fraction of events containing at least one cluster, $f_{\rm cluster} \equiv   N_{\rm evts}(N_{\rm cluster} \geq 1)/N_{\rm all} \approx \epsilon_{\rm total} / \epsilon_{cut}(\rm{MET})$ to roughly show the major dependence of the efficiency ($\epsilon_{\rm tot}$) on the theory parameters. In appendix~\ref{app:additional_plots}, we explicitly show the MET and the boost ($\beta\gamma$) distributions for a specific benchmark.

\subsection{Efficiency as a function of \texorpdfstring{$\ld$}{Lambda} and \texorpdfstring{$\mpl$}{mpl}}
%
\begin{figure}[h!]
\centering
\includegraphics[width=0.48\textwidth]{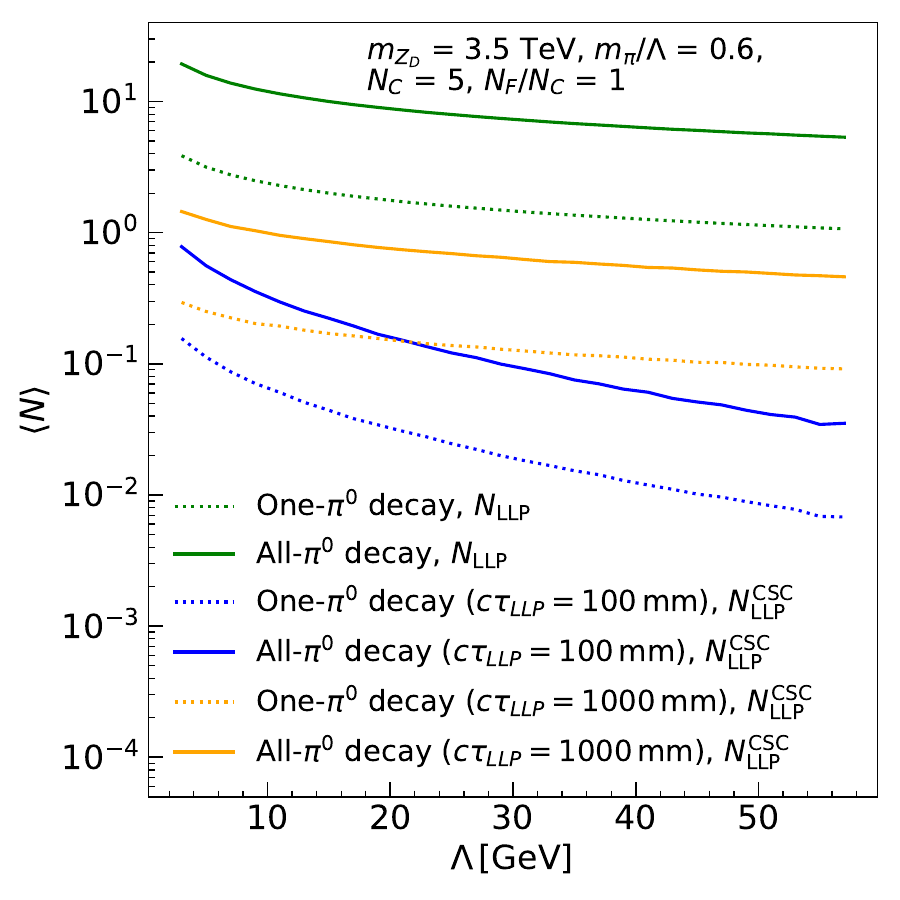}
\includegraphics[width=0.50\textwidth]{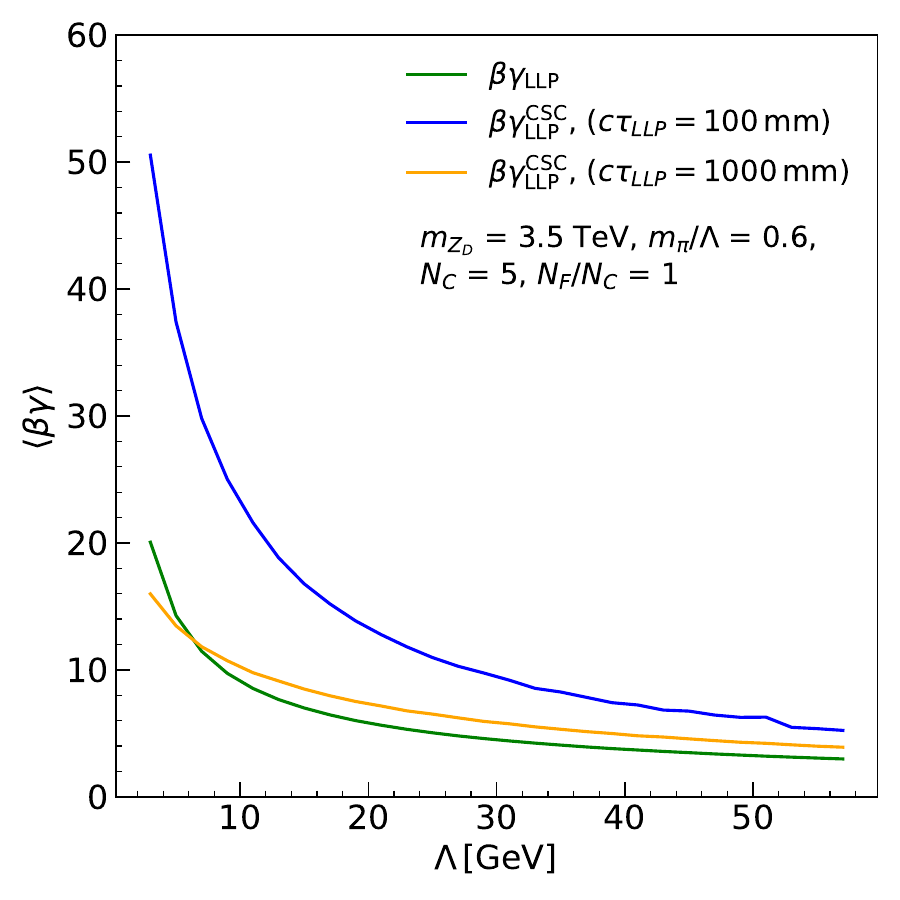}
\caption{The LLP multiplicity (left panel) and boost (right panel) as a function of $\ld$ with $\mzp = 3.5\,\rm{TeV}, \fc = 1, \mpl = 0.6, \nc = 5$. }
\label{fig:multiplicity_and_boost_lamda}
\end{figure}
Following the discussion of the final efficiency, we demonstrate the effects of $\ld$ on $N_{\rm LLP}$, $N_{\rm LLP}^{\rm CSC}$, $\beta \gamma$, $E_{\rm had}^{\rm CSC}$ and $f_{\rm cluster}$, for fixed $N_C = 5, \fc = 1$. We augment the two benchmarks discussed in table~\ref{tab:benchmarks}, by adding two values of proper decay length ($c \tau_{\rm LLP}$ =) 100 and 1000 mm. We therefore have four possible scenarios. 

In fig.~\ref{fig:multiplicity_and_boost_lamda} (left panel) we show the average generator level HV/DS pion multiplicity $N_{\rm LLP}$ as a function of HV/DS scale $\ld$ while keeping other parameters fixed as specified. The LLP multiplicity is inversely proportional to $\ld$. This is expected since the HV/DS pion mass is directly proportional to $\ld$, which leads to a shorter shower and thus decreases the multiplicity. The multiplicity for one-$\pidon$ decay scenario is $\nf$ times smaller compared to the case of all-$\pidon$ decays as expected. Along with this we also show the number of LLPs reaching the muon endcap detector ($N^{\rm CSC}_{\rm{LLP}}$), which strongly depends on the LLP lifetime. For smaller lifetime fewer LLPs reach the muon endcap detector while the number increases for large lifetimes. Finally we note that for a given $\ld$, $N^{\rm CSC}_{\rm{LLP}}$ is directly proportional to the LLP lifetime as expected. 

This proportionality is closely related to the average boost factor ($\langle\beta\gamma\rangle$) shown in fig.~\ref{fig:multiplicity_and_boost_lamda} (right panel). $\langle\beta\gamma\rangle$ is also helpful in understanding the geometrical efficiency $\epsilon_{\rm geo}$ which directly affects $N^{\rm CSC}_{\rm{LLP}}$. The dependence of boost factor on $\ld$ is non-trivial. The pion mass and momentum is directly proportional to $\ld$. One therefore expects the boost to be directly proportional to $\ld$ as well, while the opposite is observed in fig.~\ref{fig:multiplicity_and_boost_lamda} (right panel). This is because the pion mass increases faster with $\ld$ is compared to the momentum, correspondingly the average boost\footnote{Boost is defined as $\beta \gamma = p_{\pid}/m_\pid$}  shown in fig.~\ref{fig:multiplicity_and_boost_lamda} (right) decreases. We show additional distributions to this effect in fig.~\ref{fig:boost}. 

At small lifetimes, particles need large boost $\approx$ 40 -- 110 to reach the endcap detector, correspondingly, $\langle\beta\gamma^{\rm CSC}_{\rm LLP}\rangle$ is larger for $c\tau_{\rm LLP} =$ 100 mm than $c\tau_{\rm LLP} =$ 1000 mm. At small $c\tau_{\rm LLP}$ the tail of the $\beta\gamma$ is sampled and thus $\epsilon_{\rm geo}$ is small, however large $N_{\rm LLP}$ at small $\ld$ compensates for this penalty. Overall, it means for small $c\tau_{\rm LLP}$ pions corresponding to small $\ld$ have a better chance to reach the endcap detector. In other words, $\epsilon_{\rm geo}$ is inversely proportional to $\ld$ for small lifetimes. 

For large lifetimes, $c\tau_{\rm LLP} =$ 1000 mm, pions need $\beta\gamma \approx 4-11$ to reach the endcap detector, which can be easily realized for all $\ld$. Correspondingly, for large lifetimes $\epsilon_{\rm geo}$ is a rather constant function of $\ld$. This is reflected in the $N^{\rm CSC}_{\rm LLP}$ shown in left panel. $N^{\rm CSC}_{\rm LLP}$ decreases strongly as $\ld$ grows for small lifetime ($c\tau_{\rm LLP} =$ 100 mm), and is roughly flat for large lifetime ($c\tau_{\rm LLP} =$ 1000 mm).

\begin{figure}[h!]
\centering
\includegraphics[width=0.49\textwidth]{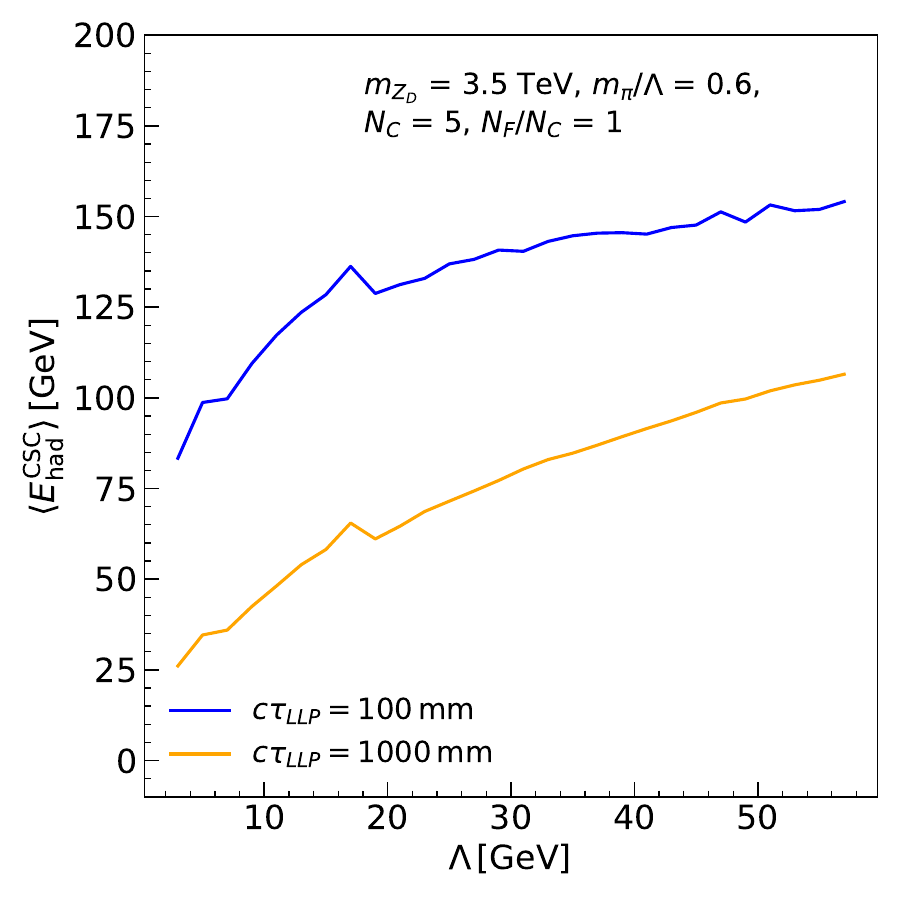}
\includegraphics[width=0.49\textwidth]{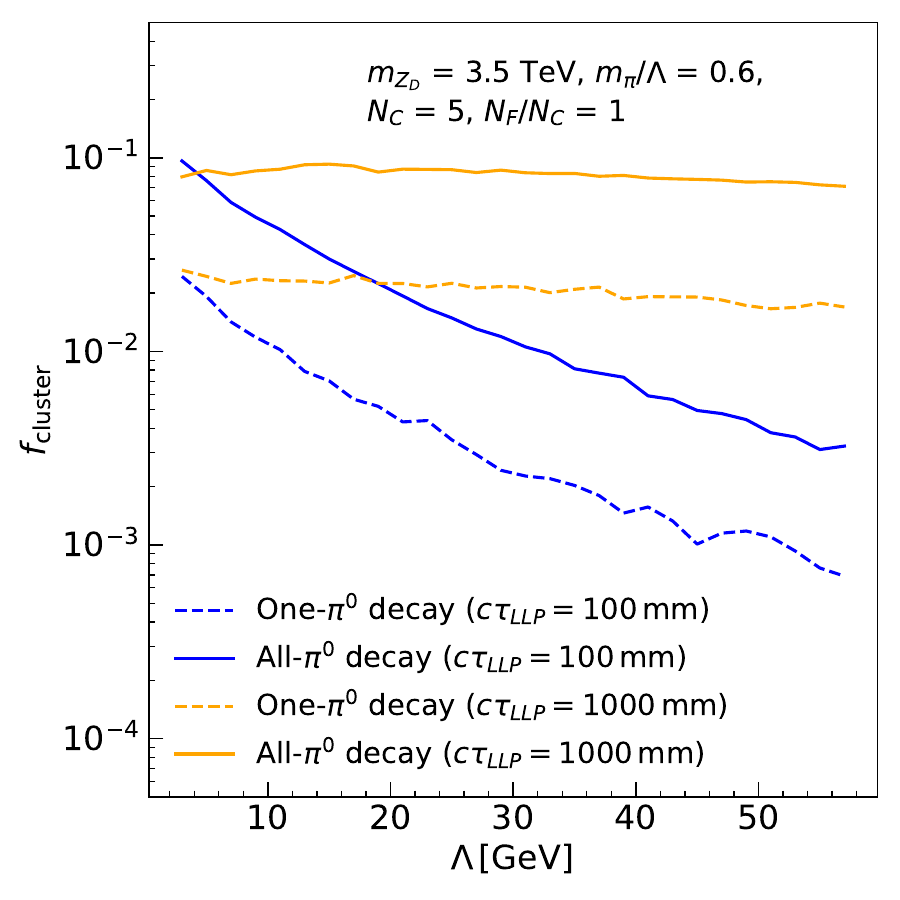}
\caption{Average hadronic energy deposited in the muon endcap detector per HV/DS pion~(left panel), and fraction of events containing at least one cluster (right panel) for $c \tau_{\rm LLP}$ = 100~and 1000~mm~(blue and orange lines) as a function of $\ld$. $\mzp = 3.5\,\rm{TeV}, \fc = 1, \mpl = 0.6, \nc = 5$ are kept fixed. }
\label{fig:eff_lambda}
\end{figure}

In fig.~\ref{fig:eff_lambda} (left panel), we show $E^{\rm CSC}_{\rm had}$ -- the hadronic energy per LLP deposited in the endcap detector --  as a function of $\ld$. $E^{\rm CSC}_{\rm had}$ directly affects reconstruction efficiency $\epsilon_{\rm reco}$. As is clear from eq.~\eqref{eq:Ehad_CSC_def}, this is a function of LLP lifetime. The amount of deposited energy is larger for $c\tau_{\rm LLP} = 100 \,\rm{mm}$ compared to $c\tau_{\rm LLP} = 1000 \,\rm{mm}$. As discussed previously, small $c\tau_{\rm LLP}$ forces selection of large momentum events required for LLPs to reach the endcap detector. Therefore, for a fixed $\ld$, the hadronic energy for $c\tau_{\rm LLP} = 100 \,\rm{mm}$ is larger than for $c\tau_{\rm LLP} = 1000 \,\rm{mm}$. Secondly, for a given lifetime the deposited energy is directly proportional to $\ld$. This can be understood as follows. Fixing pion lifetime leads to boost selection, e.g., $c\tau_{\rm LLP} = 100\,\rm{mm}$ needs $\beta\gamma = 40 - 100$, for pions to reach the CSC. However, larger $\ld$ leads to heavier pions, thus larger momentum is required to achieve the boost required, which in turn leads to larger $E^{\rm CSC}_{\rm had}$. Finally, the $E_{\rm had}^{\rm CSC}$ is independent of the number of decaying LLPs because it corresponds to the energy per particle which enters the muon endcap detector. Hence for a given lifetime, the reconstruction efficiency $\epsilon_{\rm reco}$ is directly proportional to $\ld$ while for a given $\ld$, it is inversely proportional to $c\tau_{\rm LLP}$.

Finally, in fig.~\ref{fig:eff_lambda} (right panel) we show $f_{\rm cluster}$, i.e. the fraction of events containing at least one CSC cluster for two different lifetimes. This is equivalent to $\epsilon_{\rm total}/\epsilon_{\rm cut} \approx N_{\rm LLP} \times \epsilon_{\rm geo} \times \epsilon_{\rm reco}$. This quantity decreases with $\ld$ for small lifetimes ($c\tau_{\rm LLP} = 100\,\rm{mm}$), while it is almost constant for large lifetimes ($c\tau_{\rm LLP} = 1000\,\rm{mm}$). This is because $f_{\rm cluster}$ is dominated by $\epsilon_{\rm geo}$. For small lifetime, $\epsilon_{\rm geo}$ decreases as $\ld$ increases, thus the $f_{\rm cluster}$ sharply decreases. However, for longer lifetime, a constant $\epsilon_{\rm geo}$ implies a constant $f_{\rm cluster}$. Finally, $f_{\rm cluster}$ depends on number of decaying pions, all-$\pidon$ decay case has $\nf$ times larger $N_{\rm LLP}$ by definition. 

To complete our discussion, in fig.~\ref{fig:mpl_multiplicity_and_boost} -- \ref{fig:eff_mpl} (appendix~\ref{app:eff_mpi_lam}), we also show same distributions as a function of $\mpl$ for fixed $\ld =$ 5 GeV. While the multiplicity stays rather flat, increasing $\mpl$ also leads to heavier HV/DS pions, thus smaller average boost. Hence, the dependence of $E_{\rm had}^{\rm CSC}$ on $\mpl$ is quite similar but weaker, and the final $f_{\rm cluster}$ also stays constant for $c \tau_\pid =$ 1000 mm case, but is inversely proportional to $\mpl$ for $c \tau_\pid =$ 100 mm.

\subsection{Efficiency as a function of \texorpdfstring{$\fc$}{nf/nc}}
\begin{figure}[h!]
\centering
\includegraphics[width=0.49\textwidth]{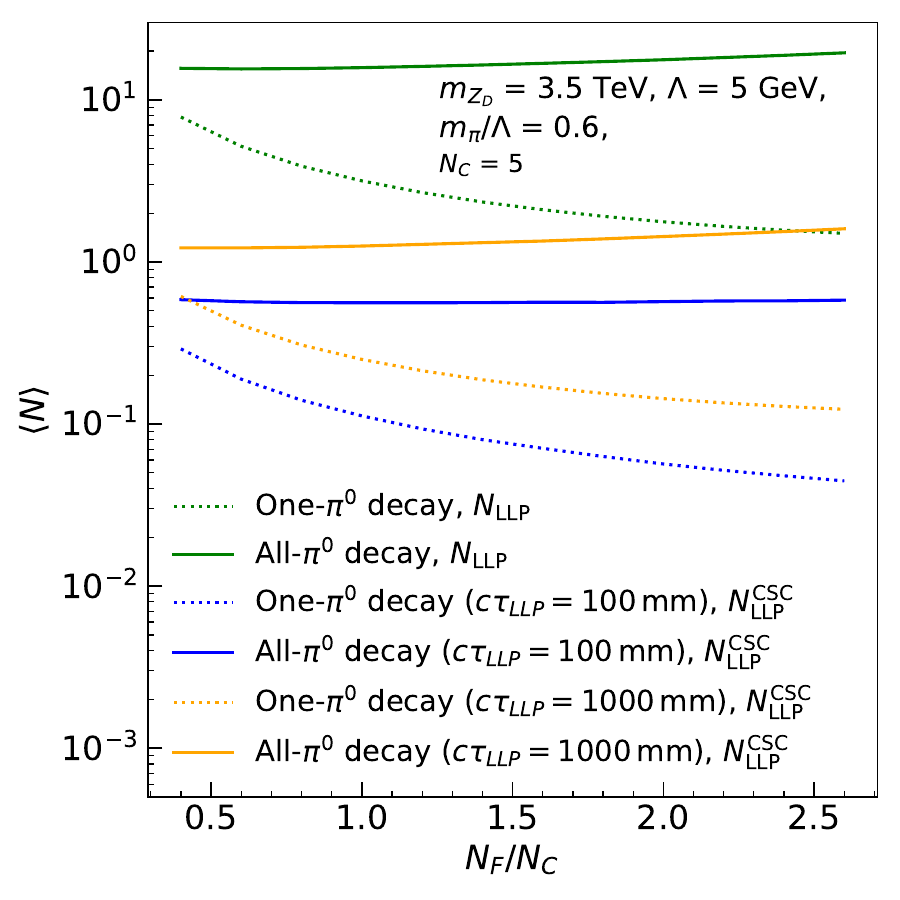}
\includegraphics[width=0.49\textwidth]{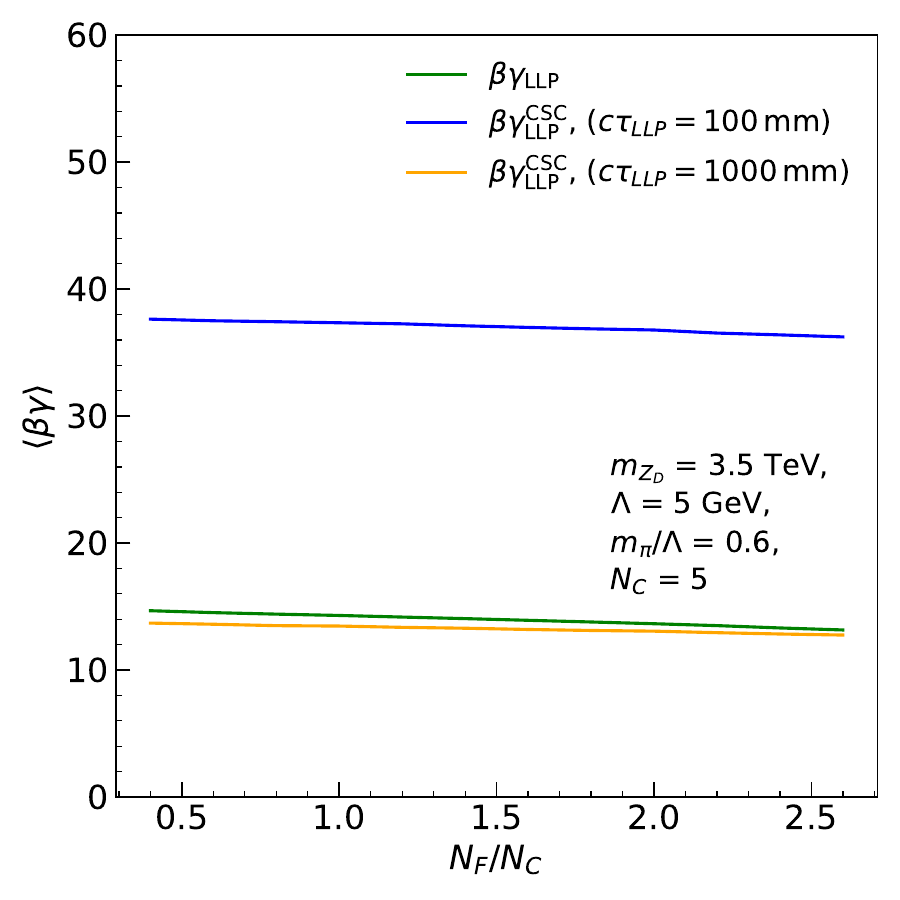}
\caption{The multiplicity (left panel) and boost (right panel) of the diagonal pions in the events a function of $\fc$, with $\mzp = 3.5\,\rm{TeV}, \ld = 5\,\rm{GeV}, \mpl = 0.6, \nc = 5$ kept fixed. }
\label{fig:multiplicity_and_boost_nfnc}
\end{figure}
\begin{figure}[h!]
\centering
\includegraphics[width=0.49\textwidth]{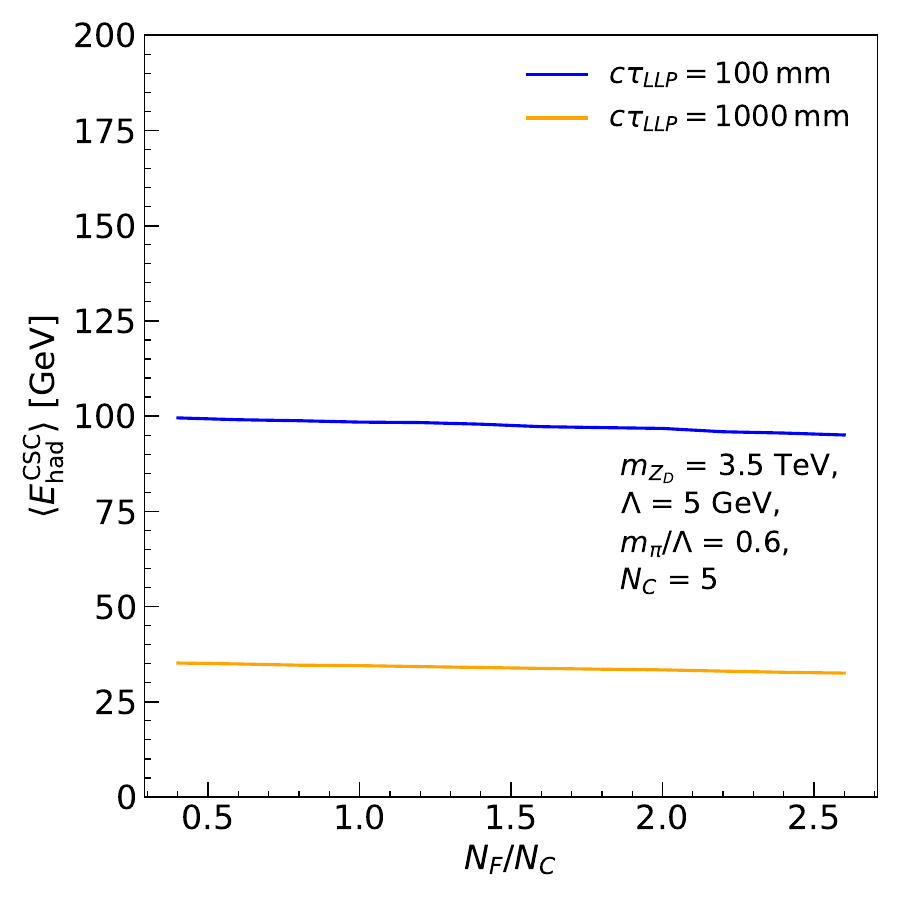}
\includegraphics[width=0.49\textwidth]{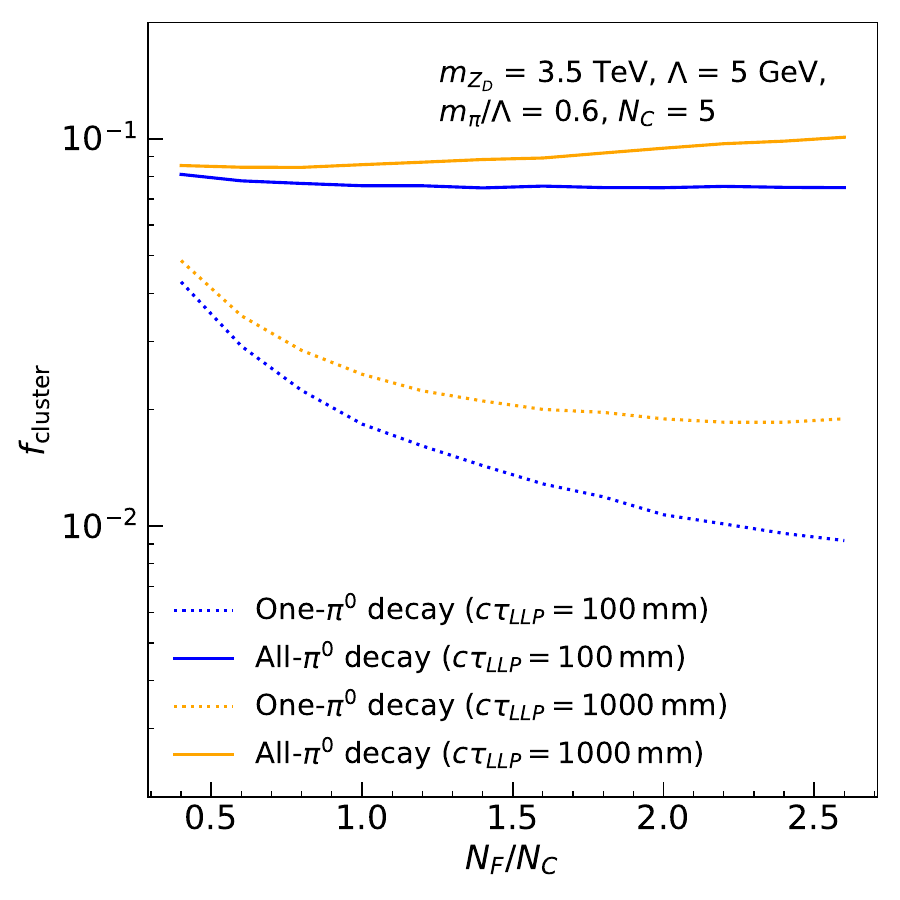}
\caption{The hadronic energy per HV/DS pion deposited in the muon system~(left panel), and fraction of events which has at least one cluster (right panel) with $\mzp = 3.5\,\rm{TeV}, \ld = 5\,\rm{GeV}, \mpl = 0.6, \nc = 5$ kept fixed. }
\label{fig:llp_nfnc}
\end{figure}
In fig.~\ref{fig:multiplicity_and_boost_nfnc} (left panel), we show the dependence of the average multiplicity on $\fc$. This is well understood within the QCD-like region~\cite{Ellis:1996mzs}. The exponential rise in the multiplicity is also reflected in fig.~\ref{fig:multiplicity_and_boost_nfnc}. As the multiplicity increases the boost decreases by almost 20\% as shown in the right panel. Since the pion mass is independent of $\fc$\footnote{In principle, the meson masses for HV/DS scenarios can be a function of $\fc$, however this dependence is neglected in the Snowmass fits and correspondingly in this study.}, $E^{\rm LLP}_{\rm had}$ is independent of $\fc$.

For all-$\pidon$ decay, we therefore expect $N_{\rm LLP}$, $\epsilon_{\rm geo}$ and $\epsilon_{\rm reco}$ to be approximately constant for given lifetime, as reflected in fig.~\ref{fig:llp_nfnc} (right panel). For one-$\pidon$ decay, the situation is different. In this case, only $1/\nf$ fraction of diagonal pions decay, $f_{\rm cluster}$ is inversely proportional to $\fc$. For a fixed $\fc$, comparison of efficiencies for two different lifetimes ($c\tau_{\rm LLP} = 100, 1000$ mm) shows that $\epsilon_{\rm geo}$ is larger for larger lifetimes, however $\epsilon_{\rm recon}$ is smaller. Ultimately the two effects balance each other resulting in approximately constant $f_{\rm cluster}$ for all-$\pidon$ decay scenario. In case of $f_{\rm cluster}$ for one-$\pidon$ decay, $1/\nf$ factor is applicable. 

Given above discussion, we expect the final limits to be a strong function of $\ld$ and $\mpl$ with a weaker dependence on $\fc$. We stress here that our expectation stems from the inclusive nature of the search described in~\cite{CMS:2021juv}. This is not a generic expectation and we recommend drawing such conclusions on a case-by-case basis.

\section{Sensitivity}
\label{sec:sensitivity}
In this section, we will discuss the model independent cross-section upper limits as well as model dependent $Z_D - q_D$ coupling ($g_D^{tot}$) limits obtained for our scenario as a function of various free parameters. The upper limits on the cross section, $\sigma^{up}$, place constraints on $Z_D$ resonance production with $q_D$ final states. We choose to show these upper limits as for a fixed $c\tau_{\rm LLP}$, in the 2D-plane of $\mpl$ and $\ld$ and finally as a function of $\fc$. 

For the original CMS displaced shower search at LHC Run--2, $\mathcal{L} = 137 $ fb$^{-1}$, we require $N_{\rm signal}^{\rm up} \approx$ 6 events at 95\% confidence level (CL), since the number of background events is $ 2.0 \pm 1.0$, and observed events is 3~\cite{CMS:2021juv}. For the prospects at HL-LHC, we consider $\mathcal{L} = 3000 $ fb$^{-1}$. Since a dedicated Level-1 and High Level Trigger targeting displaced shower is installed since Run--3~\cite{Mitridate:2023tbj}, the trigger requirement on the missing transverse energy can be lowered down to MET $>$ 50 GeV, the background events can be negligible after requiring a larger cluster with more hits, while the number of signal events is reduced by 60\%~\cite{Cottin:2022nwp}. We refer to this analysis as `soft trigger', with $N_{\rm signal}^{\rm up} \approx$ 3 events.

\subsection{Sensitivity to \texorpdfstring{$c\tau_{\rm LLP}$}{decay}}
%
\begin{figure}[h!]
\centering
\includegraphics[width=0.49\textwidth]{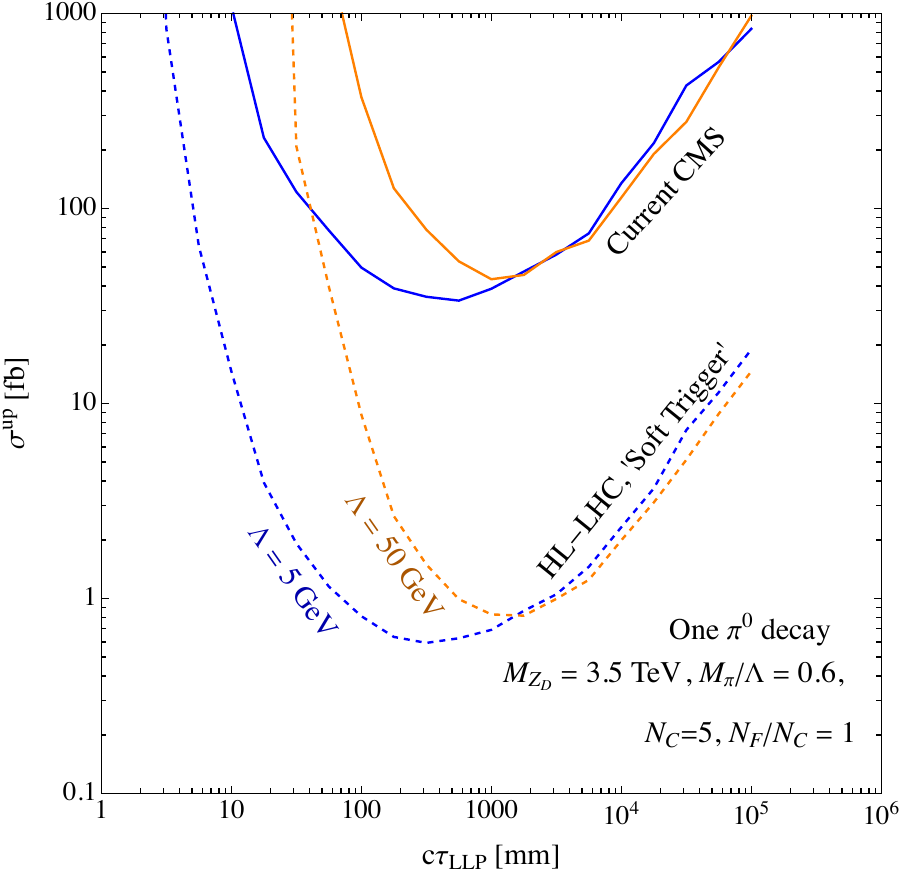}
\includegraphics[width=0.49\textwidth]{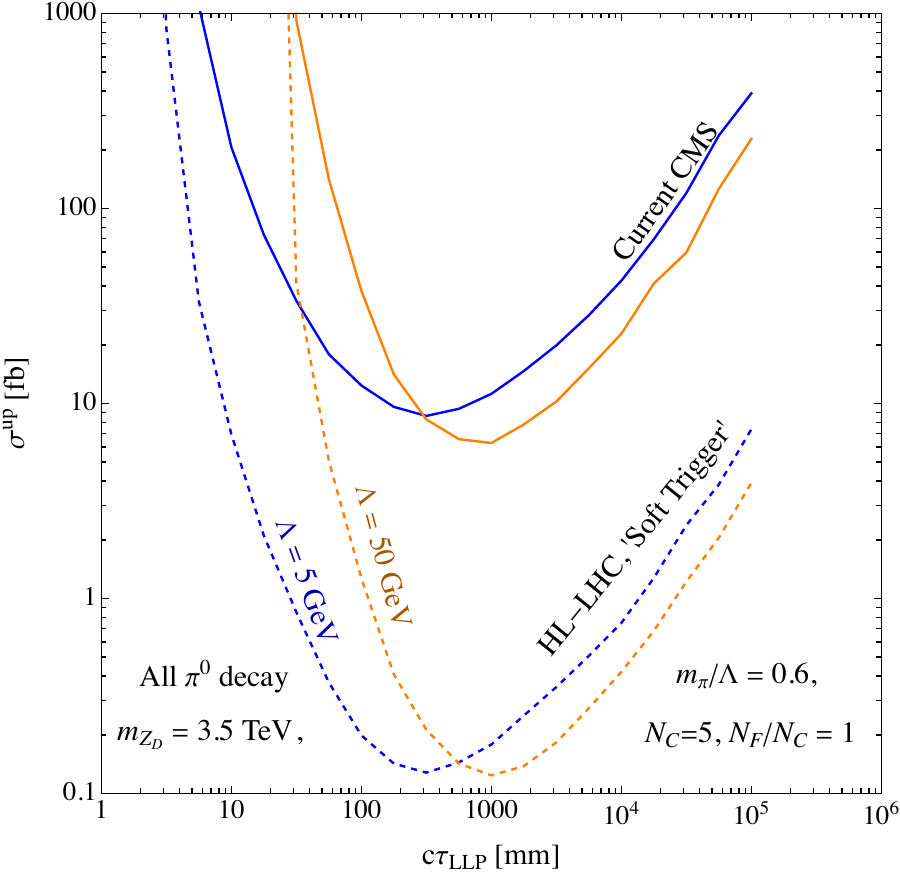}
\caption{Production cross-section upper limits for one-$\pidon$ decay scenario (left panel) and all-$\pidon$ decay scenario (right panel) as a function of proper decay length, $c\tau_{\rm LLP}$, for $\ld$ = 5~(blue) and 50~(orange) GeV, $\mpl=0.6$, $\nc=5, \fc = 1$, $\mzp = 3.5$ TeV. The solid~(dashed) curve represents using the current CMS analysis~(HL-LHC `soft trigger').}
\label{fig:eff_ctau}
\end{figure}

Starting with this section, we first present the model-independent cross-section upper limits as a function of several HV/DS parameters before moving on to model dependent limits in section~\ref{subsec:model_dependent_limits}.

In fig.~\ref{fig:eff_ctau}, we analyze the current and the projected HL-LHC upper limits ($\sigma^{\rm up}$) on the production cross-section for one-$\pidon$ decay (left panel) and all-$\pidon$ decay scenario (right panel) as a function of LLP lifetime for $\ld = 5, 50 \,\rm{GeV}$ (corresponding to $m_\pid = 3, 30\,\rm{GeV}$) while fixing all other parameters. In case of one-$\pidon$ decay, the maximum sensitivity is obtained for $c\tau_{\rm LLP} \sim 600~(2000)$ mm for current CMS reinterpretation and $\ld = 5\, (50)\,\rm{GeV}$. Recent ATLAS search for s-channel emerging jet production~\cite{ATLAS:2025bsz} also targets searches for displaced pions. The search is most sensitive for lifetimes smaller than 100 mm, and provides a complementary sensitivity to our results. Other emerging jet searches which would in principle be sensitive, consider a t-channel production mechanism and hence rely on large $H_T$ or hadronic activity in the event which is absent in our case. In addition, there is an ATLAS search for hadronically-decaying neutral long-lived particles in association with jets or leptons~\cite{ATLAS:2024ocv}. The search targets a displaced jet associated with an unusual calorimeter energy ratio and is accompanied by two resolved jets. This search is able to potentially put limits for dark shower models with $c \tau_{\rm LLP}\approx$ 100 mm, since the detector is located $\approx$ $1-30$~m.

These upper limits can be improved by up to two orders of magnitude at HL-LHC with soft trigger due to much larger $\epsilon_{\rm cut}$ and $\mathcal{L}$, while the qualitative dependence on lifetime remains the same. Correspondingly, at HL-LHC the maximum sensitivity shifts to $c\tau_{\rm LLP} = 320~\rm{mm}, 1800~\rm{mm}$ for the two $\ld$ values.  At larger $\ld$ the maximum sensitivity is obtained for larger lifetimes due to larger pion mass. Since the boost is inversely proportional to pion mass (c.f. fig.~\ref{fig:multiplicity_and_boost_lamda}), heavier pions  need larger $c\tau_{\rm LLP}$ in order to reach the muon endcap detector. For all-$\pidon$ decay scenario, the corresponding limits are stronger by up-to a factor of $1/\nf$. 

Given that the upper limits are strongest at $c\tau_{\rm LLP} = \mathcal{O} (100)\,\rm{mm}$, from now on, we fix $c\tau_{\rm LLP} = 100\,\rm{mm}$ and discuss the sensitivity to HV/DS parameters.

\subsection{Current sensitivity to  \texorpdfstring{$\mpl$}{mpl} and \texorpdfstring{$\ld$}{Lambda}}

\begin{figure}[h!]
\centering
\includegraphics[width=0.49\textwidth]{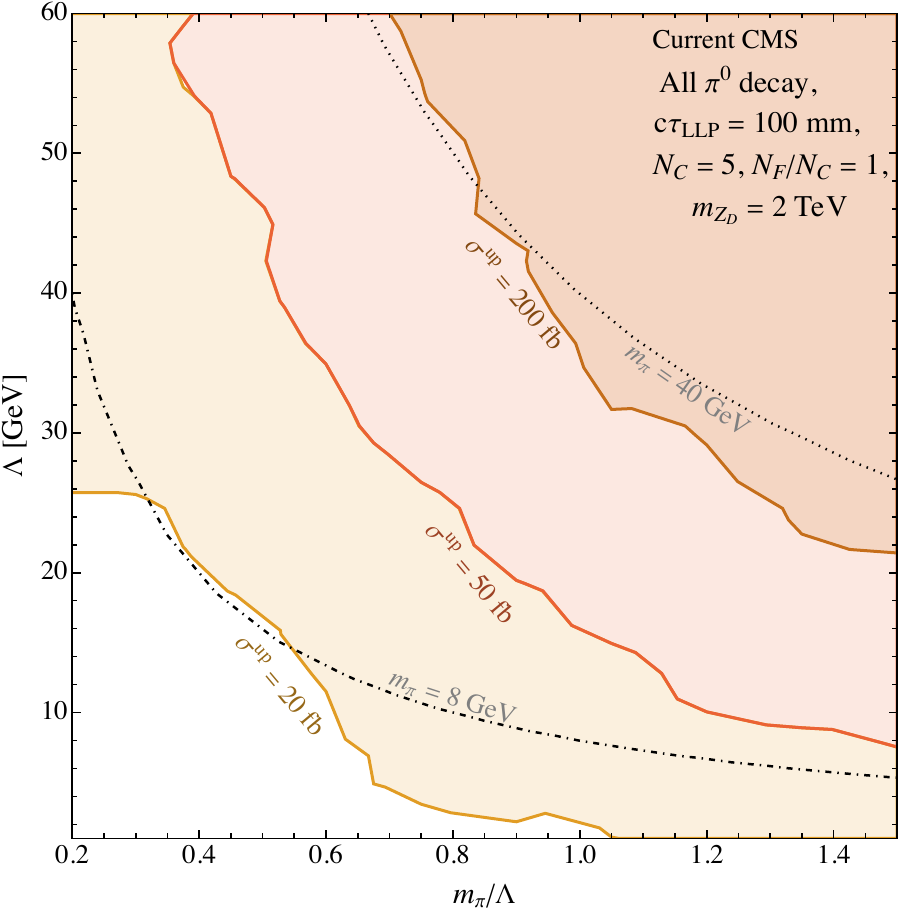}
\includegraphics[width=0.49\textwidth]{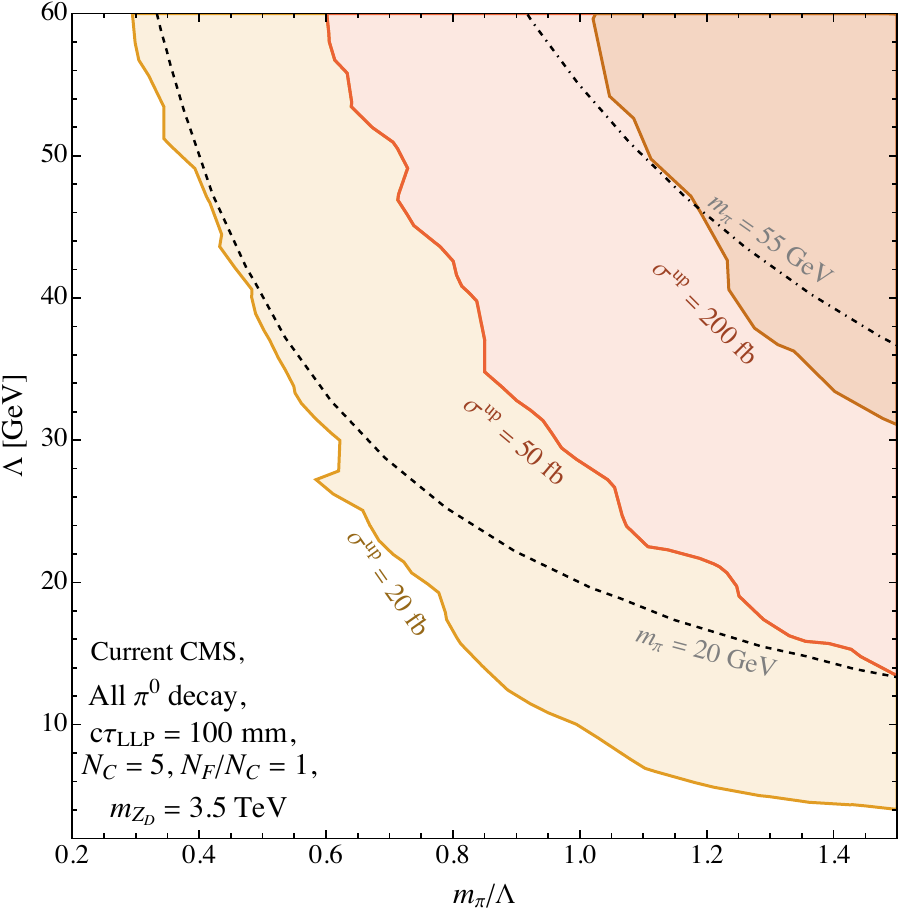}
\caption{ Production cross-section upper-limits for all-$\pidon$ decay scenario with $c\tau_{\rm LLP} = 100\,\rm{mm}$, $\nc=5, \fc = 1$, $\mzp = 2\,\rm{TeV}$ (left panel), and $\mzp = 3.5\,\rm{TeV}$ (right panel) with $\rm{MET} > 200 \,\rm{GeV}$.}
\label{fig:sen_2D_mpilam_lam}
\end{figure}

Having understood the behavior of upper-limits as a function of $c\tau_{\rm LLP}$, we now turn to the upper-limits in 2D plane of $\mpl, \ld$. In fig.~\ref{fig:sen_2D_mpilam_lam}, we show upper-limits on the production cross-section derived using current CMS analysis for all-$\pidon$ decay scenario along with contours for fixed pion mass as a function of $\mpl$ and $\ld$ for fixed $c\tau_{\rm LLP} = 100\,\rm{mm}$ and $\nc = 5$ and $\fc = 1$. As can be seen from fig.~\ref{fig:eff_ctau}, the upper-limits change mildly as a function of pion mass for $c \tau_{\rm LLP} > \mathcal{O}(100)\,\rm{mm}$, therefore we do not consider larger $c\tau_{\rm LLP}$. For $c \tau_{\rm LLP} < \mathcal{O}(100)\,\rm{mm}$, we expect limited reach in pion mass as not many LLPs will reach the muon system. We take two values of $\mzp = 2,\, 3.5 \, \rm{TeV}$ and consider only the all-$\pidon$ decay scenario to illustrate the strongest upper limits obtained. We will later comment on the expected change in the upper-limits in case of one-$\pidon$ decay scenario. As anticipated from the change in the efficiency discussed in section~\ref{sec:eff_dependence}, for a fixed $\mzp$, the upper-limits are strongest for small $\ld$ and $\mpl$.

\begin{figure}[h!]
\centering
\includegraphics[width=0.49\textwidth]{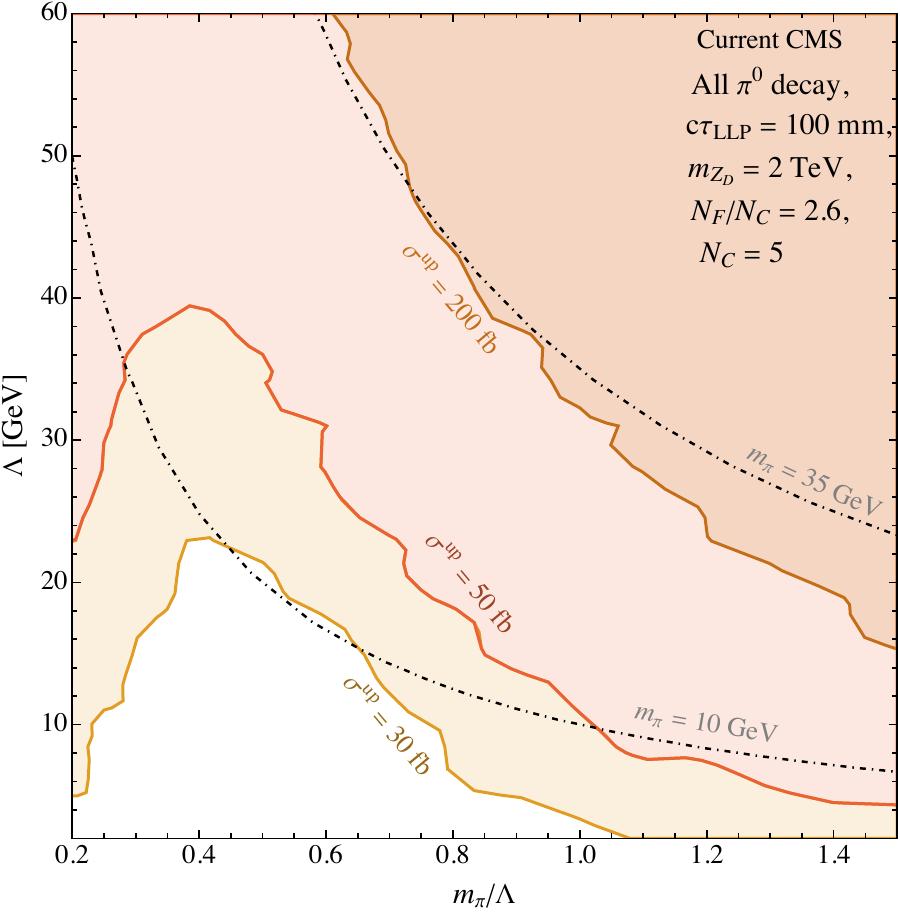}
\includegraphics[width=0.49\textwidth]{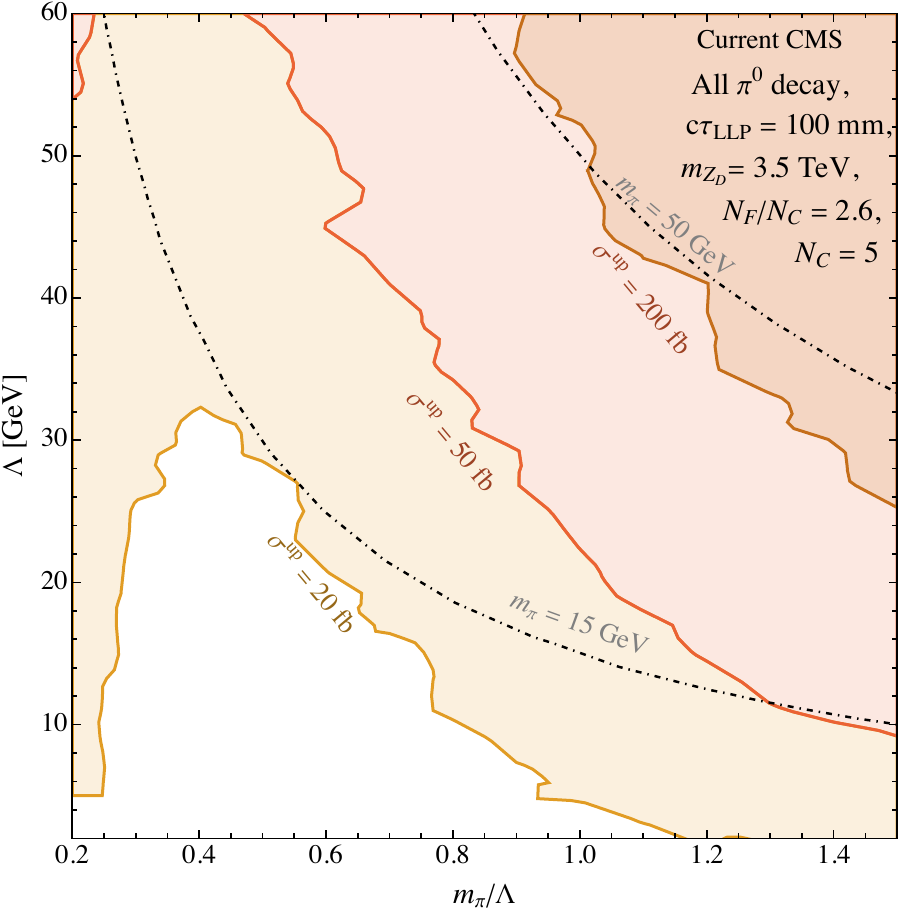}
\caption{Similar to fig.~\ref{fig:sen_2D_mpilam_lam}, but for $\fc = 2.6$. }
\label{fig:sen_2D_mpilam_lam_largefc}
\end{figure}

Comparing fig.~\ref{fig:sen_2D_mpilam_lam} left and right panels, we conclude that heavier $Z_D$ leads to a stronger upper limit as is a direct result of change in the pion boost. This is also in accordance with fig.~\ref{fig:eff_ctau}, where for a fixed $\mzp$, change in $\ld$ also leads to change in the boost\footnote{As noted previously, the quantity of interest should be $\mzp/\ld$, which controls the overall pion boost.}.  

\begin{figure}[h!]
\centering
\includegraphics[width=0.49\textwidth]{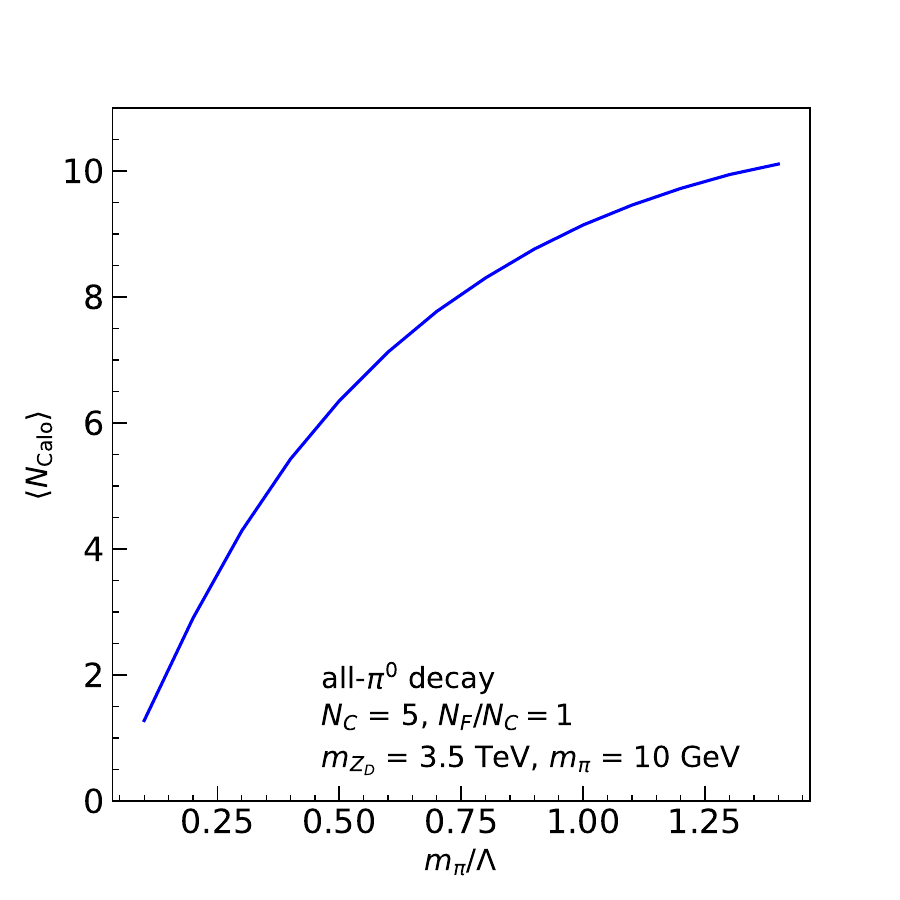}
\includegraphics[width=0.49\textwidth]{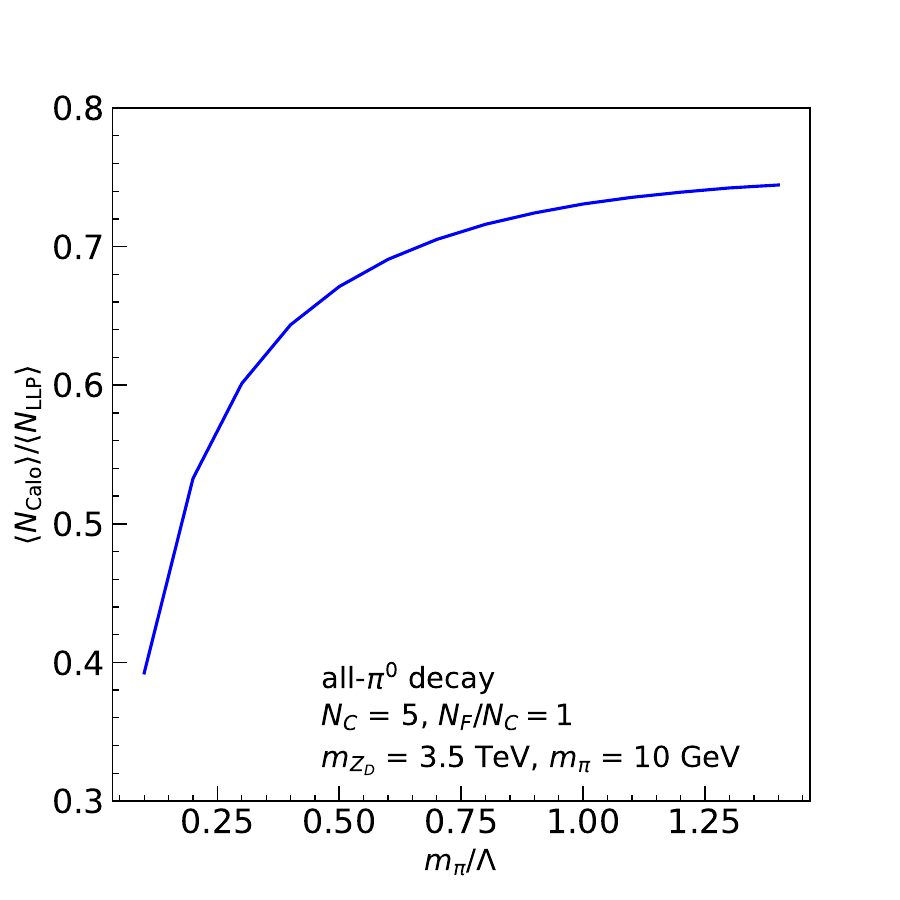}
\caption{Average number of diagonal pions which decay inside the calorimeter system, 1 mm $<l<$ 1 m, for fixed $m_{\pid}$ = 10 GeV, for all-$\pidon$ decay scenario.}
\label{fig:calo}
\end{figure}

In fig.~\ref{fig:sen_2D_mpilam_lam_largefc}, we show the upper limits similar to fig.~\ref{fig:sen_2D_mpilam_lam}, but set $\fc = 2.6$~\footnote{In fig.~\ref{fig:sen_2D_mpilam_lam_largefc} left, the lowest value of cross-section upper-limit contour is 30 fb instead of 20 fb as in the other figures. This is understandable because in this scenario, the 2 TeV $Z^\prime$ leads to smaller average boost for the dark pions, and $N_F/N_C =$ 2.6 makes the pion multiplicity smaller, so the probability for at least one of them to decay inside the muon system is smaller, so the upper limits are larger.}. This choice is motivated by observations in fig.~\ref{fig:mpl_nf_probvec}, where we discussed the characteristic dependence of HV/DS pion multiplicity on $\mpl$ due to competing factors related to large $\nf$ and $\mpl$ dependent {\tt probVector}. The turnover behavior of cross-section upper limit for $\mpl \sim 0.4$ is a direct consequence of this change in the multiplicity. Comparing fig.~\ref{fig:sen_2D_mpilam_lam} and fig.~\ref{fig:sen_2D_mpilam_lam_largefc}, we also see that the upper limits change more drastically at small $\ld, \mpl$ as $\fc$ increases. This change is less dramatic for large $\ld, \mpl$. Comparing the left and right panels of fig.~\ref{fig:final_meson_multiplicity}, it can be seen that the $\pidon$ multiplicity changes drastically at low $\mpl$, while it approaches similar value at as $\fc \to 2.6$. This drastic change in the pion multiplicity explains the rapid changes in the corresponding upper limits for small $\ld, \mpl$ when comparing fig.~\ref{fig:sen_2D_mpilam_lam} and fig.~\ref{fig:sen_2D_mpilam_lam_largefc}. 

Both fig.~\ref{fig:sen_2D_mpilam_lam} and fig.~\ref{fig:sen_2D_mpilam_lam_largefc} demonstrate that the shape of the sensitivity contour coincides with a fixed pion mass unless $\ld$ is comparatively small. At small $\ld$ events contain a jet close to clusters and hence are vetoed. This is because smaller $\ld$ results in larger multiplicity, therefore there is a larger probability for enough number of pions to decay in the calorimeters to form an energetic jet aligned with the CSC cluster. 

This is shown in fig.~\ref{fig:calo}, where we vary $\mpl, \ld$ simultaneously such that the dark pion mass is fixed. The results are shown as a function of $\mpl$, although it should be kept in mind that $\ld$ also varies. We show absolute average number of HV/DS pions  (left panel) as well as the fraction of all pions (right panel) decaying within the calorimeter. The number of dark pions decaying within the calorimeter increases sharply for $\mpl \gtrsim 0.4$, explaining the deviation from mass contour seen in fig.~\ref{fig:sen_2D_mpilam_lam}. 

To get greater sensitivity for such regions, improved analysis techniques are necessary. One possibility would be to reconstruct the decaying pion mass using clustered calorimeter deposits. This can serve as a discriminant between the punch-through jets which is the main background and signal. Another possibility could be to use track based discriminant. Depending on the HV/DS meson mass spectrum and multiplicities, it is possible that the signal creates more number of tracks in the calorimeter/tracker which can also be used as a discriminant against background. We do not study these suggestions any further within this work. 

Finally, we comment on the expected change the upper limits due to one-$\pidon$ decay scenario. The first observation is that the LLPs for one-$\pidon$ decay scenario will be smaller by a factor of $1/\nf$, compared to all-$\pidon$ decay scenario. Therefore, when the upper-limits follow HV/DS pion mass contour, the upper limits for one-$\pidon$ decay scenario will be worse by a factor of $1/\nf$. We also expect the upper limits to follow HV/DS pion mass contours even for larger $\mpl$ because less number of HV/DS pions will decay within the calorimeter and hence $\Delta R (\rm{jet/muon, cluster}) > 0.4$  will be more easily satisfied.

\subsection{Current sensitivity to \texorpdfstring{$\fc$}{nf}}
%
\begin{figure}[h!]
\centering
\includegraphics[width=0.49\textwidth]{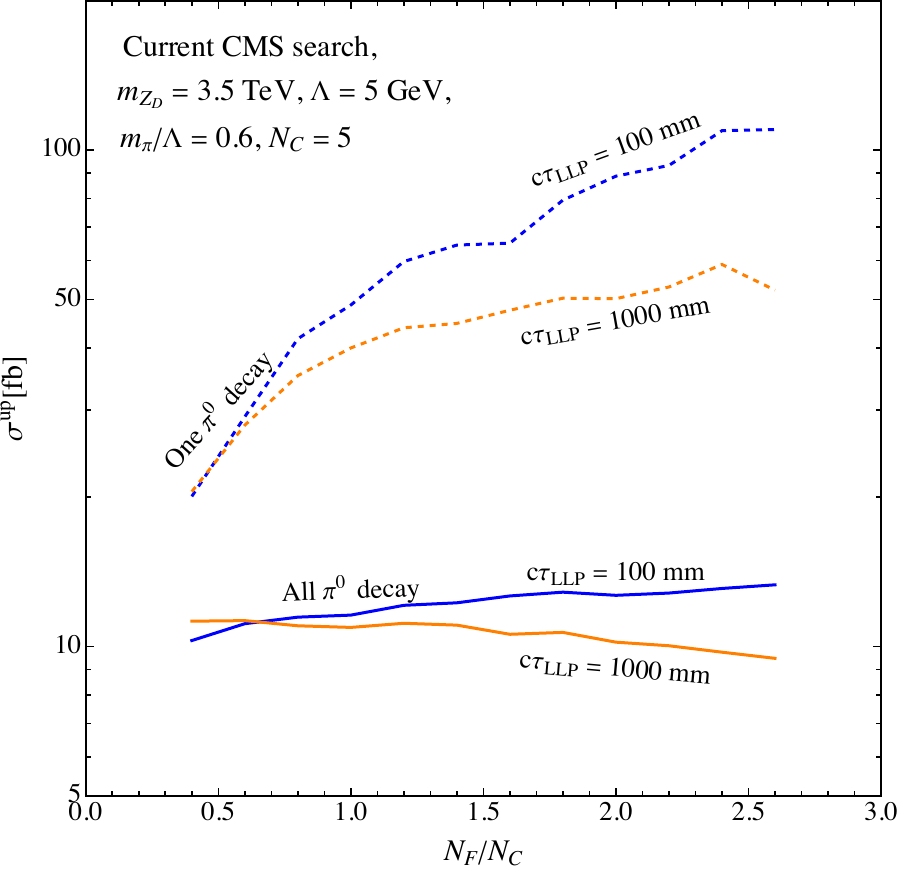}
\includegraphics[width=0.49\textwidth]{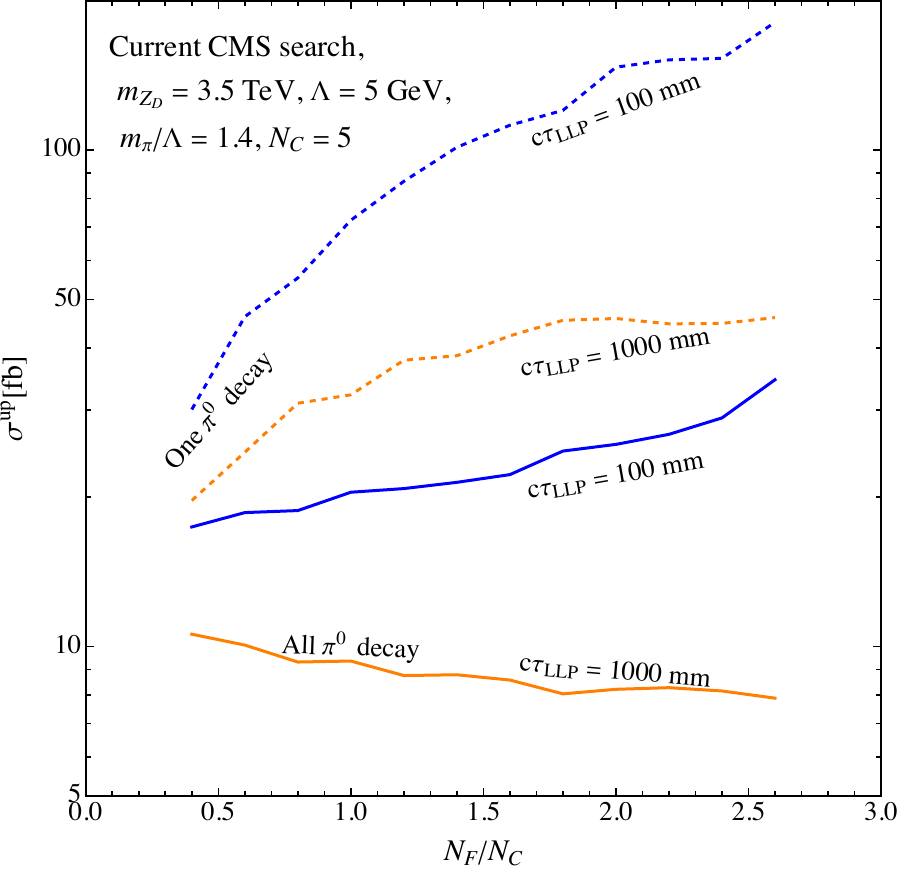}
\caption{Production cross-section upper limits as a function of $\nf$ for current CMS search (top panel) and HL-LHC projection with soft-trigger (bottom panel) for one-$\pidon$ decay scenarios(blue) and all-$\pidon$ decay scenario (orange) with two different LLP lifetimes $c\tau = 100\,\rm{mm}$ (solid lines) and 1000\,$\rm{mm}$ (dashed lines) and two different choices of $\mpl$.}
\label{fig:nfnc}
\end{figure}
In fig.~\ref{fig:nfnc}, we show the cross-section upper limits at the current CMS search as a function of $\fc$ for fixed $\mzp, \ld, \nc$ and $c\tau_{\rm LLP} = 100 \,\rm{mm}$. We consider two LLP lifetimes and two different $\mpl$. The limits get weaker as $\mpl$ increases. This is directly related to the dependence of multiplicity on $\mpl$ and $\fc$ though {\tt probVector}, as discussed in section~\ref{sec:models}. This increase in the $\pidon$ multiplicity means more pions decay in the tracker for $c\tau_{\rm LLP} = 100\,\rm{mm}$, thus vetoing more events due to jet, CSC cluster alignment. For larger lifetime, this effect is not drastic. The shape of the upper-limits is determined by {\tt probVector} fit as well. Therefore the shape of upper-limits changes for the two values of $\mpl$. 

Apart from {\tt probVector}, for which we use eqn.~\eqref{eq:exponential_fit}, we fix all other hadronization parameters to the values discussed in section~\ref{sec:sim_setup}. However, these values are not exactly known and can introduce additional, potentially uncontrolled uncertainties in the cross-section upper limits presented here. To this extent in Appendix~\ref{app:hadronization_parameters}, we discuss the effect of variation of hadronization parameters primarily on the number of HV/DS $\pidon$ as well as the total HV/DS final state meson multiplicities. We expect variation of these two quantities to be of central importance to limit setting as it controls the number of LLPs reaching the muon endcap detector. We find that generally the hadronization uncertainties do not lead to large variation of the number of HV/DS $\pidon$. 

\subsection{HL-LHC projections for  \texorpdfstring{$\mpl$}{mpl} and \texorpdfstring{$\ld$}{Lambda}}
Along with the current sensitivity, we also derive the projected cross-section upper limits at the HL-LHC. At the HL-LHC, we assume an optimistic trigger of MET $> 50$ GeV, and increase the $N_{\rm hits}$ to 370 demanding a larger cluster. With these two improvements, one can assume a background free scenario~\cite{Cottin:2022nwp} to derive the upper limits. Similar to the previous subsection, we show them in the $\mpl$--$\ld$ plane for $\fc = 1, 2.5$ and also as a function of $\fc$. 

\begin{figure}[h!]
\centering
\includegraphics[width=0.49\textwidth]{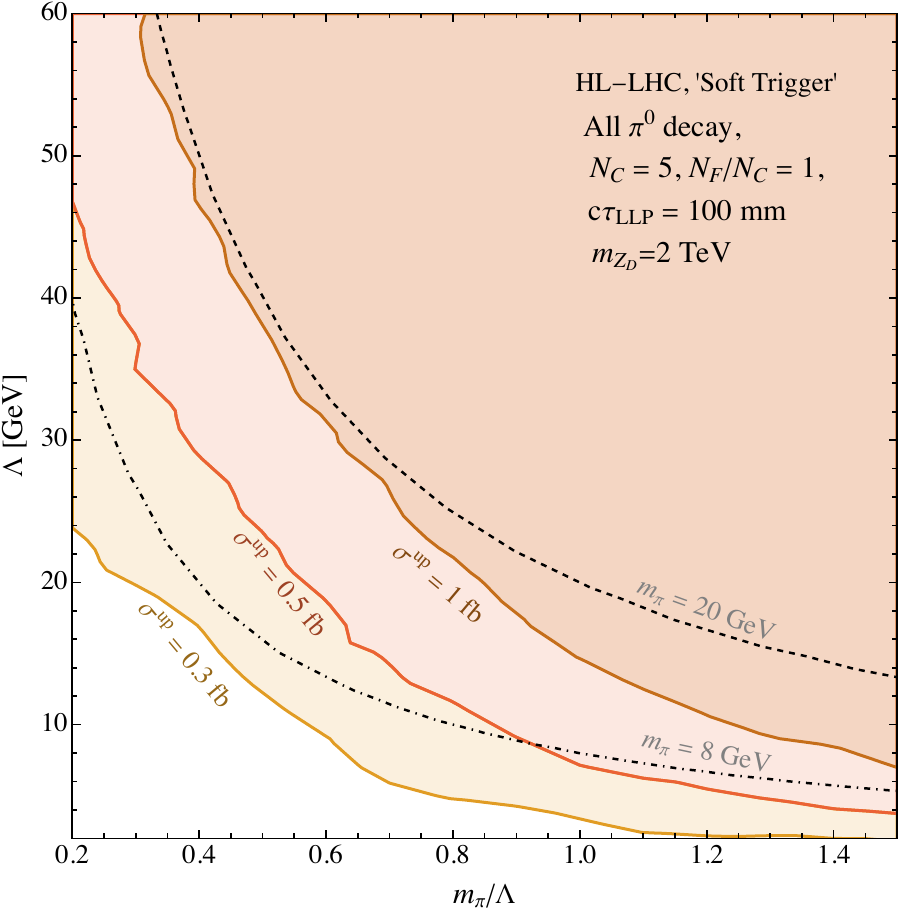}
\includegraphics[width=0.49\textwidth]{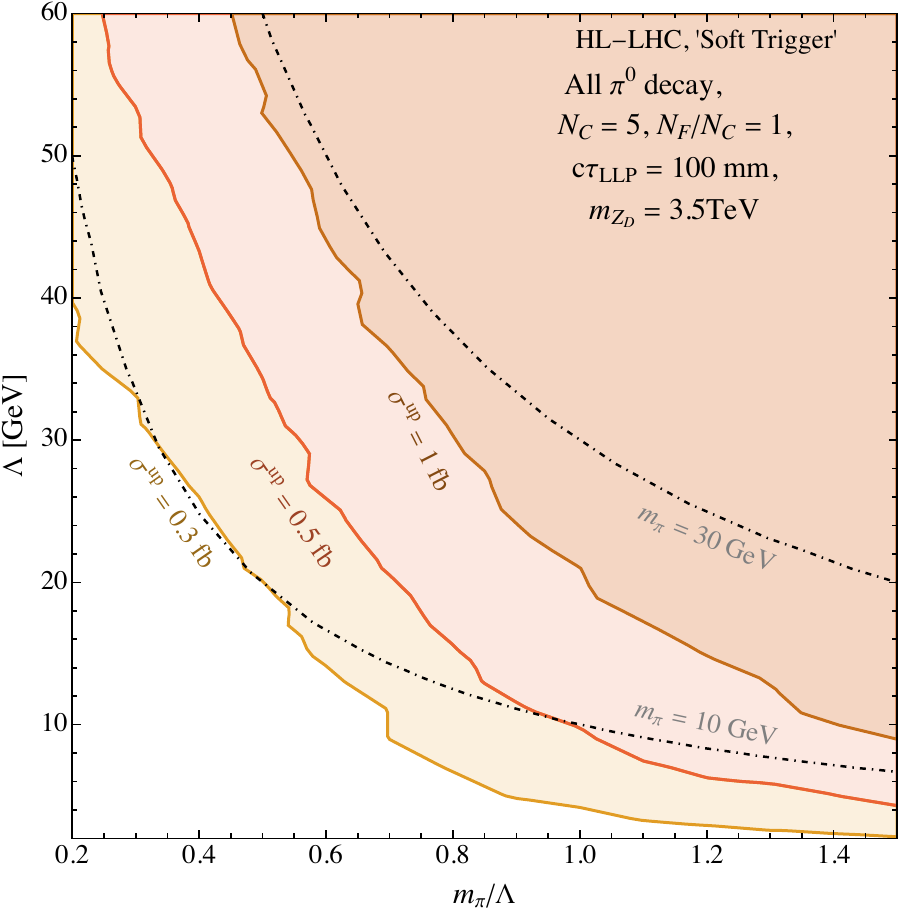}
\caption{Similar to fig.~\ref{fig:sen_2D_mpilam_lam} but HL-LHC cross-section upper limits projections using `soft trigger'.}
\label{fig:mpi_lam_HLLHC}
\end{figure}
\begin{figure}[h!]
\centering
\includegraphics[width=0.49\textwidth]{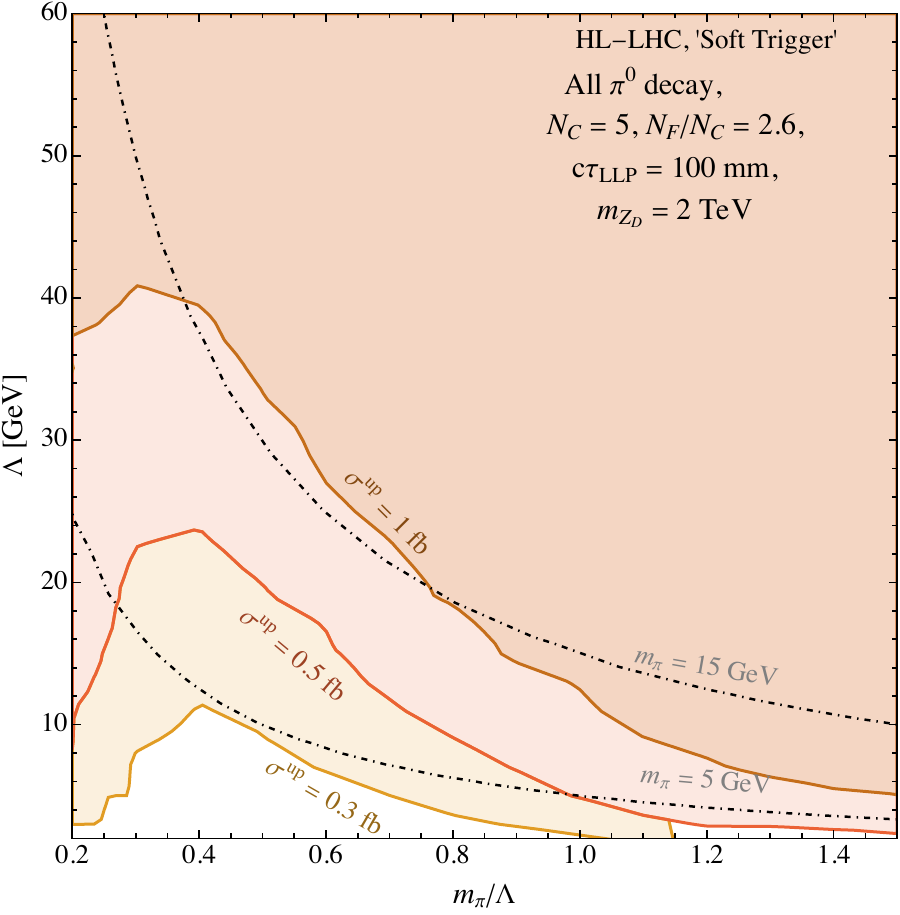}
\includegraphics[width=0.49\textwidth]{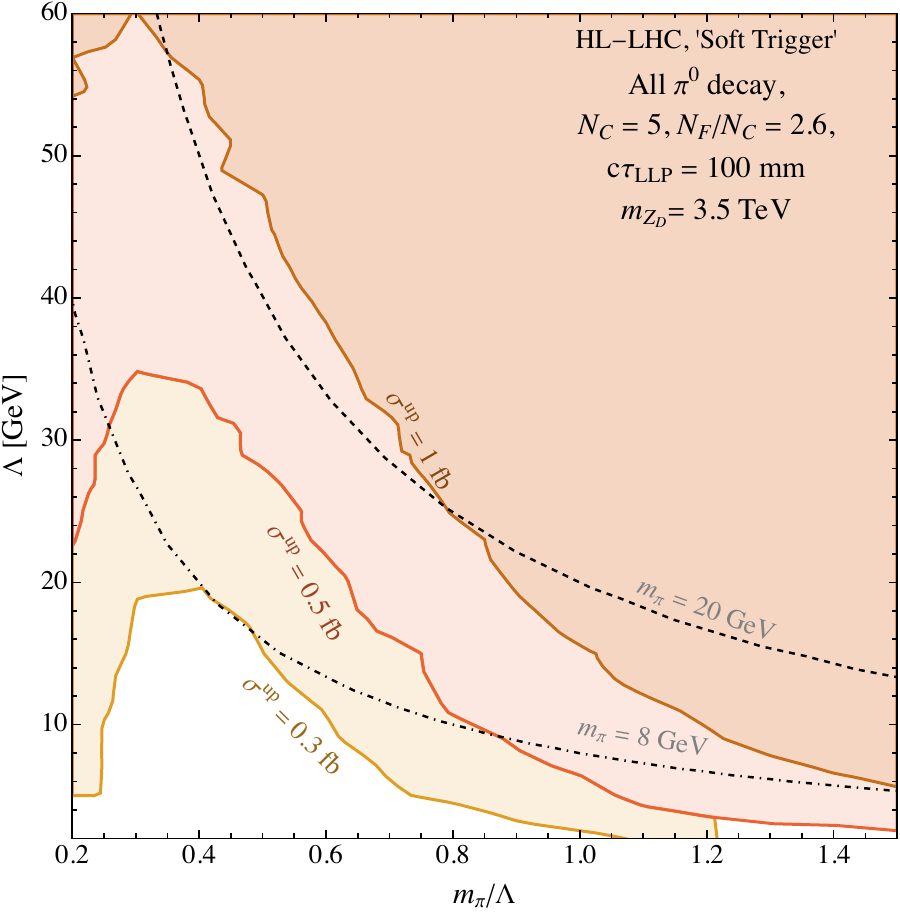}
\caption{Similar to fig.~\ref{fig:sen_2D_mpilam_lam} but HL-LHC cross-section upper limits projections using `soft trigger' and $\fc = 2.6$.}
\label{fig:mpi_lam_HLLHC_26}
\end{figure}

We show the projected cross-section upper limits in the $\mpl$ -- $\ld$ in fig.\ref{fig:mpi_lam_HLLHC} and fig.\ref{fig:mpi_lam_HLLHC_26} for $\fc = 1$ and 2.6 respectively. We note that the upper limits get stronger by up to an order of magnitude. We also notice that for $\fc = 1$ and $\mzp = 2\,\rm{TeV}$, the theoretically cautioned area representing $g_D > 1$ is absent, while at 3.5 TeV, only a small area features $g_D > 1$. This is expected as the upper-limits are very small, correspondingly the value of $g_D$ probed at any point is also small. 

The general behavior of the upper-limits remains the same for $\fc = 2.6$, except, the upper limits get somewhat weaker compared to $\fc = 1$, as also noted in the previous subsection. 

Finally, in fig.~\ref{fig:nfnc_HL_LHC} we show the projected cross-section upper-limits. The general qualitative behavior remains similar to that shown in fig.~\ref{fig:nfnc}, with the limits getting stronger by up to an order of magnitude.

\subsection{HL-LHC projections for \texorpdfstring{$\fc$}{nf}}

%
\begin{figure}[h!]
\centering
\includegraphics[width=0.49\textwidth]{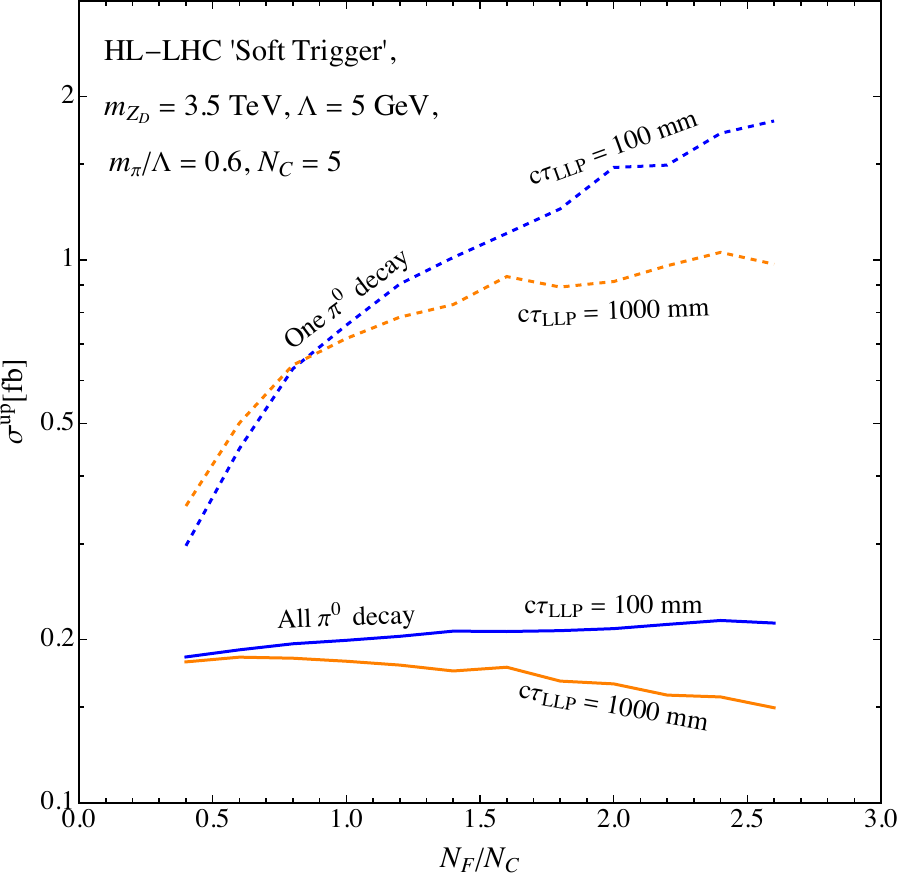}
\includegraphics[width=0.49\textwidth]{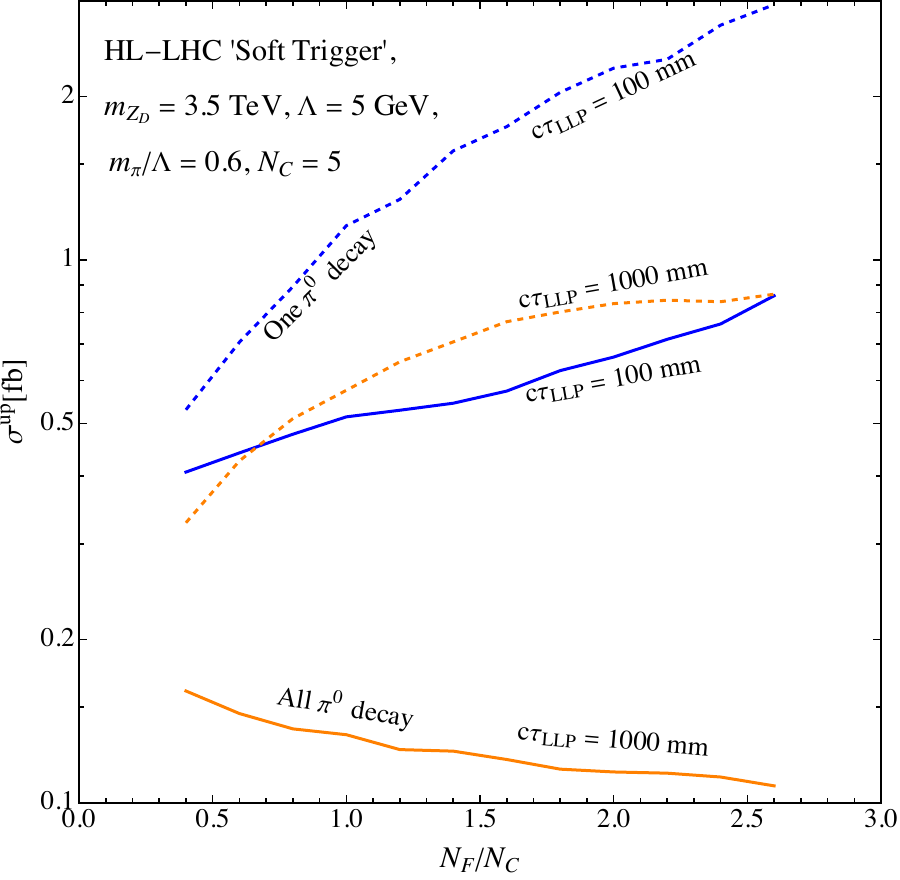}
\caption{Similar to fig.~\ref{fig:nfnc} but HL-LHC cross-section upper limits projections using `soft trigger'.}
\label{fig:nfnc_HL_LHC}
\end{figure}

Similar to fig.~\ref{fig:nfnc}, in fig.~\ref{fig:nfnc_HL_LHC} we show the HL-LHC projected cross section upper limits as a function of $\fc$. As expected the upper limits get stronger however the general trends seen in fig.~\ref{fig:nfnc} remain valid. 
\subsection{Model dependent exclusions}
\label{subsec:model_dependent_limits}

We are now in a position to derive and discuss model dependent exclusions for our scenarios. We will reinterpret fig.~\ref{fig:sen_2D_mpilam_lam}, ~\ref{fig:sen_2D_mpilam_lam_largefc}, ~\ref{fig:mpi_lam_HLLHC}, ~\ref{fig:mpi_lam_HLLHC_26} and derive the maximum allowed $Z_D$ coupling to all HV/DS quarks ($g^{tot}_D$). We will take into account the upper limits as derived fig.~\ref{fig:sen_2D_mpilam_lam}, ~\ref{fig:sen_2D_mpilam_lam_largefc}, ~\ref{fig:mpi_lam_HLLHC}, ~\ref{fig:mpi_lam_HLLHC_26} as well as limits on $Z_D$ to SM quarks as obtained from the SM dijet resonance searches as described in Appendix~\ref{app:theory_analysis}. Using the limits from dijet resonance, we fix the value of $Z_D \to q\bar{q}$ coupling such that the dijet bounds are saturated. As the HV/DS pion lifetime depends on the difference between the HV/DS quark axial vector couplings ($\rm{Tr}[(\mathcal{Q}^{D}_{A})\cdot T_i]$), a desired lifetime can always be obtained by appropriate charge differences.

\begin{figure}[h!]
\centering
\includegraphics[width=0.49\textwidth]{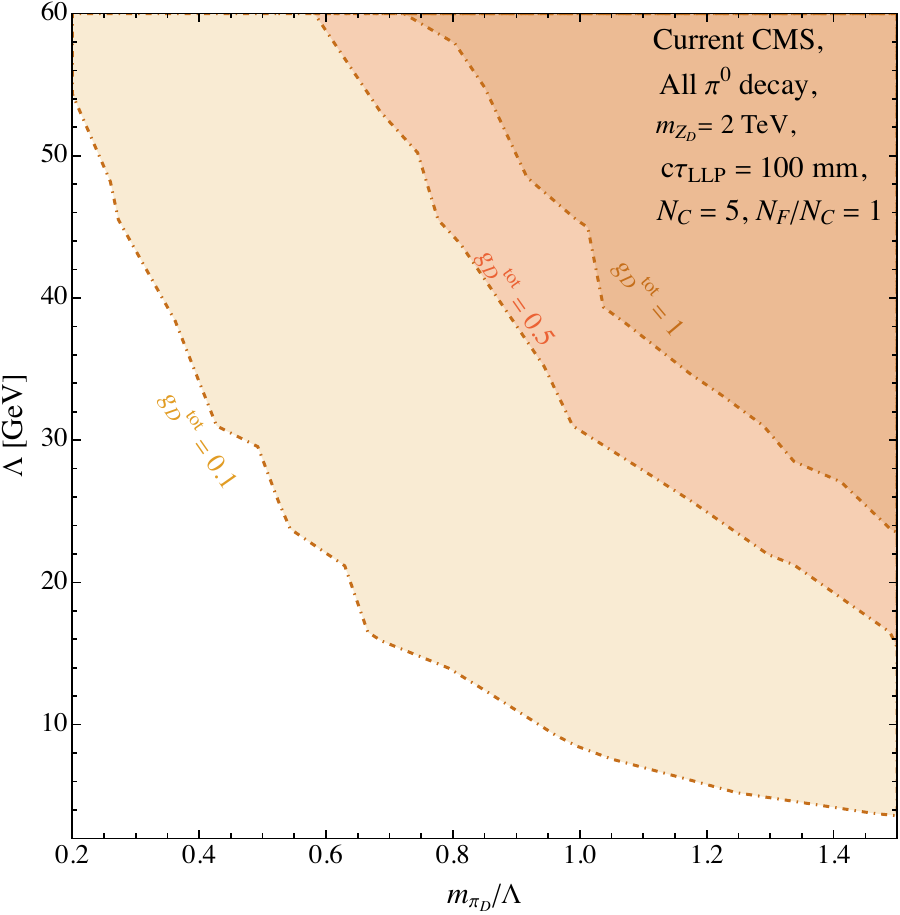}
\includegraphics[width=0.49\textwidth]{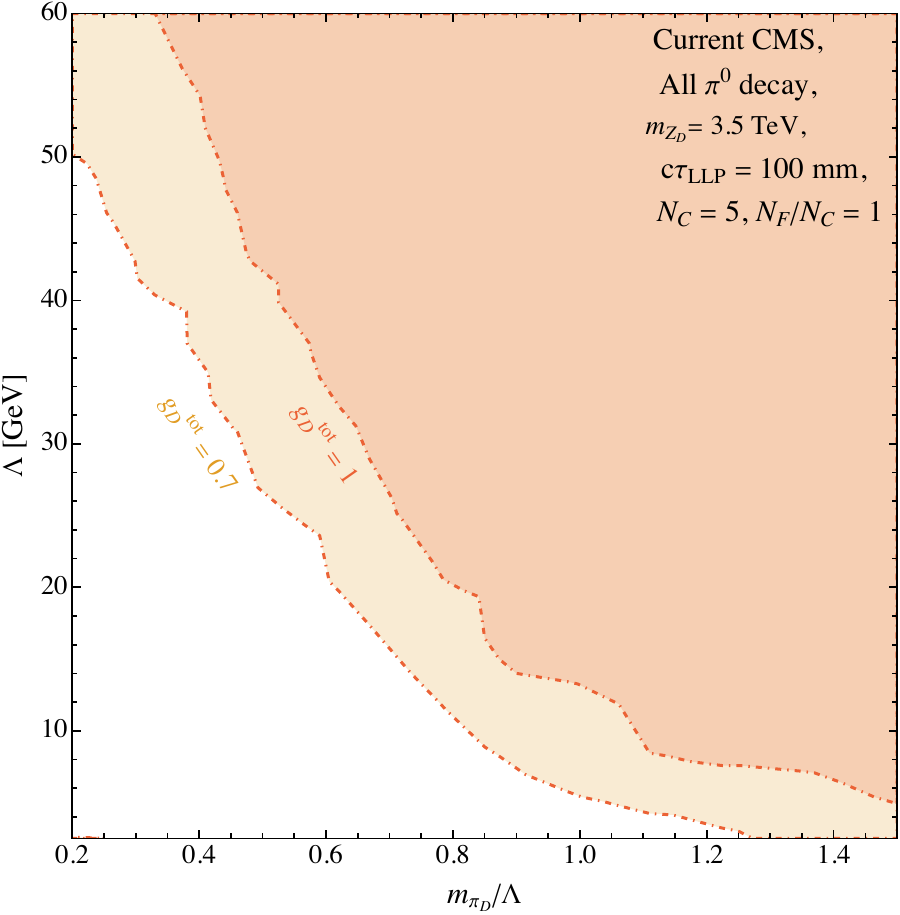}
\caption{Current exclusions on $\mpl$ -- $\ld$ for fixed $c\tau_{\rm LLP} = 100\,\rm{mm}$, $\fc= 0.4$ (left panel) and 2.6 (right panel), as well as show the exclusions for $\mzp = 2$, 3.5 TeV derived by solving Eq.~\ref{eq:gd_coupling_limit}.}
\label{fig:Exclusion_current_fc_1}
\end{figure}
\begin{figure}[h!]
\centering
\includegraphics[width=0.49\textwidth]{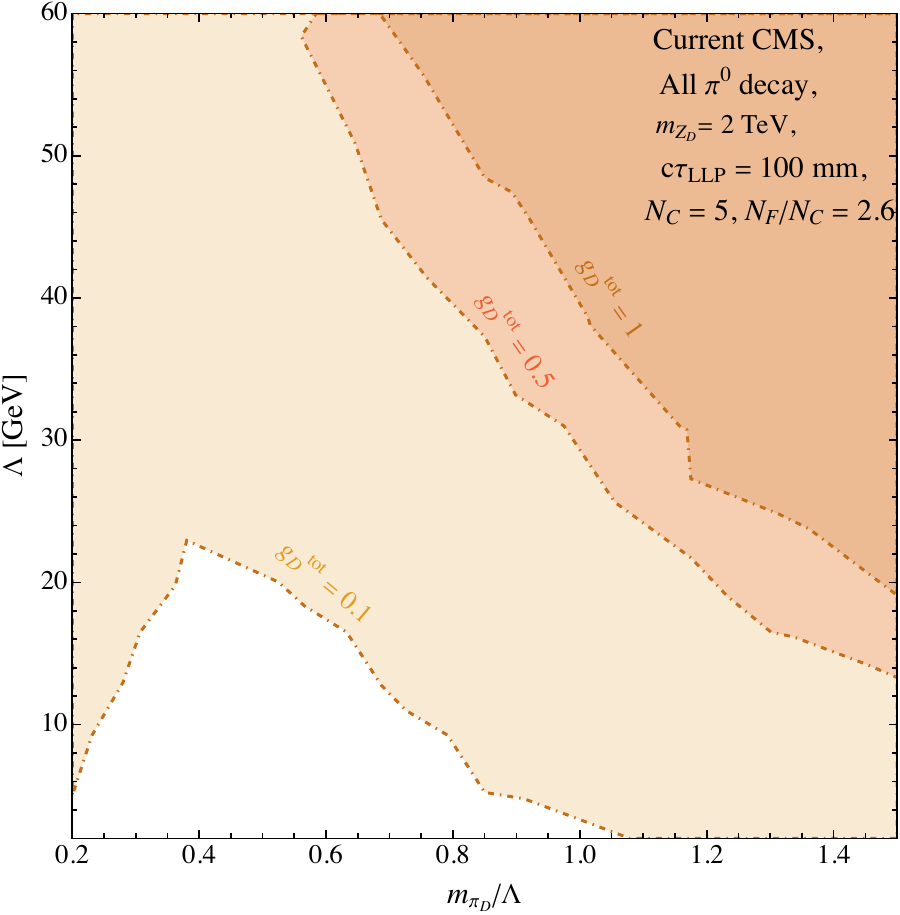}
\includegraphics[width=0.49\textwidth]{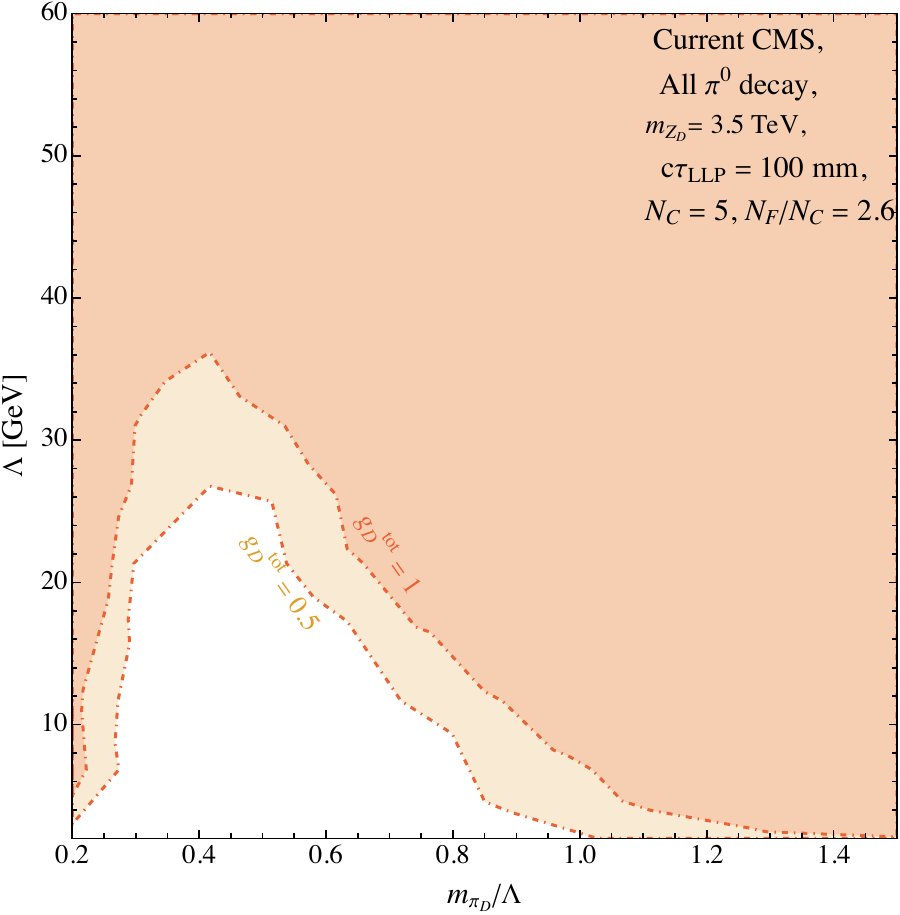}
\caption{Same as fig.~\ref{fig:Exclusion_current_fc_1} but for $\fc = 2.6$}
\label{fig:Exclusion_current_fc_26}
\end{figure}
\begin{figure}[h!]
\centering
\includegraphics[width=0.49\textwidth]{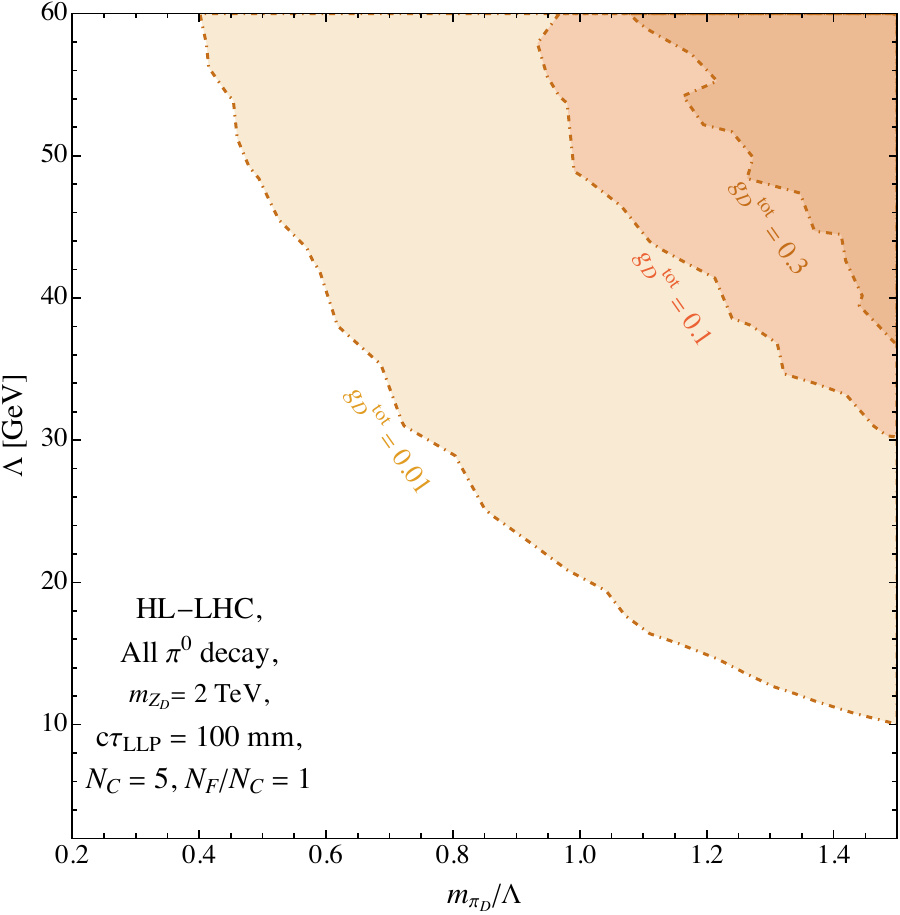}
\includegraphics[width=0.49\textwidth]{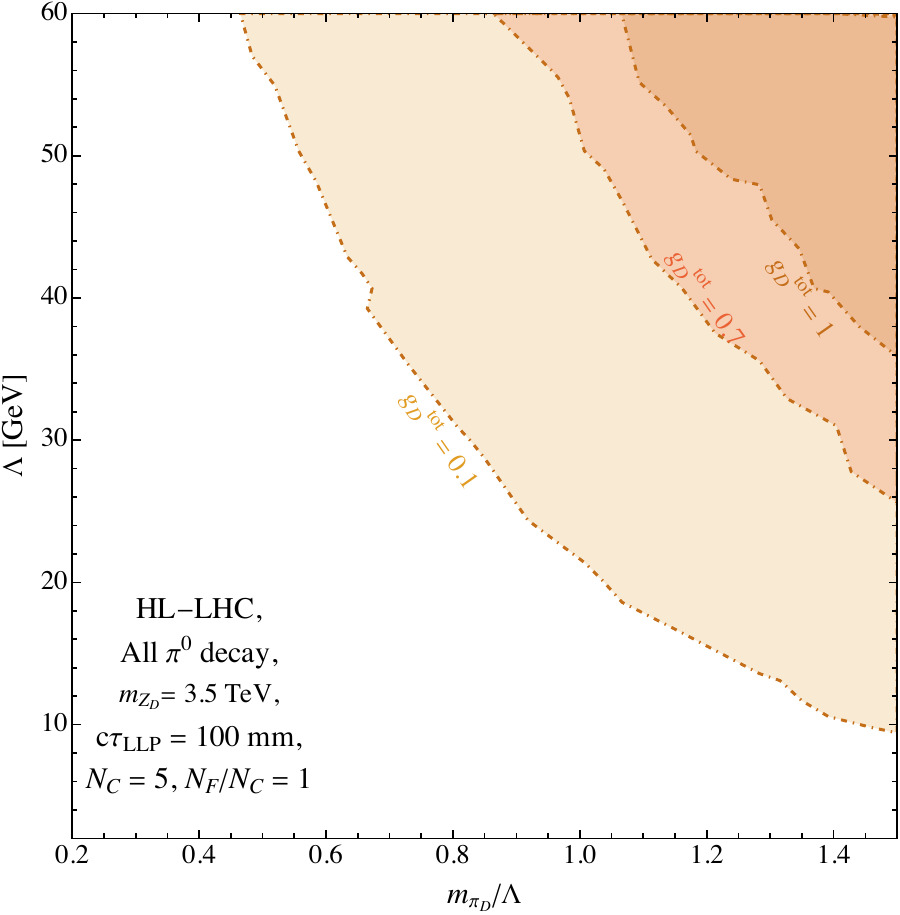}
\caption{Same as fig.~\ref{fig:Exclusion_current_fc_1} but for HL-LHC with $\fc = 1$.}
\label{fig:Exclusion_HL_LHC_fc_1}
\end{figure}
\begin{figure}[h!]
\centering
\includegraphics[width=0.49\textwidth]{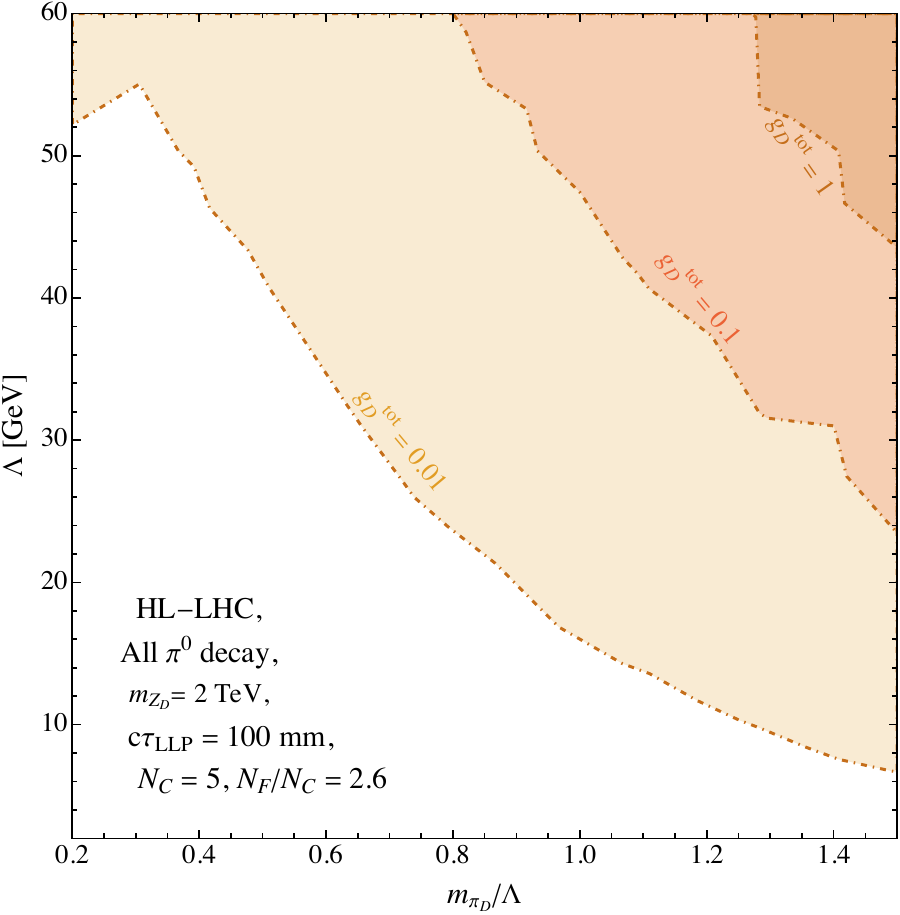}
\includegraphics[width=0.49\textwidth]{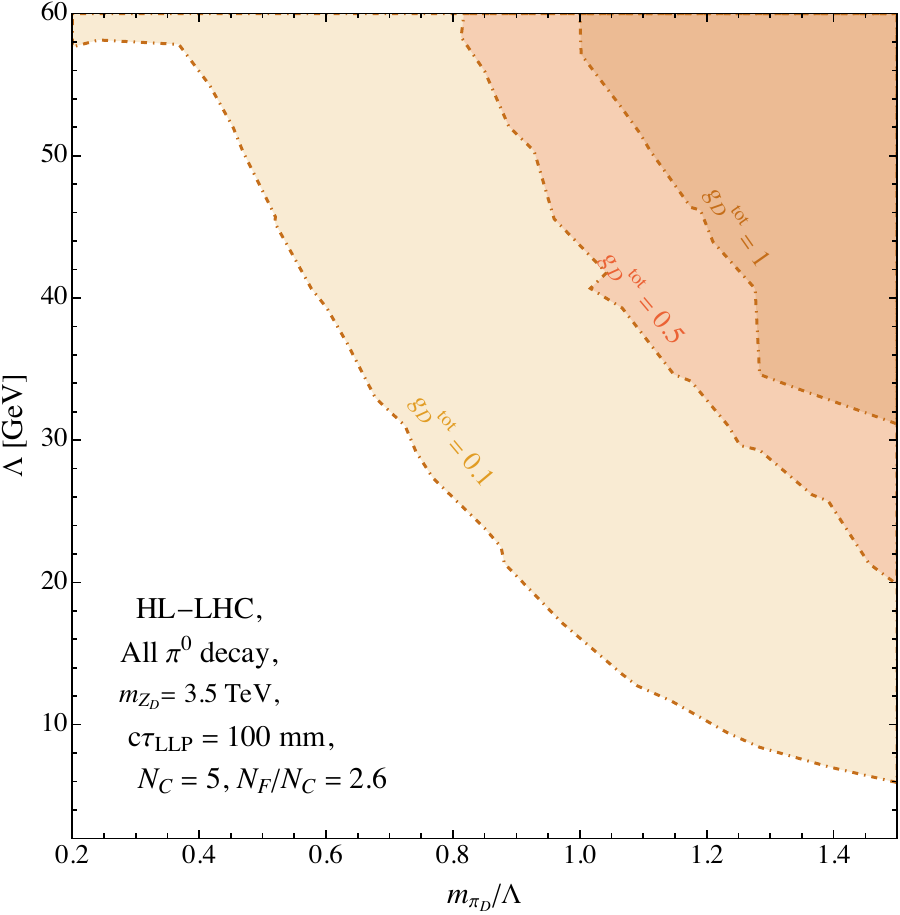}
\caption{Same as fig.~\ref{fig:Exclusion_current_fc_1} but for HL-LHC for $\fc = 2.6$.}
\label{fig:Exclusion_HL_LHC_fc_26}
\end{figure}

These limits are shown in fig.~\ref{fig:Exclusion_current_fc_1},\ref{fig:Exclusion_current_fc_26},\ref{fig:Exclusion_HL_LHC_fc_1},\ref{fig:Exclusion_HL_LHC_fc_26} for all-$\pidon$ decay scenario with a fixed LLP lifetime $c\tau_{\rm LLP}  = 100\,\rm{mm}$, for two different values of $\mzp = 2, \,3.5\,\rm{TeV}$, for two different values of $\fc = 0.4 $ and 2.6. Throughout the  $\ld, \mpl$ plane, the $Z_D$ to SM quark coupling limit as derived from resonance searches remains unchanged. The limits on $g_D^{tot}$ get stronger for small $\ld, \mpl$. This is because the upper-limits are stronger for smaller $\ld, \mpl$. This combined with the constant $Z_D$ to SM quarks coupling leads to stronger limits on $g_D^{tot}$ for small $\ld, \mpl$. We do not show $g_D^{tot} > 1$ contours. This is because once  $g_D^{tot} > 1$ the $Z_D$ starts to couple with the HV/DS sector more and more strongly. Such strong coupling may lead to HV/DS pion mass splitting through radiative corrections. While the exact value of  $g_D^{tot}$ at which such effects matter is unknown, we do not show larger values to remain conservative. It is clear that the  $g_D^{tot}$ contours follow the behavior of the cross-section upper limits described in the previous subsections. At HL-LHC the upper limits get stronger by order of magnitude as expected. 

We re-iterate here several caveats, which must be kept in mind. First, we emphasis that these are model dependent results. We assume that the HV/DS pions decay via mixing with the Goldstone mode of $U(1)_D$ Higgs. Second, and more importantly, we remind the reader that we replace the non-perturbative scale ($\tilde \Lambda_D$) associated with the scale of chiral symmetry breaking with the scale $\ld$, where perturbative one-loop running coupling diverges. This affects the relative separation between $\ld$ and the HV/DS meson masses. The associated uncertainty due to these approximations can not be easily addressed without further in depth investigation of lattice results, which is beyond the scope of this work. Never-the-less, these limits exemplify the ability of the CMS analysis to constrain the total $Z_D$ HV/DS quark coupling. Collectively these results show that strongest limits on the couplings are found at small $\ld, \mpl$. The limits for $\fc = 2.6$ are weaker as compared to $\fc = 1$. As expected the limits at HL-LHC are up to an order magnitude stronger than the existing search limits. 
\section{Conclusion}
\label{sec:conclusion}

The growing interest in HV/DS theories warrants an in-depth thinking about their characteristics and associated effects on experimental searches. This includes the search design, reinterpretation and strategies for presentation of experimental results. From this point of view, in this work, we reinterpreted the CMS search for LLPs decaying in the muon endcap detector, creating the so-called CSC clusters and provided HL-LHC projections for HV/DS scenarios. We concentrated on presenting the results in as model independent fashion as possible and provided one model dependent exclusion as an example. 

In this context, the HV/DS scenarios we considered contained $\rhod \to \pid \pid$ decays and the LLPs comprised of diagonal pions. We considered two extreme scenarios, first where one diagonal pion decays and second where all diagonal pions decay. These two extremes help illustrate the possible limits obtained via reinterpretation. We insist here that depending on the model building efforts these choices can be diversified. We leave this for future investigations. 

Rather than adopting usual strategy of fixing the HV/DS parameters or using some ``effective parameters'', we varied the fundamental HV/DS parameters, namely $\mpl, \ld,$ and $\fc$. We showed that these parameters have a complex interplay which affects the LLP multiplicities as well as kinematic distributions. In particular, the multiplicity is affected due to the dependence of hadronization parameter known as {\tt probVector}. To this end, we analyzed in detail the effect of various {\tt probVector} fits, inspired by the Lund string model and discussed their effect on the LLP multiplicity. We showed that as long as {\tt probVector} is an increasing function of $\mpl$, the LLP multiplicity changes in a non-trivial way as a function of both $\mpl$ and $\fc$. This further has implications of the analysis reinterpretation. 

Our results show that the search sensitivity depends on the LLP mass and the multiplicity. The multiplicity has important consequences for current search design as the promptly decaying fraction of LLPs leave hadronic deposits in the inner part of the detector which is identified as a jet. The alignment of the jet with the CSC cluster vetoes events with high LLP multiplicity, affecting the reach. 

We demonstrate that under certain plausible simulation assumptions -- including a Pythia-based dark hadronization model -- the CMS displaced shower search can be sensitive to Hidden Valley parameter space ($\ld$, $\mpl$) and that flavor/color choices can modulate sensitivity. However, we caution that due to significant uncertainties in modeling dark-sector hadronization, the quantitative constraints on Hidden Valley theory space should be viewed as illustrative. Extracting or constraining underlying theory parameters in a direct way remains challenging, in absence of better control over non-perturbative modeling. Nevertheless, our approach highlights how searches for displaced showers can be used to probe Hidden Valley scenarios in a more theory-informed way, so long as the theoretical uncertainties are explicitly acknowledged.

Finally, we exemplified the reach of the search via one model dependent exclusion. This exclusion suffers from theoretical uncertainties. These theory uncertainties arise because we equate the perturbative scale $\ld$ in the theory with the non-perturbative scale $\tilde \Lambda_D$ associated with the scale of chiral symmetry breaking. Never-the-less our results show that the existing CMS search constrains the total $Z_D$ to HV/DS quark coupling over a large part of HV/DS parameter space. These limits show the best case scenarios, and will worsen if the LLP lifetime is changed or if less number of pions decay. 

Our reinterpretation strategy thus shows potential pathways to explore HV/DS theories through LHC experimental searches. The systematic interplay of several theory parameters leads to deeper theory understanding but also identifies potential weaknesses in the search strategies. The presentation of results in the form of HV/DS parameters rather than effective parameters or mediator sector parameters may further be explored to develop new ways to present HV/DS search results in the future.
\acknowledgments
W.L. is supported by National Natural Science Foundation of China (Grant No.12205153). S.K. and J.L. are supported by the FWF research group funding FG1 and FWF project number P 36947-N.  We thank Matt Strassler and Torbj\"orn Sj\"ostrand for useful comments and discussions. S.K. thanks W. Porod, N. Evans, A. Deandrea and G. Cacciapaglia for useful discussions. 

\appendix 

\section{Meson multiplicity as a function of $\mpl$ and $\fc$}
\label{app:meson_multiplicity}
We use the {\tt Hidden Valley}~\cite{Carloni:2011kk,Carloni:2010tw} module of {\tt PYTHIA}~v8.311\footnote{We note here that a more updated {\tt PYTHIA} version~{\tt v8.313} is available, however has not been used in this work. The results presented here will not change in case {\tt PYTHIA~v8.312} is used. See also~\cite{Kulkarni:2024okx} for Herwig darkshowers implementation. As this version of Herwig is not yet public, we do not use it in this work.}~\cite{Sjostrand:2007gs, Bierlich:2022pfr}, to simulate $pp\to Z_D \to q_D \bar{q}_D $, with associated HV/DS parton shower and hadronization. We set all HV/DS pions to be stable for this part of the study but allow the rho mesons to decay to pions as appropriately stated throughout the discussion.

\begin{figure}[h!]
\centering
\includegraphics[width=0.49\textwidth]{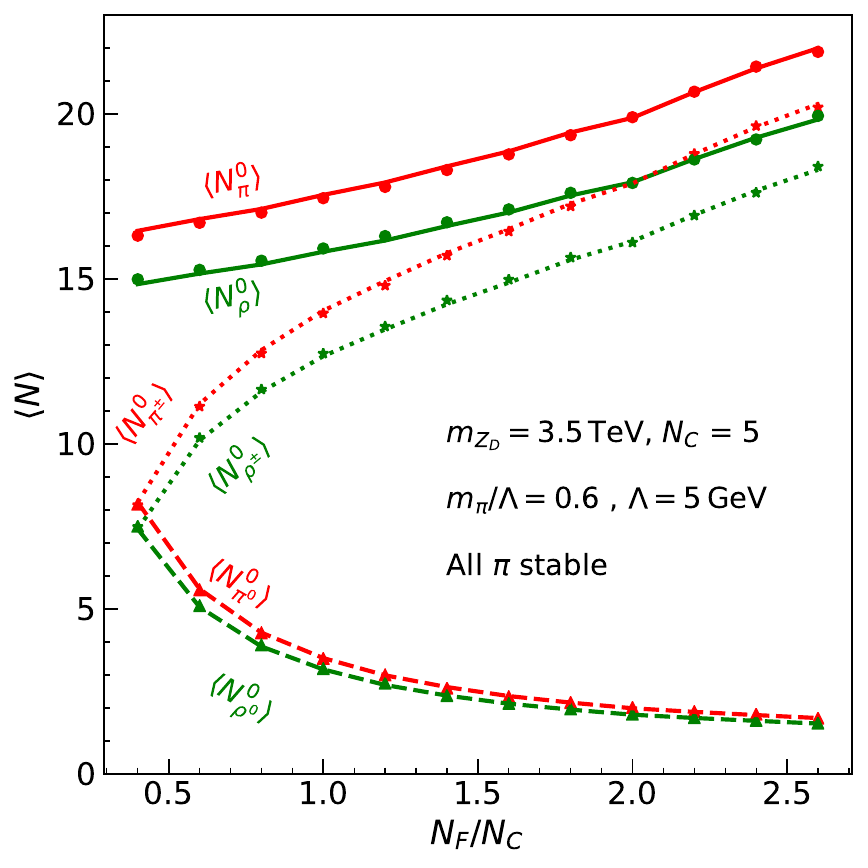}
\includegraphics[width=0.49\textwidth]{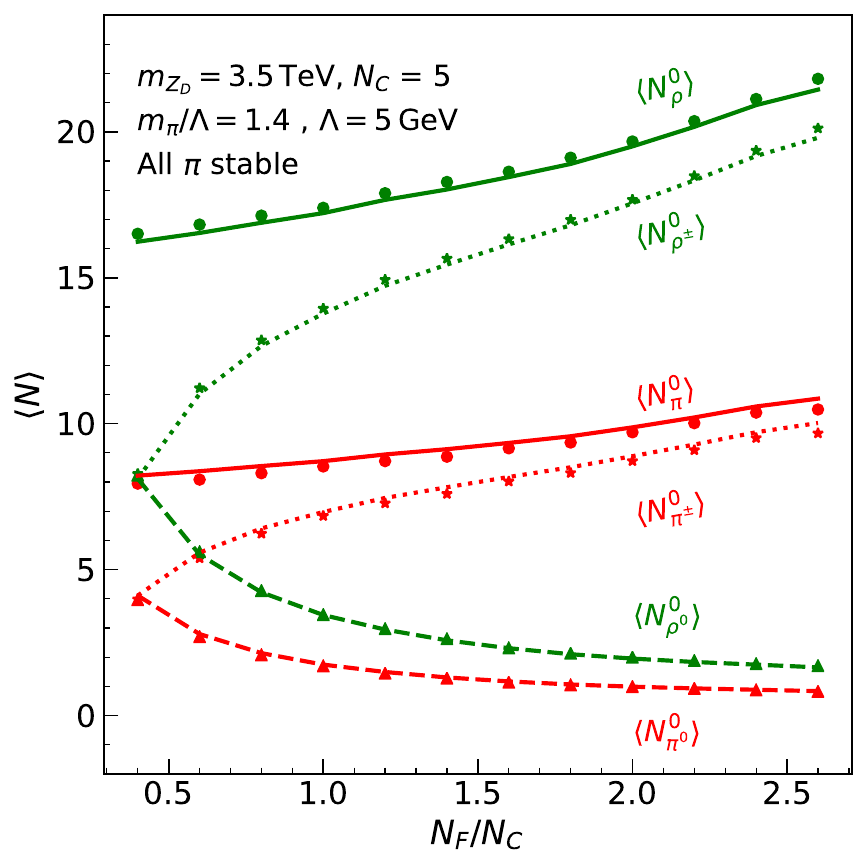}
\caption{Average multiplicities of diagonal, off-diagonal and total HV rho and pions produced after hadronization against $\fc$ in the $\rho \to \pi \pi$ open region of $\mpl$ space for $\mpl=0.6$ (left panel) and $\mpl=1.4$ (right panel). These quantities increase with $\fc$ whilst the ratio $\left<\nrinit/\npinit\right> = 0.90$ for $\mpl=0.6$ and $1.98$ for $\mpl=1.4$  for all considered $\fc$, given the \texttt{probVector} fit eq.~\eqref{eq:exponential_fit}; the details of which are described in Appendix~\ref{app:probvec_modelling}. (top right) Between $\fc=0.4$ and $2.6$, the proportion of on-diagonal $\rhod$ and $\pid$ decreases by $\sim80\%$. }
\label{fig:initial_meson_multiplicity}
\end{figure}

We begin with analyzing the dependence of HV/DS meson multiplicity on the number of flavors and colors. As discussed in section~\ref{sec:models}, $\fc$ controls the parton shower. For a fixed $\ld$ the value of running coupling at a given energy increases with $\fc$ but its rate of change with energy decreases\footnote{This has a small $\nc$ dependence which we ignore here.}~\cite{Ellis:1996mzs,Kulkarni:2025rsl}. We therefore expect the meson multiplicity to be a function of $\fc$. For our simulations, we fix $\nc = 5$, $\ld = 5\,\rm{GeV}$, $\mpl = 0.6$ and vary $\fc$. We also use one-loop running coupling and set hadronization parameters according to {\tt PYTHIA} SM QCD default-tune~\cite{pythiaTune} {\tt HiddenValley:bmqv2} = 0.087, {\tt HiddenValley:aLund} = 0.3, {\tt HiddenValley:sigmamqv} = 0.92~\cite{privcomm}. 

In fig.~\ref{fig:initial_meson_multiplicity}, we plot the initial meson multiplicities as a function of $\fc$ for two benchmarks of $\mpl=0.6$ (left) and $\mpl=1.4$ (right). For a fixed $\mpl$, the total initial meson multiplicity, i.e. the initial $\left<\ntinit\right>=\left<\npinit\right>+\left<\nrinit\right>$ increases with $\fc$. For fixed $\mpl$, the multiplicity ($\left<\ntinit\right>$) increases by around 30-35\% between $\fc=0.4$ and $2.6$. However, for a fixed $\fc$, the $\left<\ntinit\right>$ decreases with $\mpl$ by $\sim 45-50\%$ between $\mpl=0.2$ and $\mpl=1.4$, since heavier pions for a fixed $\mzp$ means a decreased multiplicity. Given the fit of \texttt{probVector} described in Appendix
\ref{app:probvec_modelling}, the total meson production begins to be dominated by the pion production in the chiral limit. Pion production overtakes rho production for $\mpl\leq0.65$, hence in the $\mpl=1.4$ benchmark, rho production remains dominant while in the $\mpl=0.6$ benchmark, $\left<\npinit\right> > \left<\nrinit\right>$.

\begin{figure}[h!]
\centering
\includegraphics[width=0.49\textwidth]{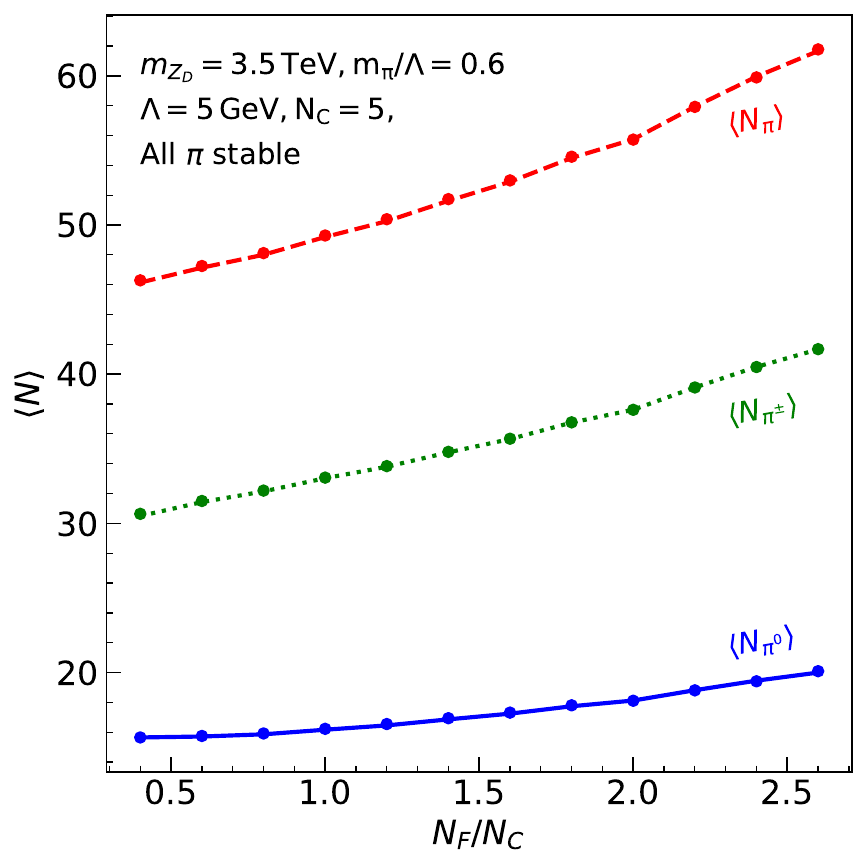}
\includegraphics[width=0.49\textwidth]{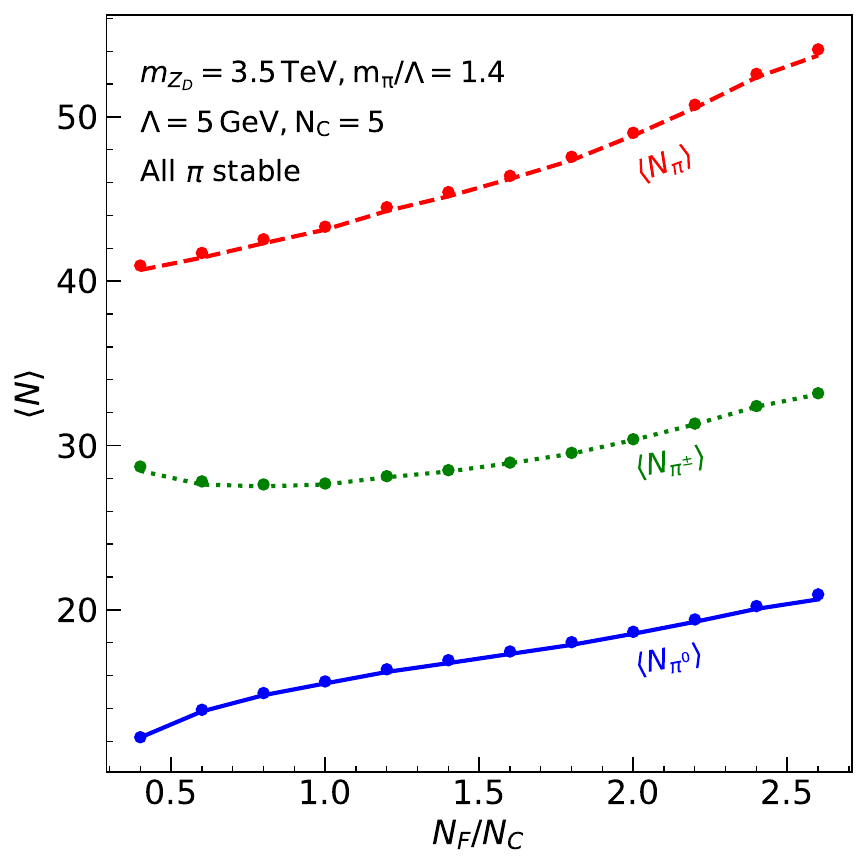}
\caption{
Average multiplicities of diagonal, off-diagonal and total HV/DS pions as a function of $\fc$ after HV rhos have decayed to HV/DS pions for two benchmarks of $\mpl=0.6$ (left panel) and $\mpl=1.4$ (right panel). The lines plotted correspond to analytical fits using eqn.~\ref{eq:initial_rhomeson}. There is a general increase for all three averages with $\fc$ though at low $\fc$ the off-diagonal multiplicity slightly decreases with $\fc$. For all $\nf > 2$, the final-state $\pid$ multiplicity is dominated by off-diagonal $\pid$. }
\label{fig:final_meson_multiplicity}
\end{figure}

Given $\left<\ntinit\right>$ it is possible to understand several additional features of fig.~\ref{fig:initial_meson_multiplicity}. An exact analytic calculation $\left<\ntinit\right>$ is beyond the scope of this work and thus is determined only by \texttt{PYTHIA} simulations for given set of hadronization parameters. In absence of exact knowledge of $\left<\ntinit\right>$, we can still determine the proportion of $\left<\ntinit\right>$ mesons that are either HV/DS rho or pions through simple considerations. For instance, if the spin-0 and spin-1 singlets are degenerate with the $\pid/\rhod$ multiplets, there are $N^2_F$ number of $\pid, \rhod$, of which $\nf$ are diagonal and $N^2_F-\nf$ are off-diagonal. Therefore $\langle \npoffinit \rangle/\langle \nponinit\rangle = \langle\nroffinit\rangle/\langle \nroninit\rangle=\nf-1$. This is reflected in  fig.~\ref{fig:initial_meson_multiplicity}, where despite $\left<\ntinit\right>$ being proportional to $\fc$, the number of diagonal $\rhod, \pid$ decrease by $\sim80\%$ between $\fc = 0.4$ and 2.6 and hence the net effect is a decrease in the diagonal meson multiplicity. On the other hand, the multiplicity of off-diagonal pions increases by $\sim150\%$. This decrease (increase) in the multiplicity of the average diagonal (off-diagonal) pions predictably follows the described $1/\nf$, ($1-1/\nf$) distributions. In other words, the fraction of diagonal and off-diagonal pions as a function of $\fc$ varies as $1/\nf$ and $1-1/\nf$, thus demonstrating that our simulations are valid for $\nf > 8$.

We now plot the cumulative effect of these decays in fig.~\ref{fig:final_meson_multiplicity} for the two aforementioned benchmarks of $\mpl=0.6$ (left panel)  and $\mpl=1.4$ (right panel). Noting that, as before, for a fixed $\nc$, $\langle \ntinit\rangle$ increases with $\fc$. The decay patterns dictate that, unlike $\langle \nponinit\rangle$, the $\langle \nponfin\rangle$ increases with $\fc$ due to the cumulative effect $\left<\ntinit\right>$ and its pre-factor from eq.~\eqref{eq:final_onpi}\footnote{It is worth noting that, given the $\texttt{probVector}$ fit, this trend reverses for $\mpl \lesssim 0.25$ and $\langle N_{\pidon}\rangle$ actually decreases with $\fc$.}. For $\langle \npofffin\rangle$, the pre-factor in eqn.~\eqref{eq:final_offpi} actually decreases with $\fc$ while $\left<\ntinit\right>$ increases with $\fc$. For $\mpl=0.6,1.4$, the net effect is an increase with $\fc$ except for small $\fc$ in the $\mpl=1.4$ benchmark where the pre-factor decreases by $4\%$ between $\fc=0.4-0.6$ while $\left<\ntinit\right>$ only increases by $1\%$ and so the net effect is a slight decrease.  
\section{Additional plots}
\label{app:additional_plots}
%
\begin{figure}[h!]
\centering
\includegraphics[width=0.49\textwidth]{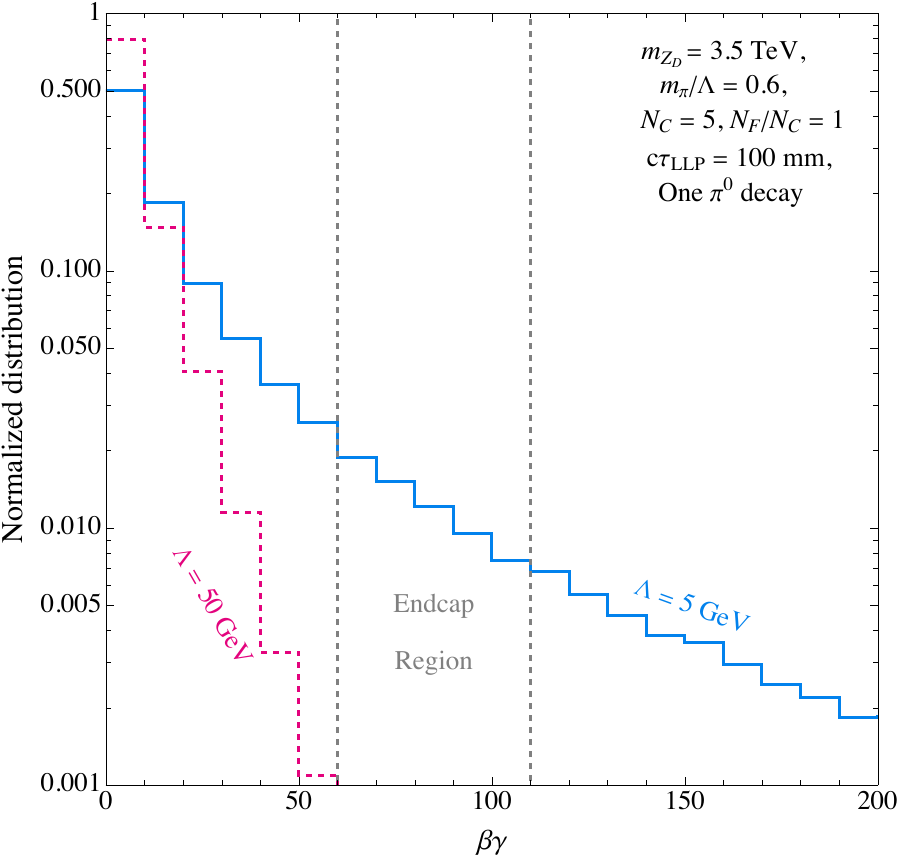}
\includegraphics[width=0.49\textwidth]{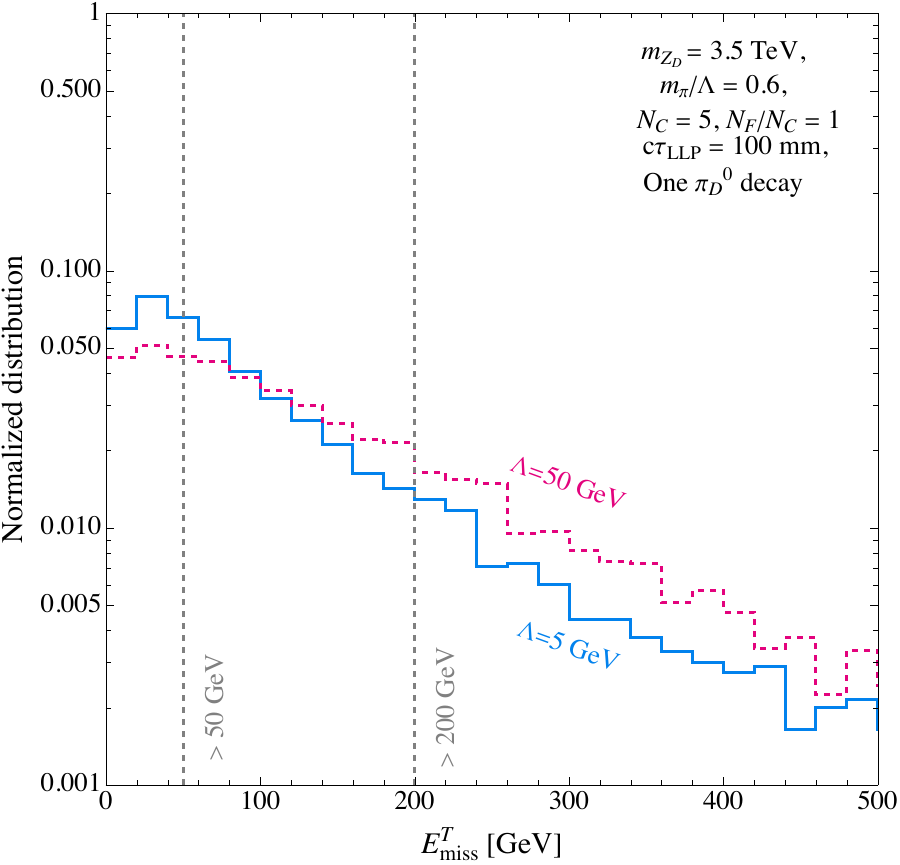}
\caption{Distribution of the boost $\beta \gamma$, for the benchmark $\fc $ = 1, $\mpl$ = 0.6, $\ld =$ 5 GeV, $c \tau$ = 100 mm, $\mzp$ = 3.5 TeV. }
\label{fig:boost}
\end{figure}
In this appendix, we show a collection of additional plots to aid the discussion in section~\ref{sec:eff_dependence}. To begin with in fig.~\ref{fig:boost}, we show the average boost of the HV/DS pions and the endcap favored regions of boost for $\mzp = 3.5\,\rm{TeV}, \mpl = 0.6, \ld = 5,~50\,\rm{GeV}, c\tau_{\rm LLP} = 100\,{\rm mm},$ $\nc = 5$ and $\fc = 1$ (left panel) as well as the missing energy distribution for the same benchmark (right panel). The boost distribution has a much longer tail for $\ld = 5\,{\rm GeV}$ compared to $50\,\rm{GeV}$, which makes it easier for HV/DS pions for smaller lifetimes to reach the muon endcap detector. We see that $\beta\gamma \sim 60 - 100$ is required for HV/DS pions with $c\tau_{\rm LLP} = 100\,\rm{mm}$ to reach the muon endcap detector. The missing energy distribution on the other hand is harder for $\ld = 50\,\rm{GeV}$, compared to $\ld = 5\,\rm{GeV}$. Therefore the MET cut efficiency is slightly higher for larger $\ld$. 

\begin{figure}[h!]
\centering
\includegraphics[width=0.49\textwidth]{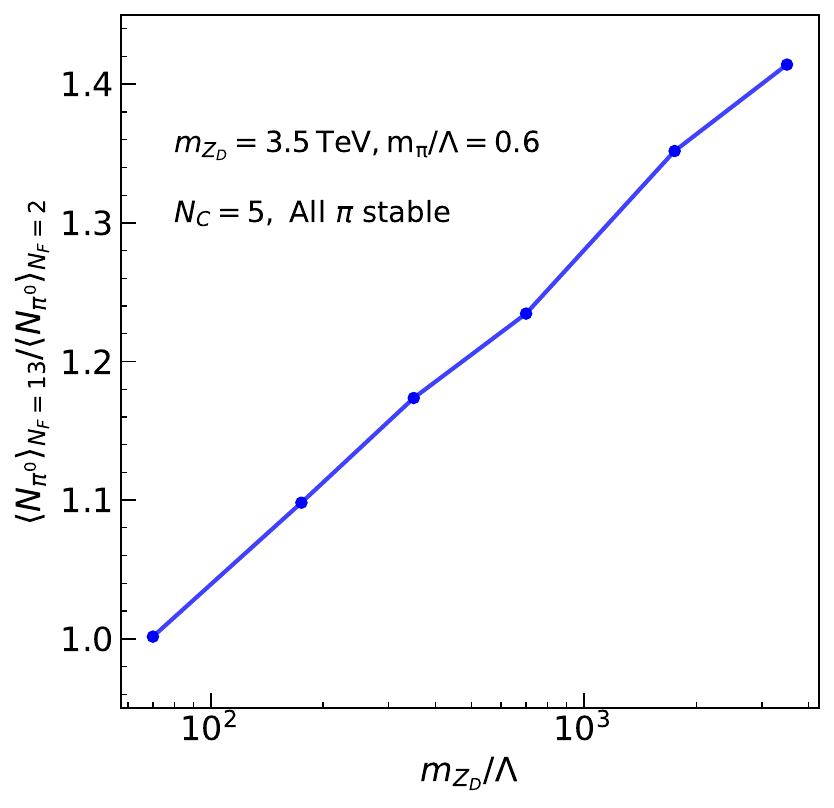}
\caption{Plot of final state $\langle N_{\pid^0}\rangle_{\fc = 2.6} / \langle N_{\pid^0}\rangle_{\fc = 0.4}$ , the ratio of average LLP multiplicity between two different flavors of  $\fc = 0.4, 2.6$, against $\mzp/\ld$. We note that this ratio increases logarithmically $\mzp/\ld$.}
\label{fig:ratio_mzplambda}
\end{figure}

Finally, we comment on the effect of $\mzp/\ld$ on the length of the shower in fig.~\ref{fig:ratio_mzplambda}. To this effect, we plot the ratio of the mean diagonal pion multiplicity for $\fc = 0.4$ and $\fc = 2.6$ as a function of $\mzp/\ld$. As expected this ratio increases logarithmically with $\mzp/\ld$. We stress here that in principle variation of $\mzp$ should be equivalent to variation of $\ld$, however $\mzp \gg \ld$ is required for HV/DS shower creation, therefore the variation of $\ld$ for a fixed $\mzp$ is restricted. At the LHC, $\mzp$ can be varied over a much wider range. While we do not completely explore the effect of this variation in this work, the $\mzp/\ld$ ratio can create additional interesting effects in the HV/DS phenomenology.

\section{Efficiency as a function of \texorpdfstring{$\mpl$}{mpl}}
\label{app:eff_mpi_lam}
%
\begin{figure}[h!]
\centering
\includegraphics[width=0.45\textwidth]{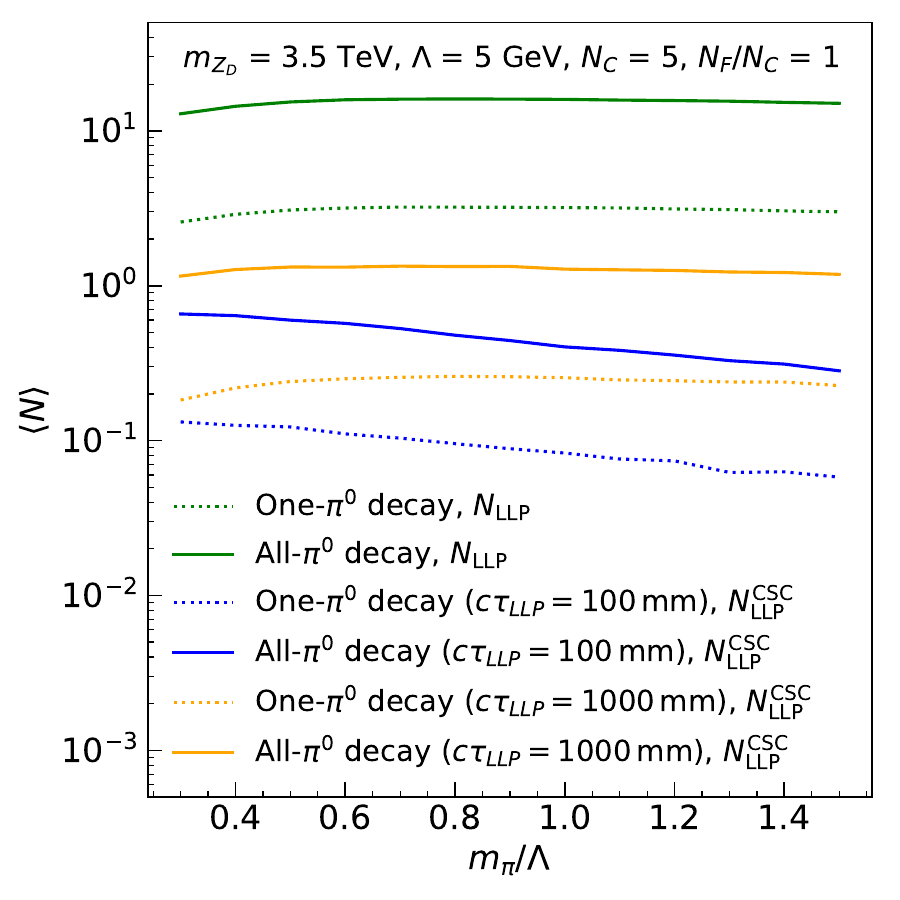}
\includegraphics[width=0.45\textwidth]{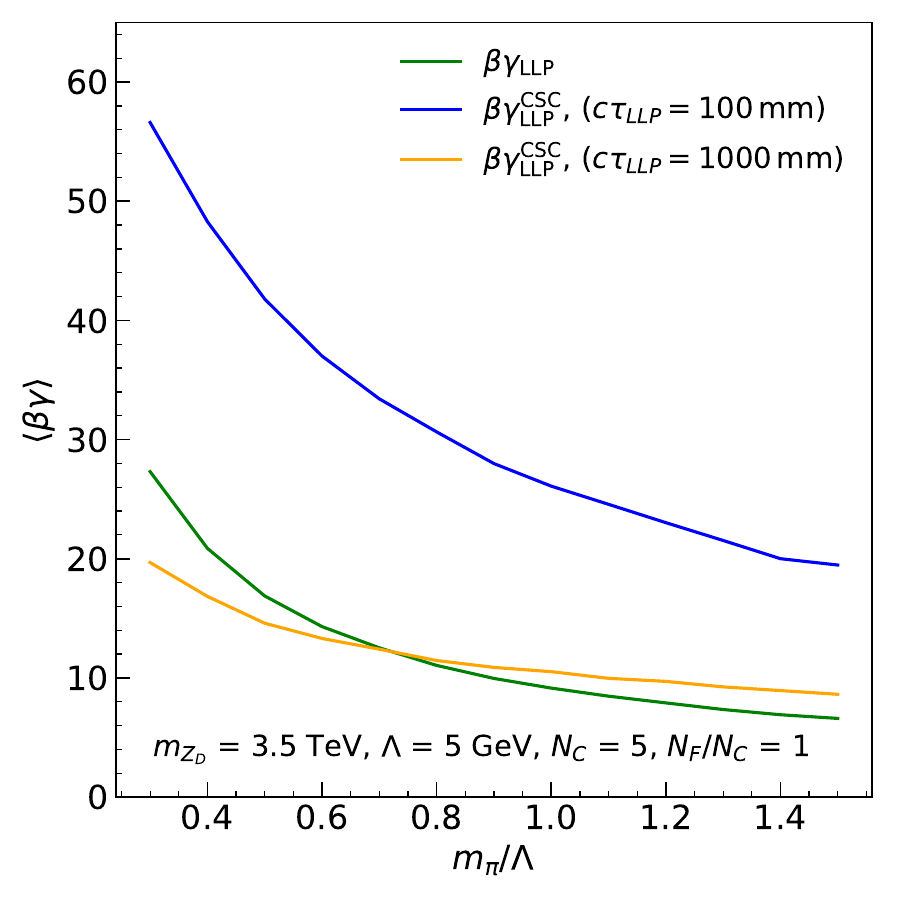}%
\caption{The multiplicity (left column) and boost (right column) of the diagonal pions in the events a function of $\mpl$, when rest of the parameters are kept constant as specified. }
\label{fig:mpl_multiplicity_and_boost}
\end{figure}
\begin{figure}[h!]
\centering
\includegraphics[width=0.49\textwidth]{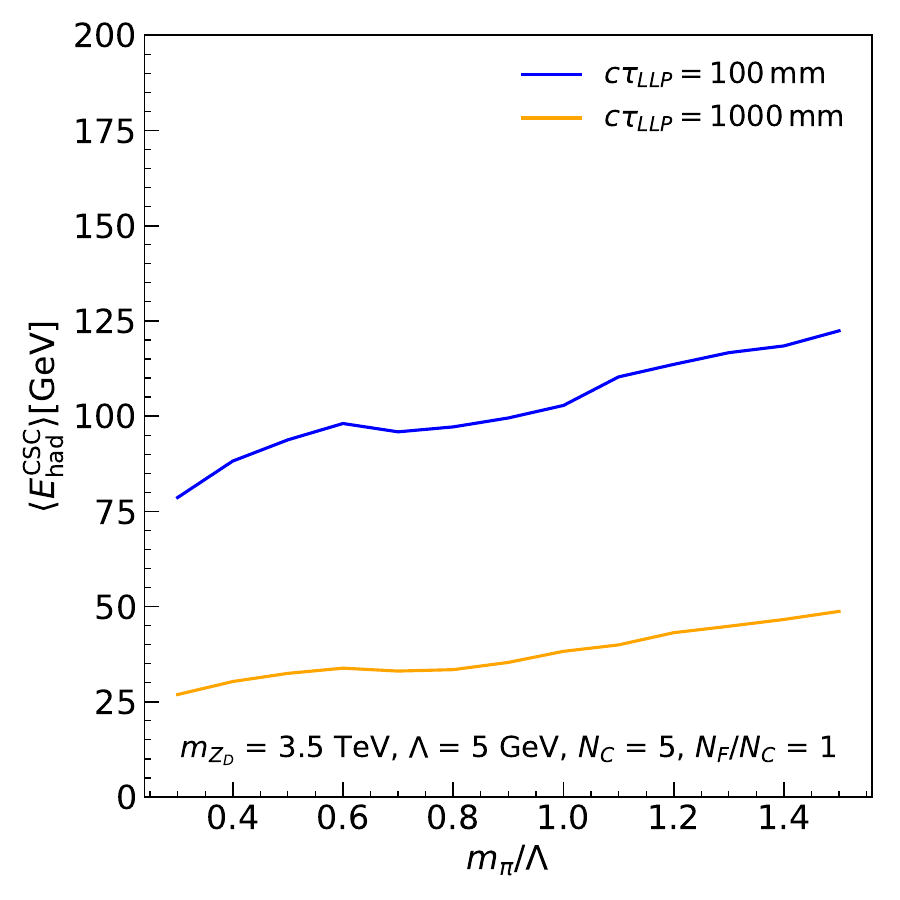}
\includegraphics[width=0.49\textwidth]{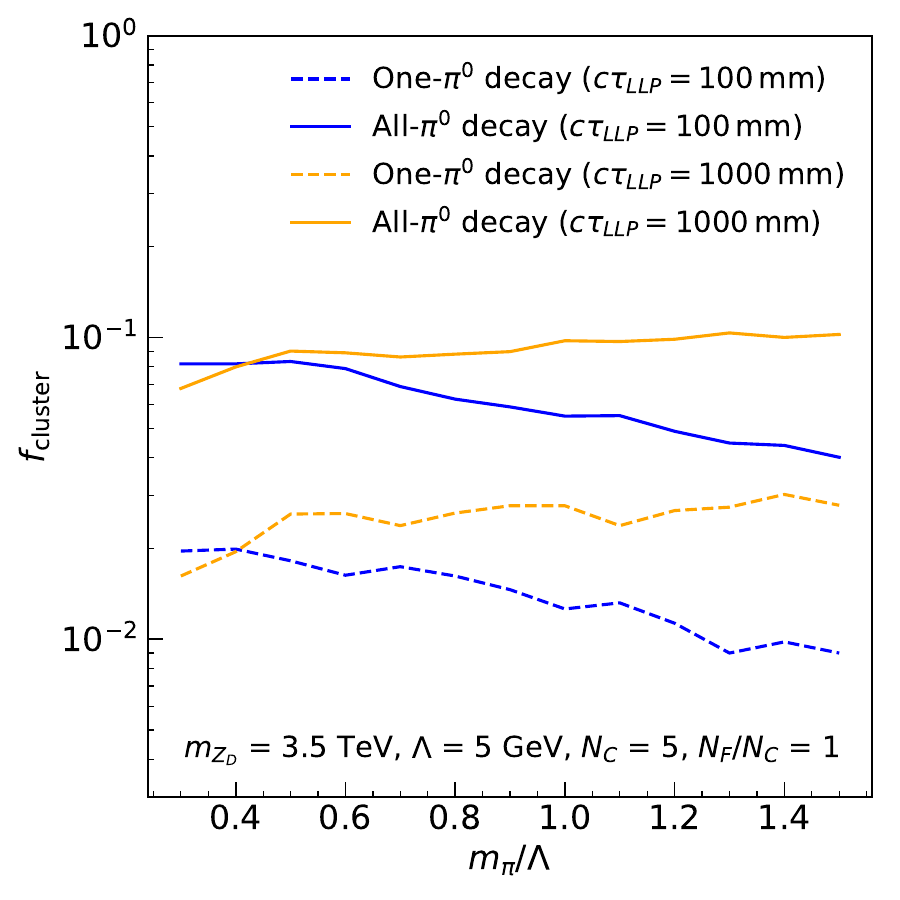}
\caption{Ehad~(left panel), fraction of events containing at least one cluster~(right panel) as a function of $\mpl$.}
\label{fig:eff_mpl}
\end{figure}

As $\mpl$ increases for a fixed value of $\ld$ the pion mass increase albeit at a much smaller rate compared to variation with $\ld$. For example, in fig.~\ref{fig:mpl_multiplicity_and_boost}, $m_{\pid}$ varies from 1.25 GeV -- 7.5 GeV over the considered range of $\mpl$. Due to this gentle pion mass dependence on $\mpl$, the number of final-state pions as shown in fig.~\ref{fig:mpl_multiplicity_and_boost} (left panel) is approximately constant. The corresponding boost (right panel) changes through the pion mass variation however the over-all behavior is similar to the one seen in fig.~\ref{fig:multiplicity_and_boost_lamda}. 

Given the understanding of $m_\pid$ as a function of $\mpl$, the trends in deposited hadronic energy and the $f_{\rm cluster}$ shown in fig.~\ref{fig:eff_mpl} are unsurprising. They follow similar trends as $\ld$ dependence, with the important difference that the change with respect to $\mpl$ is much milder compared to $\ld$.

\section{{\tt probVector} fits}
\label{app:probvec_modelling}
\begin{figure}[h!]
\centering
\includegraphics[width=0.45\textwidth]{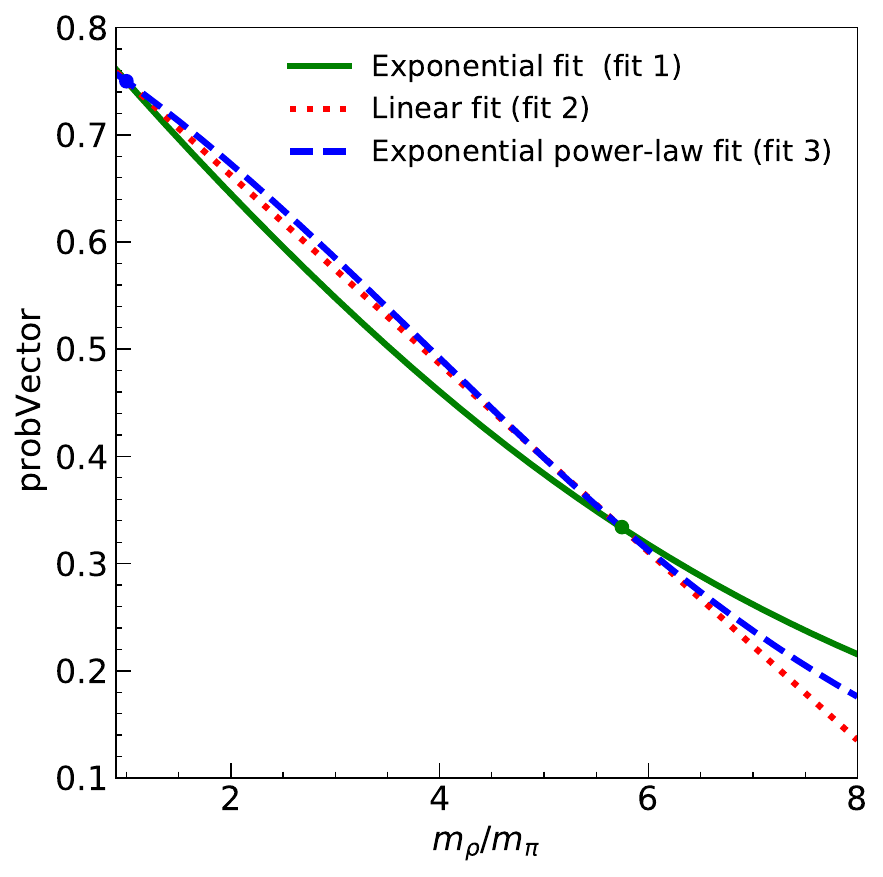}
\includegraphics[width=0.45\textwidth]{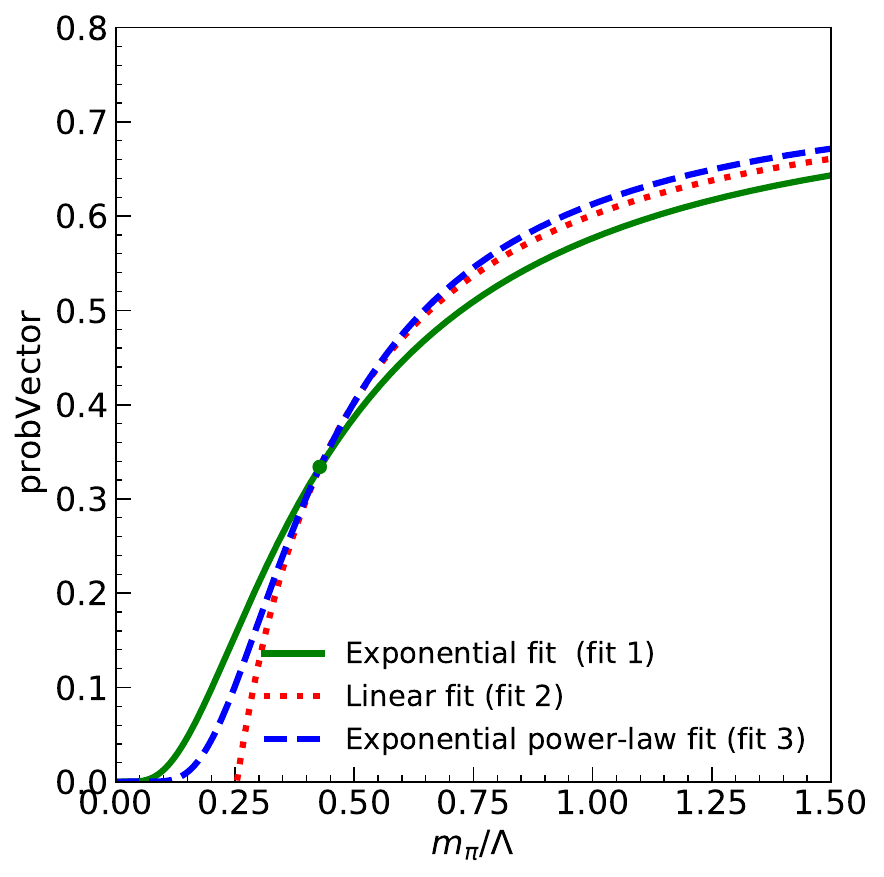}
\caption{Fit to {\tt probVector} as a function of $m_\rho/m_\pi$ (left panel) and $m_\pi/\ld$ (right panel). Two benchmark points correspond to known values of {\tt probVector} when $m_\rho = m_\pi$ and the SM benchmark.}
\label{fig:probvec_fit}
\end{figure}

Once parton showers is terminated, the produced partons are combined in order to make mesons. This process of combining partons and labeling them as hadrons with different quantum number is handled by hadronization. An important quantity that controls the number of pions and rho mesons produced during hadronization is called  $\texttt{probVector}$. $\texttt{probVector}$ controls the fraction of all mesons which are labeled as rho mesons. However, this labeling should depend on the relative rho - pion mass hierarchy. To account for this, we must introduce a mass-dependent Boltzmann factor that is a function of $m_\rhod/m_\pid$ and fit accordingly~\cite{Batini:2024zst,Becattini:1996gy}. The ratio $m_{\rhod}/m_{\pid}$ is inversely proportional to $\mpl$ as seen in eqn.~\eqref{eqn:meson_masses} thus giving $\texttt{probVector}$ a dependence on $\mpl$. 

Within \texttt{PYTHIA}, the relative abundances of $\rhod$ and $\pid$ are encoded in {\tt probVector}. This quantity controls the relative fraction of the rho mesons to pions produced after the perturbative parton shower; internally to \texttt{PYTHIA} it represents the probability that the probability that a meson will be chosen to be a $\rhod$ as opposed to a $\pid$. For a large number of Monte Carlo samples this will eventually average to $\texttt{probVector}=\langle N_{\rhod}/\ntinit\rangle$. If the $\rhod$ mesons and $\pid$ were degenerate, spin-counting would give 3 spin-1 states and 1 spin-0 state, meaning that $\rhod$ is 3 times more likely to be produced after hadronization. However, differences in mass between the two meson types causes deviation from this value, as changing their relative masses should change their relative abundances during HV/DS hadronization. Therefore {\tt probVector} is a function of $m_{\rhod}/m_{\pid}$. 

While more theory efforts are necessary to understand the complete settings for this quantity, we do have some values to compare to, namely that within the SM QCD i.e. for $\mpl \sim 0.428$, {\tt probVector} is determined to be 0.33~\cite{Skands:2014pea},\footnote{We use the world average of the SM QCD $\rho$ and $\pi$ masses of $775.45$ MeV and $134.97$ MeV respectively and find $m_{\rhod}/m_{\pid}=5.745$.}  while, as before, in the limit of $m_\rhod = m_\pid$, the {\tt probVector} should be 0.75 from spin-counting considerations~\cite{Becattini:2008tx}. In the chiral limit, since the $\pid$ will continue to get lighter for a fixed $\ld$, we expect the relative abundance of $\rhod$ decreases. Hence we suppose another boundary condition that at $m_\pid/m_\rhod = 0$, ${\tt probVector} = 0$. 

By including the chiral limit boundary condition, where rho production becomes exponentially suppressed, we can fit $\texttt{probVector}$ with an exponential. This fit supposes that, like in QCD, the $\rhod$ and $\pid$ mesons are approximately thermal, as discussed in~\cite{Andersson:1997xwk} and then the relative abundances are determined by the ratios of their two Boltzmann factors as $\texttt{probVector}\propto\exp(-(m_{\rhod}-m_{\pid})/\Gamma))$ and assume $\Gamma$ scales with $m_{\pid}$ as $\Gamma= m_{\pid}/\omega$ (as to satisfy probVector decreasing in the chiral limit)\cite{Mangano:2018sfp,Batini:2024zst,Becattini:1996gy}.\footnote{Hadron production in $e^+e^{-}$ and pp collisions has often been described using statistical models that suggest a universal thermal behavior~\cite{Andersson:1997xwk,Becattini:1996gy,Hwa_2004,Greco_2003,Fries_2003},
but whether this apparent thermality reflects genuine equilibration or merely a phenomenological fit remains debated, particularly in connection with interpretations involving the Quark–Gluon Plasma~\cite{McLerran:2003yx}.}

However, it can already be seen that $\Gamma=m_{\pid}/\omega$ is not a good assumption in the deep chiral regime as $\Gamma$  represents the temperature of the gas which can not scale with pion mass in deep chiral regime. The exact value of $\mpl$, at which the fit fails is unclear however we set this to be $\mpl = 0.2$, which determines the lowest $\mpl$ value we can simulate. The actual fit is given as follows,
\begin{equation}
\textrm{Fit 1} \qquad\qquad \texttt{probVector} = \frac{3\exp(-\omega(m_{\rhod}/m_{\pid} - 1))}{1+3\exp(-\omega(m_{\rhod}/m_{\pid} - 1))};
\label{eq:exponential_fit_app}
\end{equation}
where $\omega$=0.38. The form of this fit closely matches the definition of {\tt StringFlav:mesonUDvector} parameter as done in the {\tt PYTHIA} SM hadronization module. Inclusion of these additional conditions show that an exponential fit is better compared to a straight line.

Relaxing the chiral limit boundary condition, we can fit the other two boundary conditions with a straight line fit. The linear fit is given by,
\begin{equation}
\textrm{Fit 2} \qquad\qquad \texttt{probVector} = a_1(m_{\rhod}/m_{\pid}) + b_1
\label{eq:linear_fit}
\end{equation}
where we fit with two parameters $a_1$ = -0.088 and $b_1$ = 0.838. 

Being as generic as possible, where $m_{\rhod}$ and $m_{\pid}$ are two generic mass scales, we could also include the scenario in which $m_{\rho}=0$. This is an unrealistic scenario since in a chirally broken limit, pions are always the lightest states. Nevertheless since at $m_\rhod/m_\pid = 0$ the rho mesons are massless, thus ${\tt probVector}=1$ since $\rhod$ production is favored. We fit with an exponential-power-law form,
\begin{equation}
\textrm{Fit 3} \qquad\qquad \texttt{probVector} = \frac{3\exp(1-(m_{\rhod}/m_{\pid})^b)}{1+3\exp(1-(m_{\rhod}/m_{\pid})^b)};
\label{eq:exponential_powerlaw_fit}
\end{equation}
where $b$=0.59. 

We show these three different fits in fig.~\ref{fig:probvec_fit}, where we show the range of $m_\pid/\ld$ applicable for our simulations. We note here that {\tt probVector} changes rapidly in the chiral regime while beyond $m_\pid/\ld \sim 0.5$ the value is more or less constant within the range of our simulation. The exponential and linear fits match for most of our region of interest, but that the linear fit fails in the chiral limit and is unrealistic, due to becoming negative around $\mpl=0.2$. 

\begin{figure}[h!]
\centering
\includegraphics[width=0.4\textwidth]{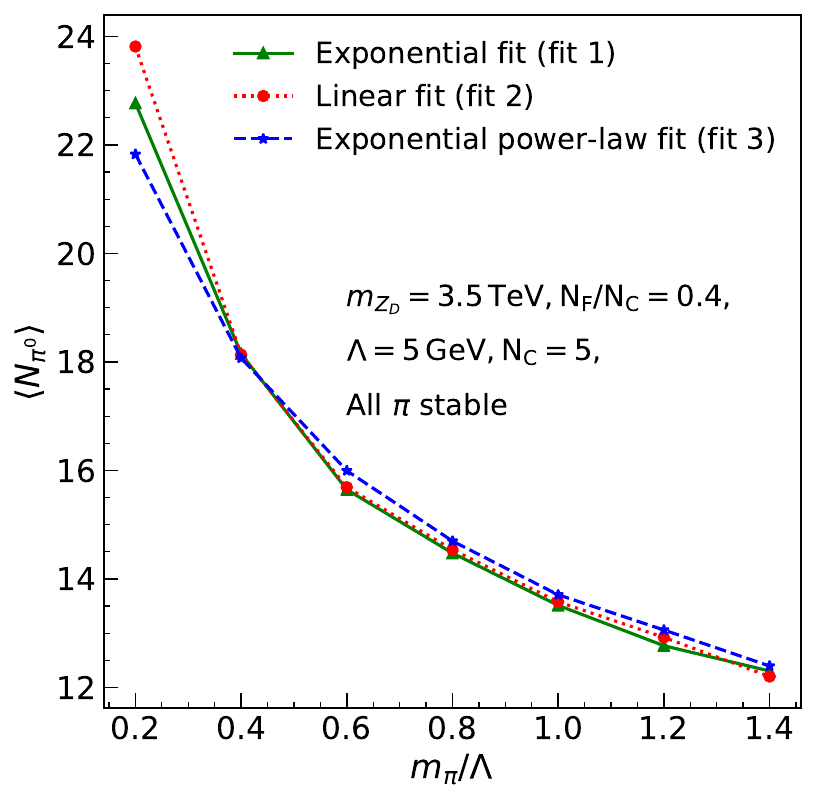}
\includegraphics[width=0.4\textwidth]{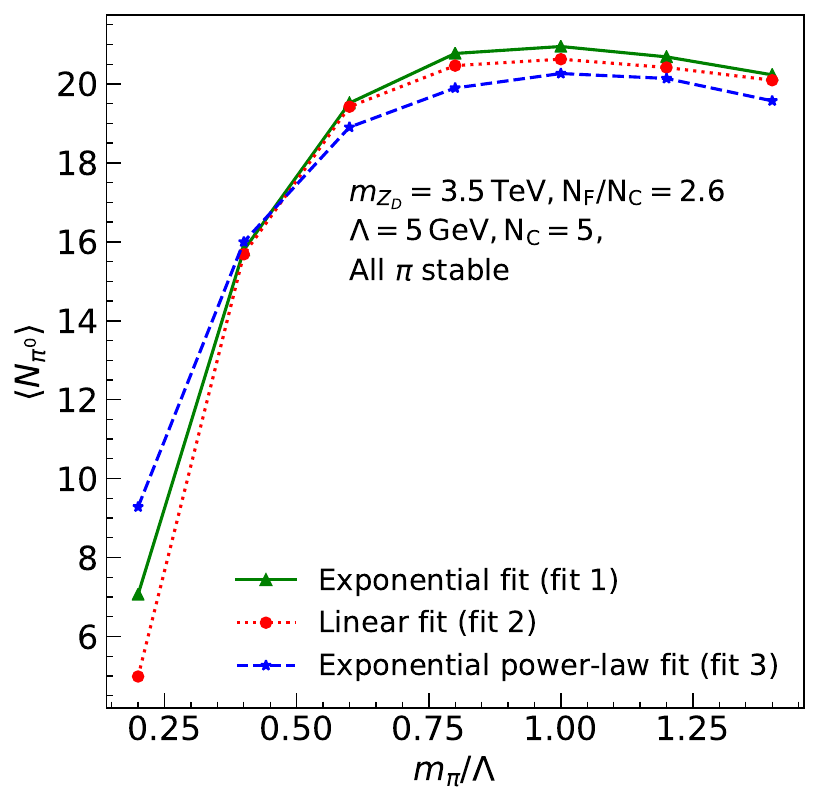}
\caption{Variation in the final diagonal dark pion multiplicity for $\fc = 0.4$ (left panel) and $\fc = 2.6$ (right panel) for the ${\tt probVector}=1$ fits described in eqn.~\eqref{eq:exponential_fit},~\eqref{eq:linear_fit} and  ~\eqref{eq:exponential_powerlaw_fit}.}
\label{fig:probvec_LLPnum}
\end{figure}

In fig.~\ref{fig:probvec_LLPnum}, we plot the average final state $\pidon$ multiplicity against $\mpl$ for two benchmark scenarios of $\fc = 0.4$ (left panel) and $\fc = 2.6$ (right panel) for the three ${\tt probVector}=1$ fits. We demonstrate much of the same effects seen in fig.~\ref{fig:probvec_fit}. Close to the chiral limit of $\mpl=0$, $\langle N_{\pidon}\rangle$ differs considerably between fits whereas for larger $\mpl$, the differences between the exponential and linear are quite small, while the differences between eqn.~\eqref{eq:exponential_fit} and ~\eqref{eq:exponential_powerlaw_fit} are at most 5\% - around the same level as uncertainties arising from hadronization.
Due to motivation from existing literature calculating relative abundances and from its behavior in the chiral limit, we decided to use eqn.~\eqref{eq:exponential_fit} throughout this paper.

\section{Dependence on hadronization parameters}
\label{app:hadronization_parameters}
There are a number of additional shower and hadronization related parameters in the {\tt PYTHIA} Hidden Valley module, which may affect the results presented. For our analysis the LLP multiplicity, which is related to the diagonal pion multiplicity is the most important quantity. In this appendix, we therefore analyze the variation of the total and the diagonal pion multiplicities as a function of several shower and hadronization parameters. 
%
\subsection{Hadronization parameter dependence}
\begin{figure}[h!]
\centering
\includegraphics[width=0.4\textwidth]{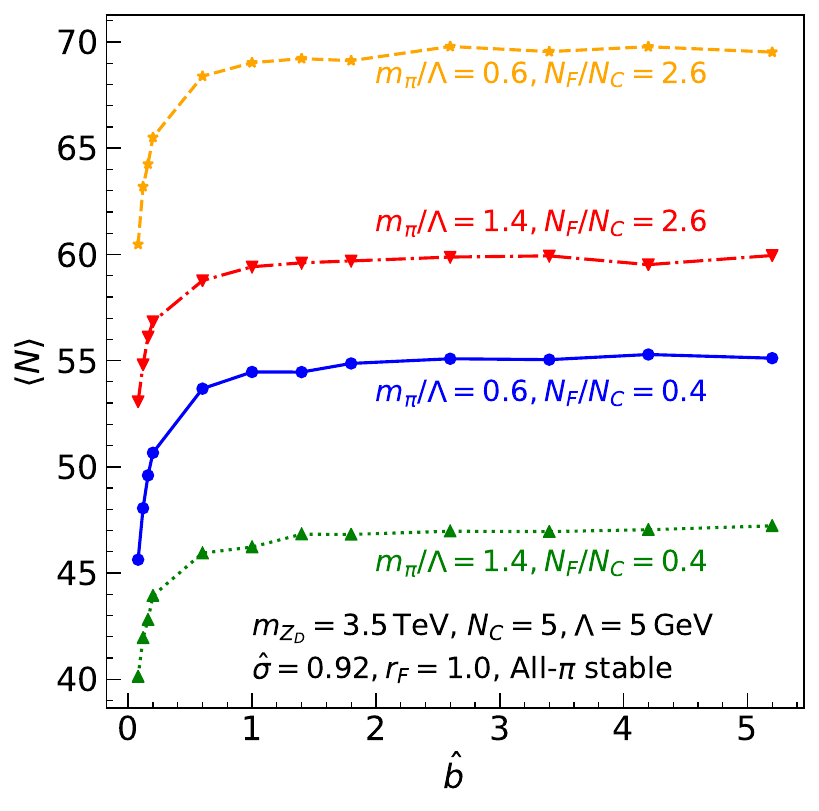}
\includegraphics[width=0.4\textwidth]{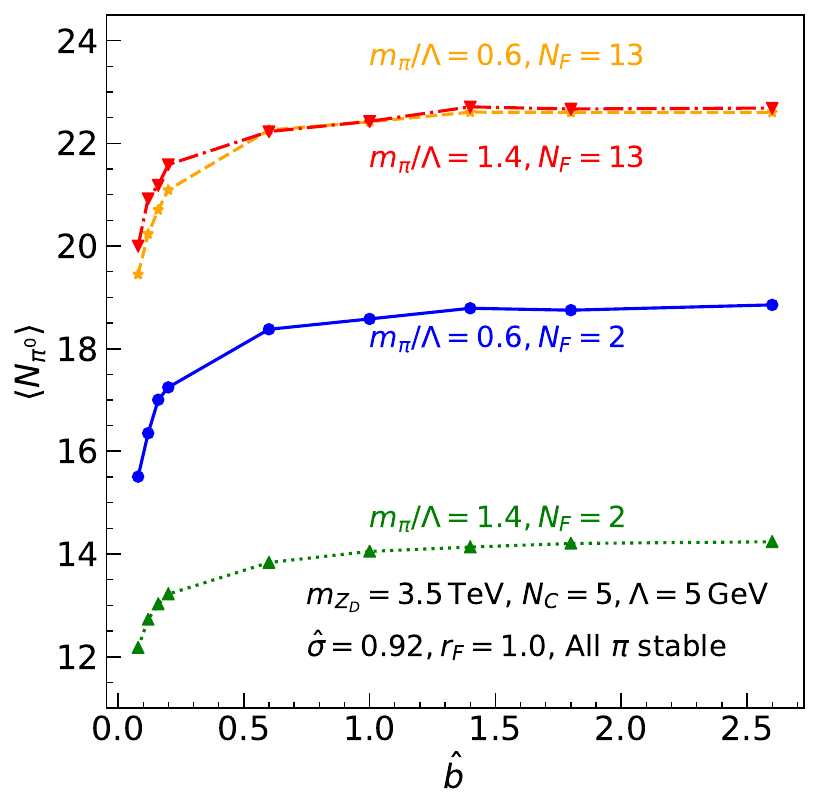}
\caption{Variation of total (left panel) and diagonal (right panel) multiplicities as a function of Lund hadronization parameter $\hat{b}$ for several $\fc$ and $\mpl$.}
\label{fig:bmqv2_variation}
\end{figure}
\begin{figure}[h!]
\centering
\includegraphics[width=0.4\textwidth]{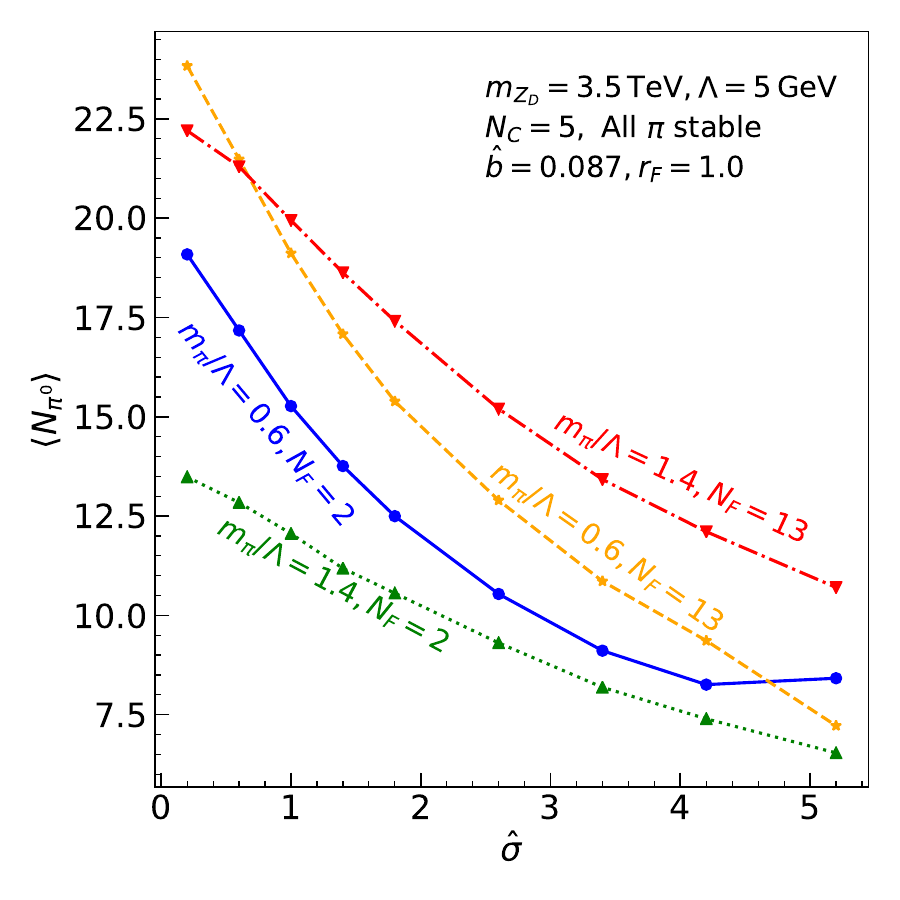}
\includegraphics[width=0.4\textwidth]{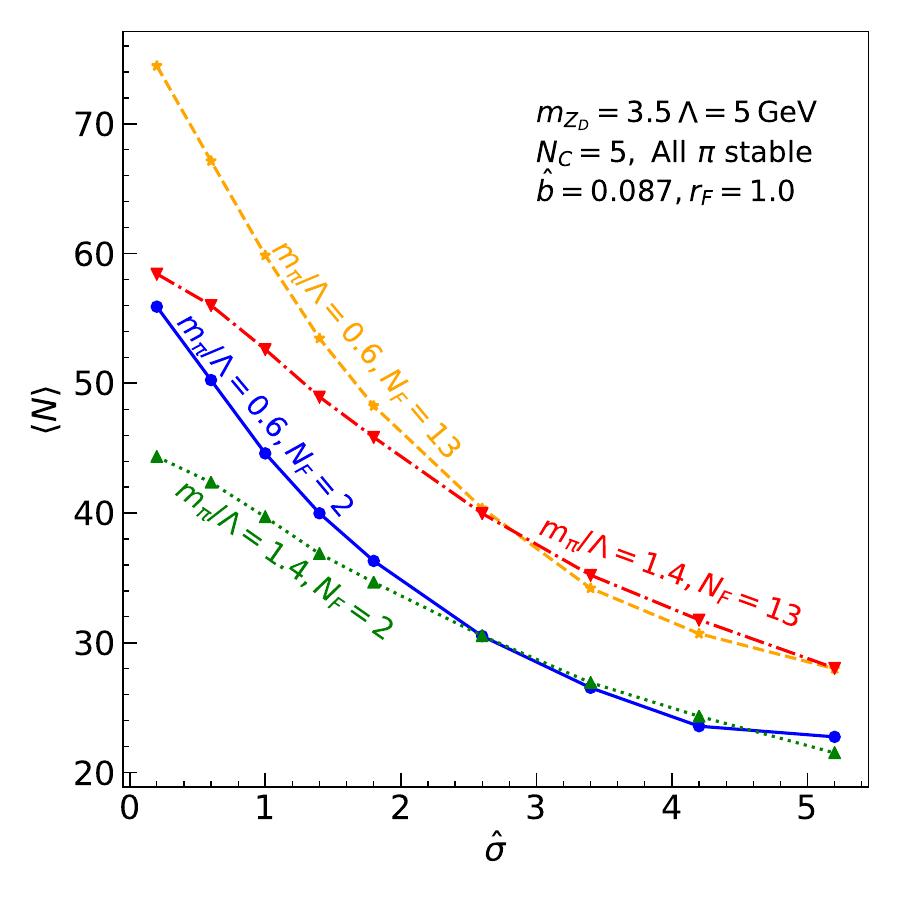}
\caption{Variation of LLP multiplicity (left) and final-state meson multiplicity (right)  with $\hat{\sigma}$.}
\label{fig:sigmamqv_variation}
\end{figure}
After a perturbative parton shower process, the \texttt{PYTHIA} Hidden Valley module performs hadronization through the Lund string model~\cite{Sjostrand:2006za}. After the shower is cutoff, the quarks, anti-quarks and gluons are portioned into color-singlet objects called strings. Observed mesons are then formed by minimizing the tension between these strings. This model uses the Lund string fragmentation function to determine the probability for a meson to be produced with momentum fraction $z$ and is given as follows,

\begin{equation}
f(z) = z^{-1}(1-z)^a \exp(-\hat{b}m_{T}^2/m_Q^2z).
\label{eq:frag_function}
\end{equation}
Here, $m_Q$ is the constituent quark mass and $a,\hat{b}$ are fixed constants. $m_T^2$ is the square-transverse mass and is given as $m_T^2=m^2+p_T^2$ where $p_T$ is the transverse momentum and $m$ is the mass of the meson. The $z^{-1}$ is often written as $z^{-1-r_Qb_{m^2_Q}m_Q^2}$ called the Bowler correction factor, where $m_Q$ is the constituent quark mass\footnote{Note that this is different than the current quark mass which enters the Lagrangian. We work with the definition $m_Q \equiv m_q + \ld$.}. It is only applicable for hadronization involving heavy quarks \cite{Bierlich:2023fmh,Bowler:1983ij}. We only consider light quarks of $m_q\ll\ld$ so $r_Q = 0$. 

The two parameters $a,\hat{b}$ enter the \texttt{PYTHIA} Hidden Valley in version 8.312 as \\ \texttt{HiddenValley:aLund} and \texttt{HiddenValley:bmqv2} respectively. $a$ is tuned using the default {\tt PYTHIA8} tune and is set to \texttt{HiddenValley:aLund}=0.3 similarly we set \texttt{HiddenValley:bmqv2}=0.087~\cite{privcomm}. 

The $a$ and $\hat{b}$ parameters are related through the average of the $\kappa\tau$ distribution~\cite{Andersson:1997xwk}, 
\begin{equation}
\langle \kappa\tau \rangle/m_Q^2 = (1+a)/\hat{b},
\label{eq:mean_breakup}
\end{equation}
where $\kappa$ is the string tension and $\tau$ is the string breakup time, both are assumed to be independent of $\nf, \nc$. We recommend the reader to carefully check this assumption for future studies as an improvement over our analysis. Such a check will require a deeper understanding of modeling of hadronization parameters, which is beyond the scope of this work. 

Hence, for a given theory at a fixed mass scale, $\langle \kappa\tau \rangle$ must remain constant and hence varying $\hat{b}$ necessitates also varying $a$. If we change $\hat{b} \to \hat{b}'$ then we find the change in $a$ to be,
\begin{equation}
a' = (\hat{b}(1+a)/\hat{b}')-1.
\label{eq:aLund_variation}
\end{equation}
The transverse momentum ($p_T$) of every hadron is obtained from the transverse momentum of the $q_D\bar{q}_D$ pair which itself is sampled from a Gaussian distribution. This Gaussian comes from the calculation of the tunneling probability of a $q\bar{q}$ in the WKB approximation~\cite{Andersson:1997xwk}, 
\begin{equation}
d\mathcal{P}/dp_T^2\propto\exp(-\pi p_T^2/\kappa)\exp(-\pi m_Q^2/\kappa).
\end{equation}
Naively, one would expect the sampling of hadron $p_T$ to be given with a variance of $\sigma_{p_T}/m_Q=1/m_Q\sqrt{\kappa/\pi} \approx 0.744$.
This produces too soft of a hadron spectrum so $\sigma_{p_T}/m_Q$, called {\tt sigmamqv} and labeled here as $\hat{\sigma}$ is left as a free parameter in the {\tt PYTHIA} Hidden Valley module~\cite{Andersson:1980vj,Bierlich:2023fmh}.

In fig.~\ref{fig:bmqv2_variation} - \ref{fig:sigmamqv_variation}, we therefore vary the Lund string fragmentation parameter $\hat{b}$ and $\hat{\sigma}$ for two different values of $\mpl$ and $\fc = 0.4, 2.6$. Fig.~\ref{fig:bmqv2_variation} shows that irrespective of the exact value of $\hat{b}$ the total final state meson multiplicity is directly proportional to $\fc$ for a fixed $\mpl$ as expected. Similar observation holds for the diagonal pion multiplicity as well. For $\hat{b} > 0.5$ the multiplicity is approximately constant, and maximum change is observed when $\hat{b}$ is small $< 0.5$. The number of final-state diagonal pions $\pidon$ change by $10-20\%$ between $0 < \hat{b} < 0.5$. Whether $\hat{b}$ is a function of $\fc,\, \ld$ or $\mpl$ is not known. Therefore we can not predict which parts of the fig.~\ref{fig:sen_2D_mpilam_lam} are maximally affected. Nevertheless, we conclude $\hat{b}$ variation may affect our limits.

In fig.~\ref{fig:sigmamqv_variation} we plot $\langle N\rangle$ and $\langle N_{\pidon}\rangle$ against $\hat{\sigma}$ for the four common benchmarks of $\fc = 0.4 ,2.6$ and $\mpl=0.6,1.4$. Unlike $\hat{b}$ there is a very significant dependence on $\hat{\sigma}$ where both final-state multiplicities decrease rapidly with increased $\hat{\sigma}$. Since $\sigma_{p_T}$ represents the variance in hadron $p_T$, increased $\sigma_{p_T}$ means a broader distribution and thus a decreased multiplicity. .

\subsection{Shower cutoff dependence}
\begin{figure}[h!]
\centering
\includegraphics[width=0.45\textwidth]{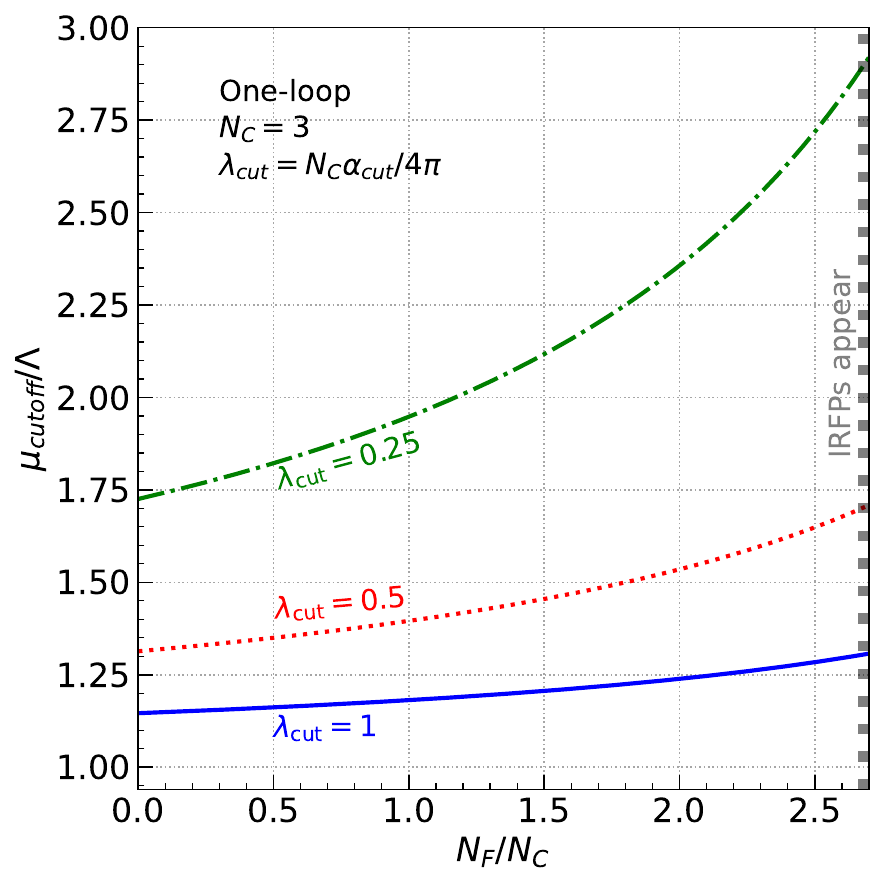}
\caption{$\mu_{\rm{cutoff}}/\ld$ as a function of $\fc$ determined for various 't Hooft coupling $\lambda_{\rm{cut}} = \nc\alpha_{\rm{cut}} = 1,0.5, 0.25$.}
\label{fig:thooft_cutoff}
\end{figure}

The perturbative description on which parton showers are based is no longer valid once the 't Hooft coupling becomes too large. Therefore, the strength of 't Hooft coupling should be the natural parameter to determine shower cut-off. In {\tt PYTHIA8} Hidden Valley module, the default shower-cutoff is determined to be $\mu_{\rm{cutoff}}/\ld  = 1.1$. We first determine $\mu_{\rm{cutoff}}/\ld$ at which the 't Hooft coupling becomes non-perturbative. We take three illustrative values $\lambda_{\rm{cutoff}} = \nc \alpha_{\rm cutoff}/4\pi = 1,0.5, 0.25$. We determine this cutoff through the formula,
\begin{equation}
\mu_{\rm{cutoff}}/\ld = \exp(\nc/4\pi\beta_0\lambda_{\rm{cutoff}}),
\label{eq:thooft_cutoff_oneloop}
\end{equation}
which is simply the one-loop running coupling at the associated 't Hooft coupling rearranged for $\mu/\ld$. For reference, the $\beta_0$ coefficient is 
\begin{eqnarray}
    \beta_{0} &=& \frac{1}{4\pi}\left(\frac{11}{3} C_A - \frac{4}{3} T_R \nf\right) ,\nonumber \\
\end{eqnarray}
with the corresponding RGE equation $\mu^2 d\alpha/d\mu^2 = -\alpha^2\beta_0$. Here $C_A = \nc$ and $C_F = {\left(\nc^2 - 1\right)}/{\left(2 \nc\right)}$ are the adjoint and fundamental Casimir invariants, while $T_R = 1/2$. 

We show the resulting $\mu_{\rm{cutoff}}/\ld$ in fig.~\ref{fig:thooft_cutoff}. These results show that if the shower is cut-off at very large values of 't Hooft couplings, $\mu_{\rm{cutoff}}/\ld$ varies slowly with $\fc$. Likewise, it varies exponentially with $\fc$ if cutoff at small values of 't Hooft couplings.

Fig.~\ref{fig:thooft_cutoff} therefore shows that while the default shower cut-off ($\mu_{\rm{cutoff}}/\ld  = 1.1$ in {\tt PYTHIA} Hidden Valley module) does change as a function of $\fc$ the default value is not too far from the one determined by $\lambda_{\rm{cutoff}} = 1$. The most extreme case is understandably observed when $\lambda_{\rm{cutoff}} = 0.25$, which results in a very early shower cut-off. 

\begin{figure}[h!]
\centering
\includegraphics[width=0.45\textwidth]{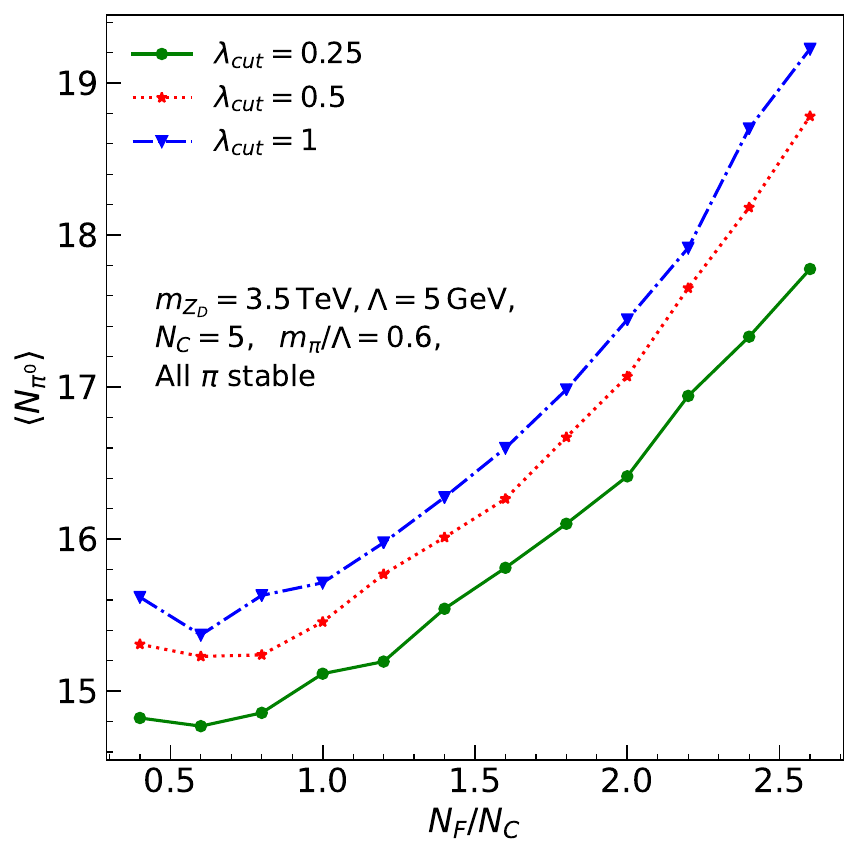}
\includegraphics[width=0.45\textwidth]{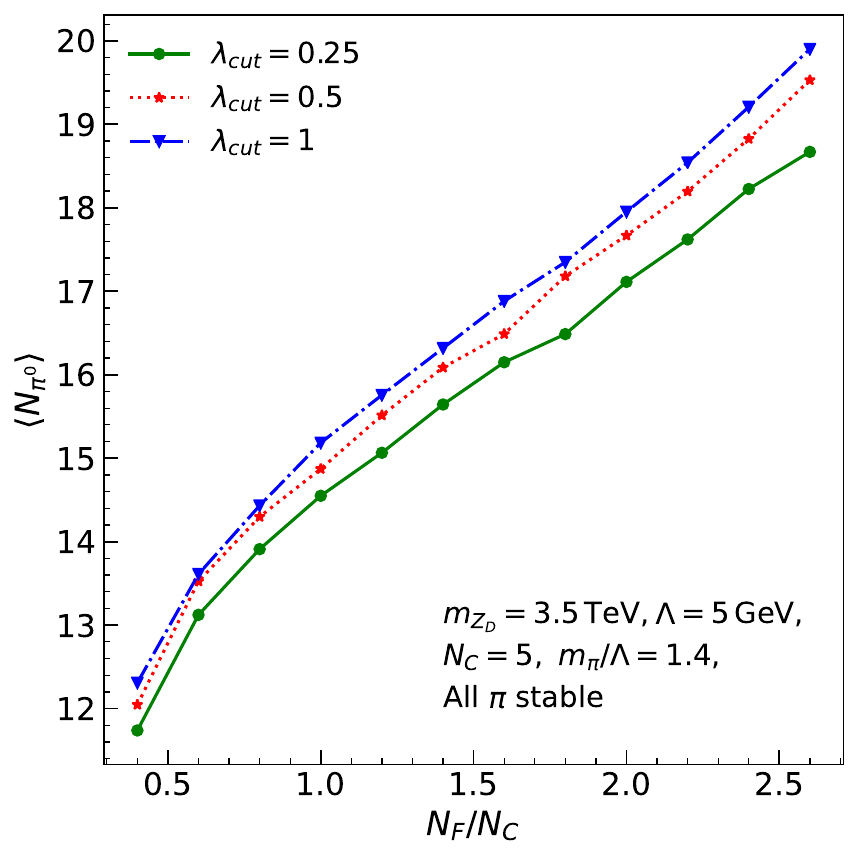}
\caption{Variation of diagonal pion multiplicity against $\fc$ for two benchmarks of $\mpl =0.6$ (left panel) and $\mpl = 1.4$ (right panel) for three values of $\lambda_{\rm{cutoff}} = 1,0.5, 0.25$. The lower the value of $\lambda_{\rm{cutoff}}$, the lower the multiplicity.}
\label{fig:thooft_cutoff_LLP}
\end{figure}
In fig.~\ref{fig:thooft_cutoff_LLP}, we correspondingly show the variation in the final-state diagonal pion multiplicity as a function of $\fc$ for the various shower-cutoffs considered in fig.~\ref{fig:thooft_cutoff} for two different values of $\mpl = 0.6$ (left panel) and $\mpl = 1.4$ (right panel). While the change in shower cut-off does affect the final-state diagonal pion multiplicity, wherein the multiplicity decreases for decreased $\lambda_{\rm{cutoff}}$, the variation is not strong as expected.

\begin{figure}[h!]
\centering
\includegraphics[width=0.46\textwidth]{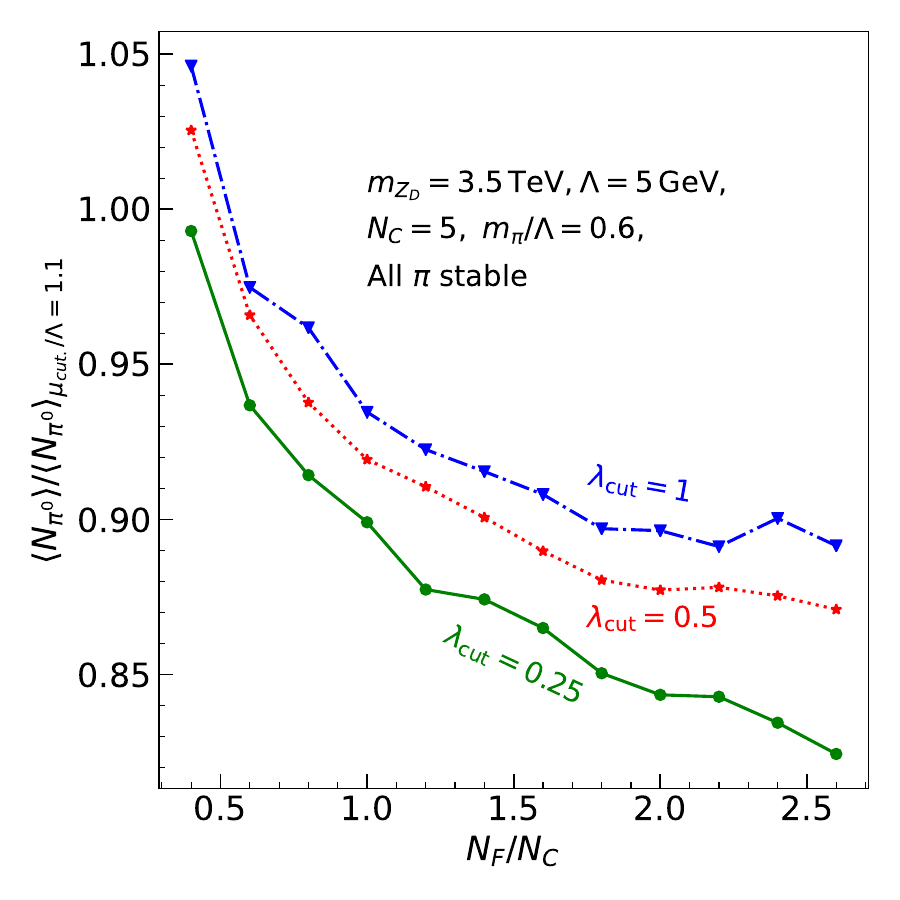}
\includegraphics[width=0.45\textwidth]{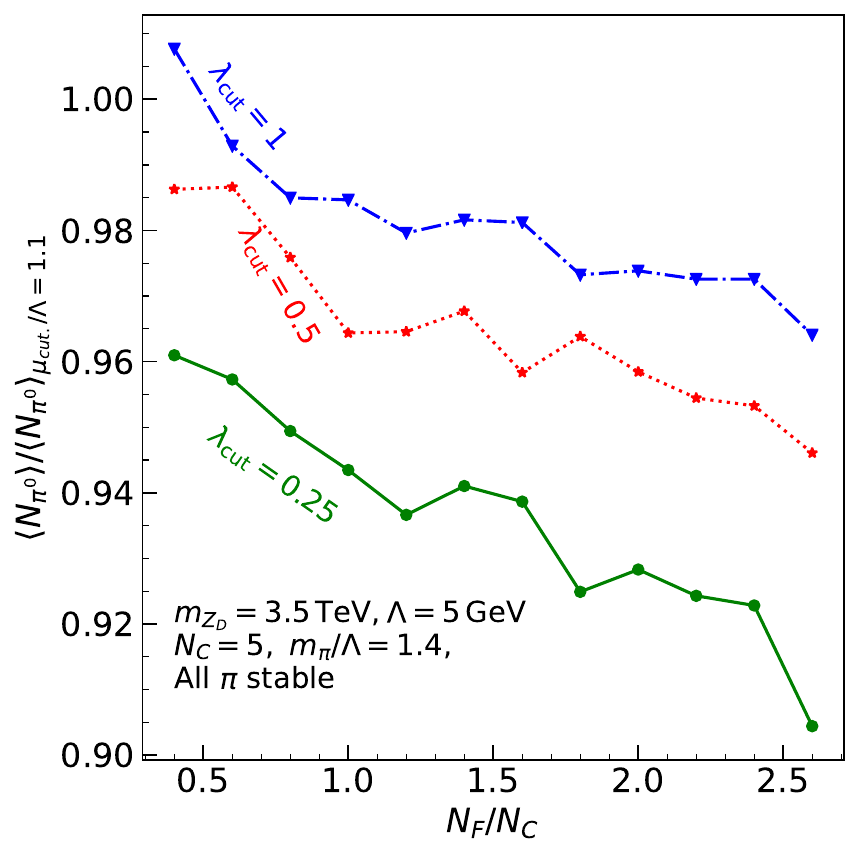}
\caption{$\langle N_{\pidon}\rangle/\langle N_{\pidon}\rangle_{\mu_{\rm{cutoff}/\ld = 1.1}}$ against $\fc$ for the various shower-cutoffs considered in fig.~\ref{fig:thooft_cutoff} for two different values of $\mpl = 0.6$ (left panel) and $\mpl = 1.4$ (right panel). Cutoff dependence only becomes significant for small $\lambda_{\rm{cutoff}}$, especially at large values of $\fc$.}
\label{fig:thooft_cutoff_LLP_relative}
\end{figure}

The actual effect of the differing $\lambda_{\rm{cutoff}}$ is hidden because even without an $\fc$ dependent cutoff, $\langle N_{\pidon}\rangle$ still increases with $\fc$ for all $\mpl\gtrsim0.25$. By dividing the results of fig.~\ref{fig:thooft_cutoff_LLP} by $\langle N_{\pidon}\rangle_{\mu_{\rm{cutoff}/\ld = 1.1}}$ -- the current \texttt{PYTHIA} default cutoff -- we can negate any increase of multiplicity with $\fc$. We show this in fig.~\ref{fig:thooft_cutoff_LLP_relative} where this is plotted against $\fc$ for the various shower-cutoffs considered in fig.~\ref{fig:thooft_cutoff} for two different values of $\mpl = 0.6$ (left panel) and $\mpl = 1.4$ (right panel). We see that, relative to $\langle N_{\pidon}\rangle_{\mu_{\rm{cutoff}/\ld = 1.1}}$, $\langle N_{\pidon}\rangle$ actually decreases with $\fc$ for both benchmarks. The decrease is significant for all considered $\lambda_{\rm{cutoff}}$, with it being most obvious for small $\lambda_{\rm{cutoff}}$ where for $\mpl=0.6$ it deviates up to $\sim15\%$. 

\begin{figure}[h!]
\centering
\includegraphics[width=0.45\textwidth]{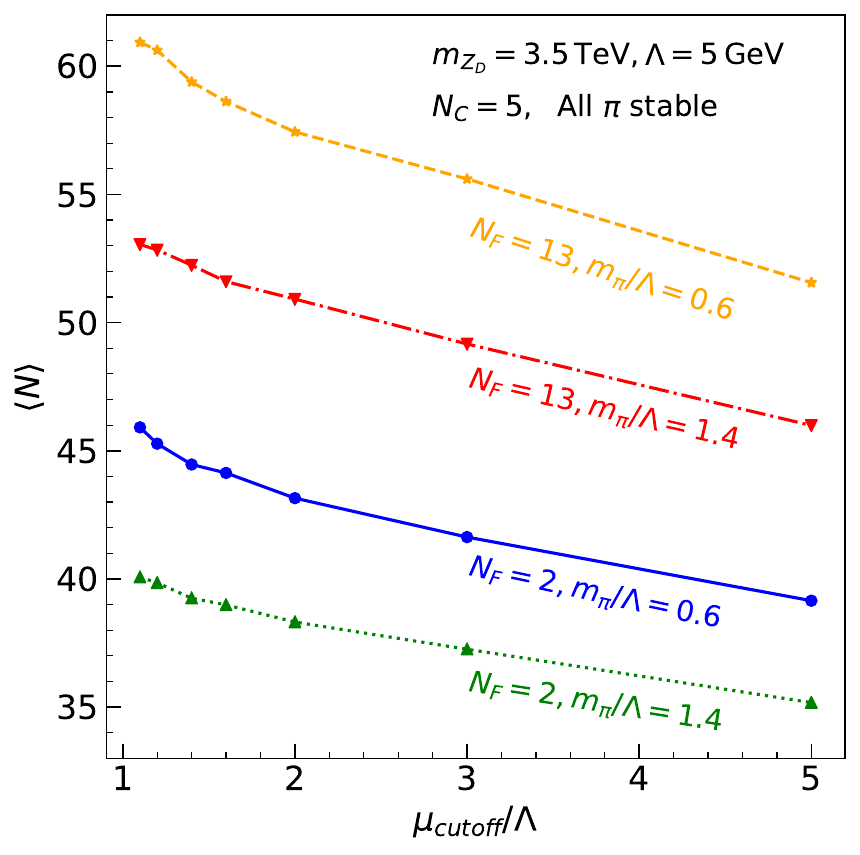}
\includegraphics[width=0.45\textwidth]{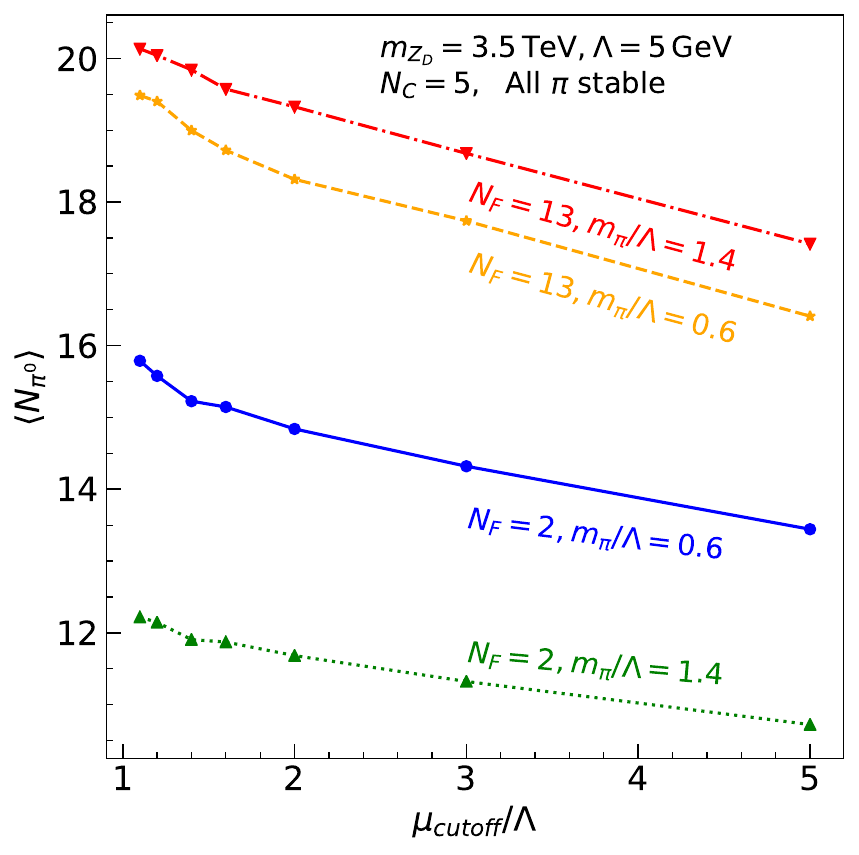}
\caption{Variation of all (left panel) and diagonal (right panel) pion multiplicity as a function of $\mu_{\rm{cutoff}}/\ld$ for various values of $\fc$ and $\mpl$.}
\label{fig:cutoff}
\end{figure}

In a full event-generation picture, the shower-cutoff is not an isolated quantity but should rather be coupled together with the variation of hadronization parameters. Such a comprehensive study is beyond the scope of this work. However, we complement our previous investigation by studying the final-state diagonal and the total dark pion multiplicities, $\langle N_{\pidon}\rangle$ and $\langle N_{\rm{tot.}}\rangle$ respectively,  as a function of $\mu_{\rm{cutoff}}/\ld$, where we vary it over a large range irrespective of the strength of the 't Hooft coupling. The results are shown in fig.~\ref{fig:cutoff} for various $\fc$ and $\mpl$. As $\mu_{\rm{cutoff}}/\ld$ increases, the shower is cutoff earlier and thus the resulting pion multiplicity decreases. The most dramatic effects are observed for small $\mu_{\rm{cutoff}}/\ld$, where the decrease is significantly more pronounced. These results are consistent with cutoff effects on multiplicities varying logarithmically.

Additionally, this implies that the uncertainties introduced by shower cutoff are substantial and should carefully be accounted for in the experimental analyses and theory studies. 

\section{Fixed probvector}
\label{app:fixed_probvec}

\begin{figure}[h!]
\centering
\includegraphics[width=0.49\textwidth]{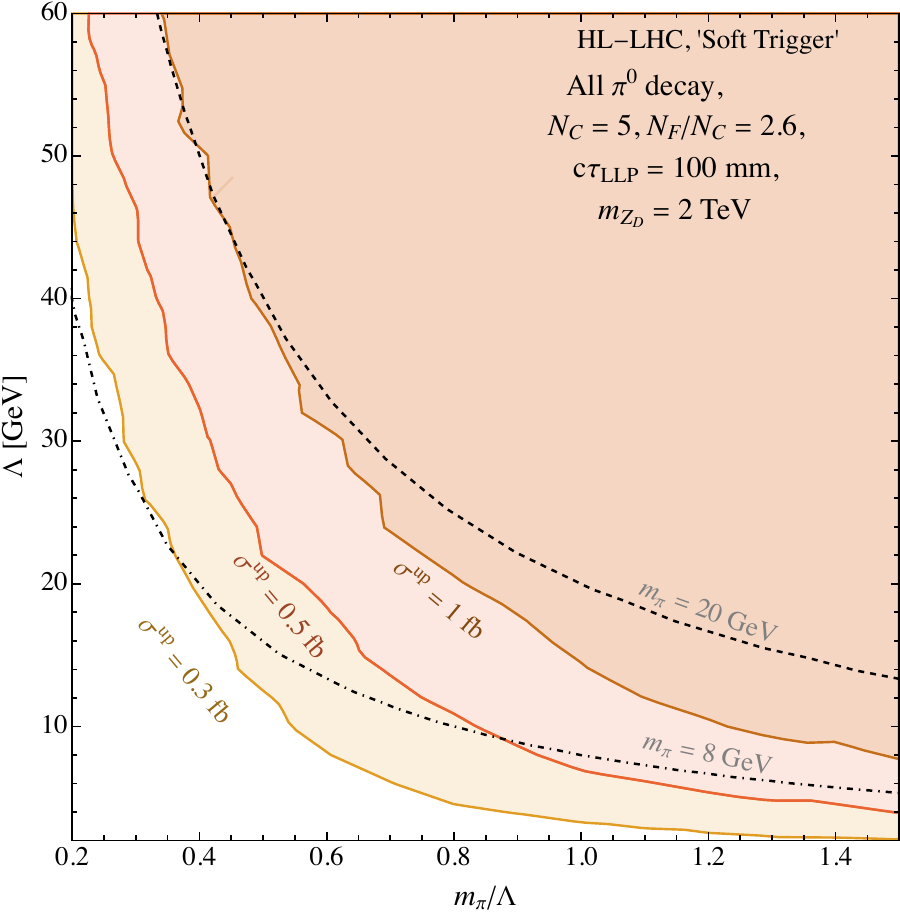}
\includegraphics[width=0.49\textwidth]{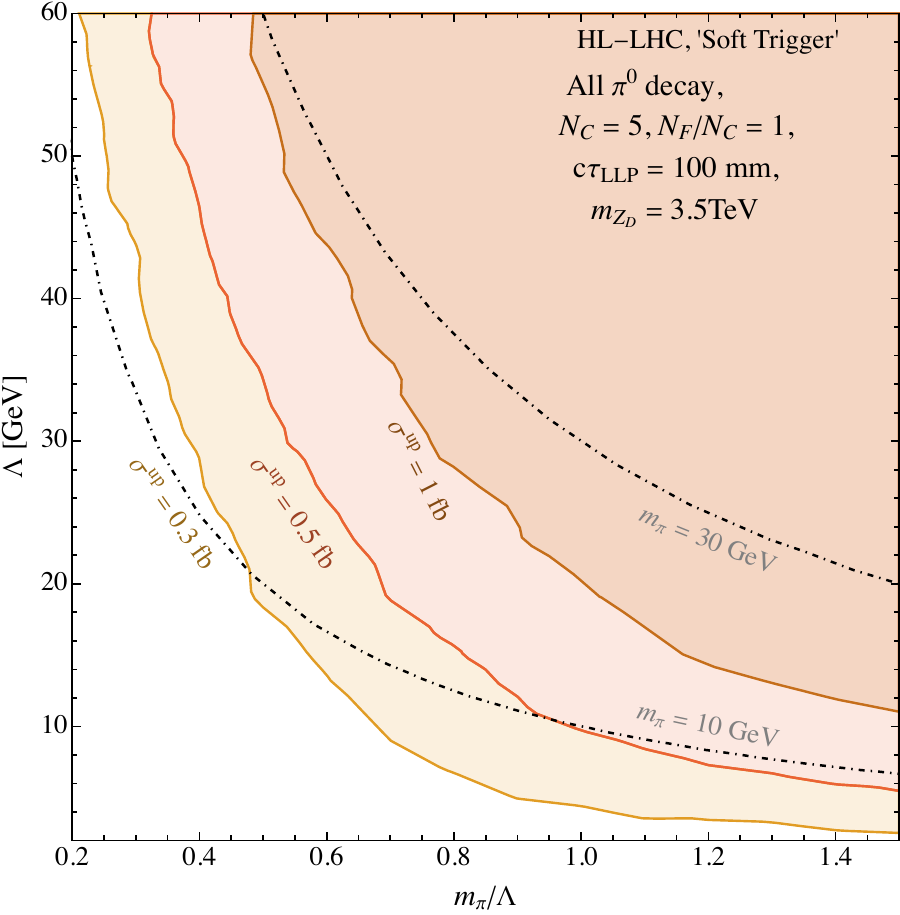}
\caption{ Same as Fig.~\ref{fig:mpi_lam_HLLHC} but for fixed {\tt probVector}.}
\label{fig:mpi_lam_HLLHC_fix}
\end{figure}

\begin{figure}[h!]
\centering
\includegraphics[width=0.49\textwidth]{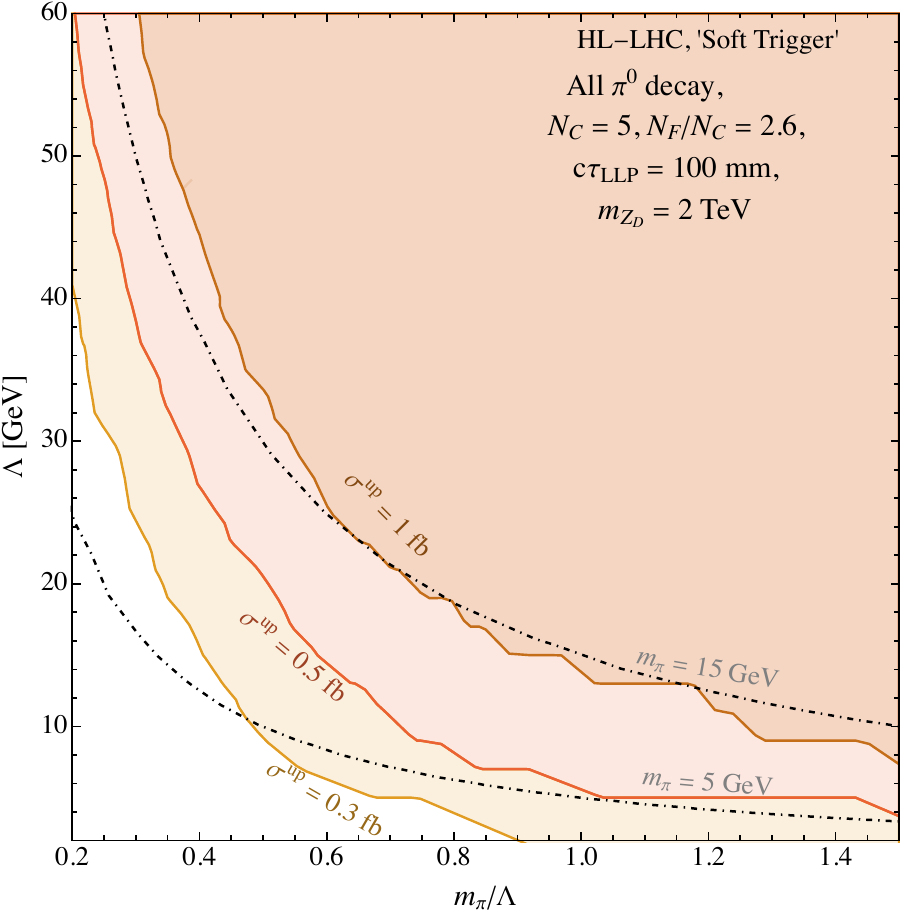}
\includegraphics[width=0.49\textwidth]{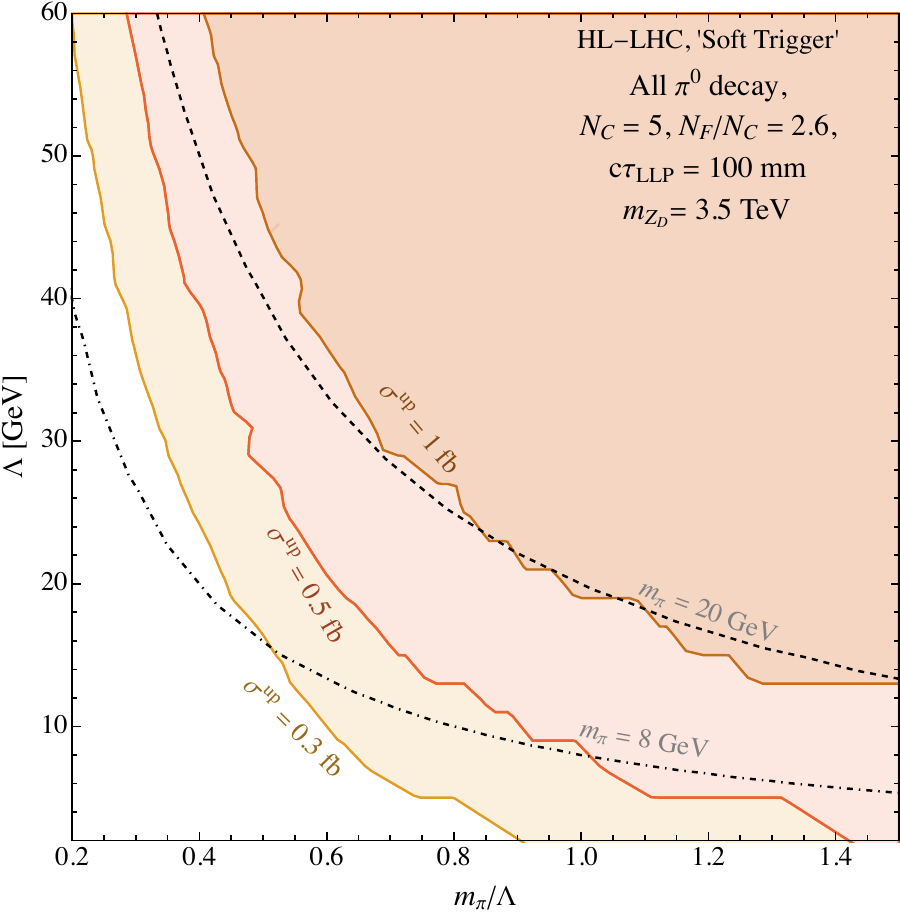}
\caption{Same as Fig.~\ref{fig:mpi_lam_HLLHC_26} but for fixed {\tt probVector}.}
\label{fig:mpi_lam_HLLHC_26_fix}
\end{figure}

Given that our {\tt probVector} fits in appendix~\ref{app:probvec_modelling} rely on the assumption of a thermal spectrum for pion and rho mesons, in fig.~\ref{fig:mpi_lam_HLLHC_fix} and ~\ref{fig:mpi_lam_HLLHC_26_fix} we show the model independent limits for HL-LHC with $\fc = 1, 2.6$ and $\mzp = 2, 3.5\,\rm{TeV}$ derived assuming a fixed {\tt probVector}. The largest change in the shape of the upper limits is understandably at small $\mpl$, where previously a turnover behavior due to fast changing {\tt probVector} was observed. At large $\mpl$ the thermal fits we use had an approximately constant {\tt probVector}, which remains true now, albeit its exact value is now reduced. This leads to minor changes in the quantitative limits derived but the qualitative conclusions at large $\mpl$ remain the same. 
%
\section{SM dijet constraints}
\label{app:theory_analysis}
We rescale the existing limits from LHC dijet searches in order to derive constrains on $Z_D$ to SM quark coupling. We sketch the associated procedure below. 

Given the Lagrangian in eqn.~\eqref{eq:lagrangian}, the total $Z_D$ width away from any thresholds is 
\begin{equation}
    \frac{\Gamma_{\rm tot}}{m_{Z_D}} = \frac{1}{12\pi}\left(N^{\rm SM}_C\times N^{\rm SM}_F \times g^2_{\rm SM} + \nc \times (g^{tot}_D)^2\right) \lesssim 0.1
\end{equation}
We have the existing limits from resonance searches on $g_q$ of the $Z_q$ which only couples to SM quarks, and $BR(Z_q \rightarrow q \bar{q}) \approx 1$~\cite{ATLAS:2024kpy}. To recast the limits into our model, the cross section must be the same,
we need to make sure 
\begin{align}
g_{\rm SM}^2 \times BR(Z_D \rightarrow q \bar{q}) = g_{\rm q}^2  \times BR(Z_{\rm q} \rightarrow q \bar q),
\label{eq:recast}
\end{align}
assuming the width is the same.

Neglecting phase space factors, and in the limit $g_D \gg g_{\rm SM}$ with $N_C \sim 3$ as in SM,
\begin{eqnarray}
BR(Z_D \rightarrow q \bar{q})  &\approx&  \frac{N^{\rm SM}_C\times N^{\rm SM}_F \times g_{SM}^2}{\Gamma_{\rm tot} } \\ 
BR(Z_D \rightarrow q_D \bar{q}_D)  &\approx& \frac{(g^{tot}_D)^2 \times N_C} {\Gamma_{\rm tot} }, 
\end{eqnarray}
where, $N^{\rm SM}_F = 6$ are the number of light SM quark flavors.

To recast the limit, from Eq.~\ref{eq:recast} we have
\begin{equation}
g_{\rm SM} \approx g_{\rm q} \times \sqrt{\frac{BR(Z_{\rm q} \rightarrow q \bar q)}{BR(Z_{D} \rightarrow q \bar{q})}}  
\approx \sqrt{\frac{3 \times g_{\rm q}^2 + \sqrt{g_{\rm q}^2 (9 g_{\rm q}^2 + 2 (g^{tot}_D)^2 N_C )} }{6}}
\label{eq:gsm}
\end{equation}
where $g_{\rm q}$ takes the upper limit from resonance searches for $Z_q$ as in Ref.~\cite{ATLAS:2024kpy}. Note that $g_D \equiv \sqrt{\mathcal{Q}^2_D}\kappa_D$, thus it contains the magnitude of the dark $U(1)_D$ charge vector.
For $g_D \lesssim 1$, our maximal allowed $g_D$ fixes the maximal allowed value of $g_{\rm SM}$.

The cross section of $\sigma(pp \rightarrow Z_D \rightarrow q_D q_D )$ can be obtained from a reference coupling,
\begin{equation}
    \sigma(pp \rightarrow Z_D \rightarrow q_D q_D ) \approx \frac{g_{SM}^2}{ g_{\rm SM,ref}^2} \times \sigma(pp \rightarrow Z_D)_{\rm ref} \times Br(Z_D \rightarrow q_D q_D).
\label{eq:cs}
\end{equation}
For example, $\sigma(pp \rightarrow Z_D)_{\rm ref}$ for $g_{\rm SM, ref} =$ 0.01 
can be obtained via Monte-Carlo simulation using {\tt  MadGraph@aMC}~\cite{Alwall:2014hca}, after implementing the Universal FeynRules Output~({\tt UFO}) files~\cite{Degrande:2011ua} of {\tt DMsimp}~\cite{Mattelaer:2015haa}.

In Tab.~\ref{tab:limit}, we show the upper limits of $\sigma(pp \rightarrow Z_D \rightarrow q_D q_D)$ following Eq.~\ref{eq:cs}.

\begin{table}[h!]
\centering
\begin{tabular}{ |c|c|c|c|c| } 
\hline
$M_{Z_{q,D}}$ & $g_{\rm q}$ & $g_{\rm SM}$ & $\sigma(pp \rightarrow Z_D)_{\rm ref}$ & $\sigma(pp \rightarrow Z_D \rightarrow q_D q_D)$  \\
\hline
2 TeV & $\lesssim 0.06~(0.03)$ & $\lesssim 0.18~(0.13)$ & 1.0 fb & $\lesssim$ 320~(170) fb  \\
\hline
3.5 TeV & $\lesssim 0.15~(0.07)$ & $\lesssim 0.30~(0.20)$ & 0.03 fb & $\lesssim$ 27~(12) fb  \\
\hline
\end{tabular}
\caption{For $M_{Z_{q,D}} = 2, 3.5$ TeV, the upper limit of $g_{\rm q}$ from resonance searches for $Z_q$ at LHC~(HL-LHC, from scaling luminosity) as in Ref.~\cite{ATLAS:2024kpy} is listed. The corresponding $g_{\rm SM}$ range which will lead to $g_D \lesssim$ 1 is also listed. $\sigma(pp \rightarrow Z_D)_{\rm ref}$ for $g_{\rm SM, ref} =$ 0.01 is provided after simulation. Finally, the range of $\sigma(pp \rightarrow Z_D \rightarrow q_D q_D)$ in order to make $g_D \lesssim 1$ derived from Eq.~\ref{eq:cs} is shown.
We fix $N_C$ = 5. }
\label{tab:limit}
\end{table}

We can also transfer the limits on $\sigma(pp \rightarrow Z_D \rightarrow q_D q_D)$ to $g_D$. Using Eq.~\ref{eq:cs}, 

\begin{equation}
g_{\rm SM,ref}^2 \times \frac{\sigma(pp \rightarrow Z_D \rightarrow q_D q_D)}{\sigma(pp \rightarrow Z_D)_{\rm ref} \times Br(Z_D \rightarrow q_D q_D)}  =  g_{\rm SM}^2. 
\label{eq:tran}
\end{equation}
As $BR(Z_D \rightarrow q_D q_D) \approx 1$, we get the upper limit on $g_{SM}$ when taking $\sigma(pp \rightarrow Z_D \rightarrow q_D q_D)$ as the upper limit. Taking $g_D \gg g_q$ and $N_C =$ 5, from Eq.~\ref{eq:gsm}, we roughly have $g_{\rm SM} \approx \sqrt{g_q g^{tot}_D/2}$. So we solve Eq.~\ref{eq:gsm} and \ref{eq:tran},
and finally we have 
\begin{equation}
 g^{tot}_D \approx \frac{2 \sigma(pp \rightarrow Z_D \rightarrow q_D q_D)}{ \sigma(pp \rightarrow Z_D)_{\rm ref}} \times \frac{g_{\rm SM,ref}^2}{ g_{\rm q}}.
 \label{eq:gd_coupling_limit}
\end{equation}
When $\sigma(pp \rightarrow Z_D \rightarrow q_D q_D)$ takes the upper limits from the displaced shower search, the corresponding upper limits of $g^{tot}_D$ is obtained. 

\bibliography{main}

\end{document}